\documentclass[useAMS,usenatbib]{mn2e}
\usepackage[dvipdfm]{graphicx}
\usepackage{psfrag, color, epsfig, rotating, pstricks, pst-node, ulem}
\usepackage{dsfont}
\usepackage{txfonts}
\usepackage{array}
\usepackage{multirow}
\input{psfig.sty}
\newcommand{\kms}{\>{\rm km}\,{\rm s}^{-1}}

\newcommand{\kpc}{\>{\rm kpc}}
\newcommand{\Msun}{\>{\rm M_{\odot}}}

\def\gtsima{$\; \buildrel > \over \sim \;$}
\def\ltsima{$\; \buildrel < \over \sim \;$}
\def\prosima{$\; \buildrel \propto \over \sim \;$}
\def\gsim{\lower.7ex\hbox{\gtsima}}
\def\lsim{\lower.7ex\hbox{\ltsima}}
\def\simgt{\lower.7ex\hbox{\gtsima}}
\def\simlt{\lower.7ex\hbox{\ltsima}}
\def\simpr{\lower.7ex\hbox{\prosima}}
\def\la{\lsim}
\def\ga{\gsim}

\newcommand{\mnras}{MNRAS}

\newcommand{\sersic}{S{\'e}rsic}

\newdimen\hssize
\newdimen\hdsize
\newdimen\figsize
\def \figsmall{2.5cm}

\def\mnras{MNRAS}

\def\beq{\begin{equation}}
\def\eeq{\end{equation}}
\def\bey{\begin{eqnarray}}
\def\eey{\end{eqnarray}}
\def\beqarray{\begin{eqnarray}}
\def\eeqarray{\end{eqnarray}}

\def\mpc{\,{\rm {Mpc}}}

\def\kpc{\,{\rm {kpc}}}

\def\kms{\,{\rm {km\, s^{-1}}}}
\def\msun{M_\odot}

\def\v200{V_{200}}

\def\my{\,{\rm M_\odot\, yr^{-1}}}

\def\L850{L_{850\rm \mu m}}

\def\sersic{\rm S{\rm \acute{e}}rsic}

\newcommand{\PreserveBackslash}[1]{\let\temp=\\#1\let\\=\temp}
\newcolumntype{C}[1]{>{\PreserveBackslash\centering}p{#1}}
\newcolumntype{R}[1]{>{\PreserveBackslash\raggedleft}p{#1}}
\newcolumntype{L}[1]{>{\PreserveBackslash\raggedright}p{#1}}

\begin{document}


\title
[SEDs and morphologies of LBGs at $z\sim 1$] {SED-inferred
properties and morphology of Lyman-break galaxies at $z\sim 1$ in
the CDF-S}
\author[Z. Chen, C. G. Shu,  D. Burgarella, V. Buat, J. -S. Huang \& Z. J. Luo]{Z. Chen$^{1,2,3}$, C. G.
Shu$^{2,1}$\thanks{E-mails: {cgshu@shao.ac.cn (CS)},
monachen@shao.ac.cn (ZC), {denis.burgarella@oamp.fr (DB)},
{veronique.buat@oamp.fr (VB)}, {jhuang@cfa.harvard.edu (JSH)},
{zjluo@shao.ac.cn}}, D. Burgarella$^{3}$,
V. Buat$^{3}$, J. -S. Huang$^{4}$ \& Z. J. Luo$^2$\\
$^{1}$Shanghai Astronomical Observatory, 80 Nandan Rd, Shanghai, 200030, China\\
$^{2}$Shanghai Key Lab for astrophysics, Shanghai Normal University,
100 Guilin Road,Shanghai,
200234,China\\
$^{3}$Observatoire Astronomique Marseille Provence, Laboratoire
d'Astrophysique de Marseille, 13012 Marseille,
France\\
$^{4}$Harvard-Smithsonian Center for Astrophysics, MS 65, 60 Garden
Street, US Cambridge, MA 02138, USA}


\date{}

\maketitle



\begin{abstract}

After carefully cross-identifying a previously discovered
GALEX-selected Lyman Break Galaxy (LBG) candidates one-to-one with
their optical counterparts in the field of the CDF-S, we re-estimate
their photometric redshifts using multi-wavelength data from UV,
optical to NIR. With the consideration of their re-estimated
photometric redshifts and SEDs, we refine a new updated sample of
383 LBGs at $0.7\la z \la 1.4$ with two confirmed AGNs being
excluded.

260 and 111 LBGs are classified as starburst and irregular types.
Ages of LBGs spread from several Myr to 1.5Gyr with a median of
$\sim50$Myr. Their dust-corrected star formation rates (SFRs) and
stellar masses ($M_*$) are from $4\my$ to $220\my$ and from
$2.3\times 10^8 \msun$ to $4 \times 10^{11} \msun$ with median
values of $\sim 25\my$ and $\sim 10^{10} \msun$. The rest-frame FUV
luminosity function of LBGs are presented with the best-fit
Schechter parameters of $\alpha = -1.61\pm 0.40 $, $\rm M^* = -20.40
\pm 0.22$ and $\phi^* = (0.89\pm 0.30)\times 10^{-3}\mpc^{-3}{\rm
dex^{-1}}$, respectively.

LBGs of irregular types mainly distribute along the ``main sequence"
of star forming galaxies while most LBGs of starburst types locate
in the starburst region. Together with previous studies, we suggest
that the star formation mode for LBGs at $z>3$ is mainly starburst
and it evolves to be more significant to the quenching mode after
$z\sim 3$.  A ``downsizing" effect is clearly found with its
physical implications and the comparisons with previous studies
being discussed in detail. LBGs with larger SFRs are on average more
compact. In the rest-frame color $(U-B)$-$M_*$ diagram, LBGs
distribute in the ``blue" cloud. We suggest that LBGs may evolve
along the blue ``cloud" from later to earlier types.


HST images in F606W ($V$ band) and F850LP ($z$ band) are taken from
the GEMS and GOODS-S surveys for morphological studies of LBGs.
SExtractor and GALFIT are applied to get their morphological
parameters. We establish an image gallery of 277 LBGs commonly
detected in both bands by visually classifying individual LBGs as
types of ``chain", ``spiral", ``tadpole", ``bulge" and ``clump",
respectively. A morphological sample of 142 LBGs with reliable
results of $\sersic$ and sizes in both bands is defined. We find
that LBGs at $z\sim 1 $ are dominated by disk-like galaxies, with
their median sizes of $2.34\kpc$ and $2.68\kpc$ in F606W and F850LP.
Correlations between photometric and morphological properties of
LBGs are investigated. Strong correlations between their half-light
radii and $M_*$, i.e., size-stellar mass relations, are found in
both bands. Physical connections between correlations and the
``downsizing" effect are discussed.

\end{abstract}


\begin{keywords}

galaxies: evolution - galaxies: Lyman break galaxies - galaxies:
SEDs - galaxies: morphology

\end{keywords}


\section{Introduction}
\label{sec:Introduction}


It is well established that massive galaxies are results of mergers
of small ones (\citealt{White91}; \citealt{Cole00};
\citealt{Somerville01}).
Disks and ellipticals form due to gas angular momentum by tidal
torques and through mergers among disks, respectively (e.g.,
\citealt{Mo98}; \citealt{E00}; \citealt{Steinmetz02};
\citealt{Governato07}; \citealt{Croft09}; \citealt{Scannapieco09}).
The epoch of major mergers ends at  $z\sim 2$ for the brightest and
most massive systems (\citealt{Conselice03} and the references
therein).
Although this hierarchical scenario of galaxy formation and
evolution has archived great successes to explain a lot of
observational phenomena in general, there still exist some poorly
known fields, such as the role of AGN feedback, star formation,
morphological evolution, etc. Details can be seen in the the recent
review by \cite{Silk12}.

%

Currently there are two regimes that we understand the global
evolution of galaxies well. One is the local universe where disk and
elliptical galaxies have been studied in very detail for several ten
years (see \citealt{Longair06} for a detailed review; \citealt{de la
Fuente Marcos09}; \citealt{van Dokkum10}; \citealt{Bell10};
\citealt{Boissier10}; \citealt{Tempel11}). Especially after the
two-degree Field Galaxy Redshift Survey (2dFGRS,
\citealt{Colless01}) and the Sloan Digital Sky Survey (SDSS,
\citealt{York00}) which provide huge multi-color and spectroscopic
galaxy samples with big sky coverages, many important and global
statistical results have been obtained and taken as the ``zero
points" to calibrate the paradigm of galaxy formation and evolution
as a whole (e.g., \citealt{Blanton03, Blanton05};
\citealt{Blanton06}; \citealt{Shen03}; \citealt{Tremonti04};
\citealt{Chang06}; \citealt{Shao07}; \citealt{Salim07}; Kauffmann et
al. 2003a,b; \citealt{Kauffmann08}; \citealt{Abazajian09};
\citealt{Li08}; \citealt{Li09}).
%
Together with deep surveys, such as COMBO 17 (\citealt{Wolf04}),
DEEP2 (\citealt{Madgwick03, Davis03}) and COSMOS
(\citealt{Scoville07}),
samples of more than $10^4$ galaxies with either photometric or
spectroscopic redshifts from local to $z\sim 1$
have shown the picture of galaxy evolution
from half-age of the universe to the present day (\citealt{Bell04,
Bell06}; \citealt{Borch06}; \citealt{Faber07}; \citealt{Gerke07};
\citealt{Conroy07}; \citealt{Scoville07}; \citealt{Yan09};
\citealt{Jahnke09}; \citealt{Salim09}; \citealt{Maier09};
\citealt{K10}; \citealt{Coppa10}; \citealt{Cucciati10};
\citealt{Cooper11}).

The other is at high redshift ($z \sim 2.5$), where color-selected
samples of galaxies, pioneering by \citet{Steidel95}, have been
established,
such as Lyman break galaxies (\citealt{Adelberger98}), sub-mm
galaxies (\citealt{Smail04}), distant red galaxies
(\citealt{Franx03}) and BzK galaxies (\citealt{Daddi04}).
For instance, more than 1000 Lyman break galaxies (hereafter LBGs)
at $z\sim 3$ have been found and confirmed by the follow-up
spectroscopic observations with further studies
(\citealt{Steidel93}; \citealt{Steidel96}; \citealt{Pettini00};
\citealt{Teplitz00}; \citealt{Giavalisco02}; \citealt{Steidel03}).
Applying the Lyman break technique to different bands, star forming
galaxies at different redshift ranges are also successfully found.
\citet{Wilkins10}, for instance, found 6 candidates of star forming
galaxies at $z\sim 7$ by comparing images of the Y (980nm) band in
WFC3 with those of z (850nm) band in ACS from the observations of
Hubble Space Telescope {\it (HST)}. Adopting the multi-wavelength
imaging data of the European Southern Observatory (ESO) Remote
Galaxy Survey, \citet{Douglas09} established a sample of 257 LBGs at
$z\sim 5$. Using $B, V, R, i'$, and $z'$ filters in the SUBARU deep
field, \citet{Ouchi04} compiled a large sample of 2600 LBGs with
$i'\la 27$ at $z\sim 4.5$.
Based on {\it FUV} and {\it NUV} data in the Galaxy Evolution
Explorer {\it (GALEX)} deep fields,
\citet{Burgarella06,Burgarella07} obtained a very deep sample of LBG
candidates at  $z\sim 1$ cross-correlated with optical data from
COMBO 17.
%
\citet{Nilsson11a} analyzed slitless grism observations by {\it HST}
ACS for 15 LBGs at $z \sim 1$ drawn from the sample of
\citet{Burgarella07} and concluded that they are young while massive
and dusty.
%
Also based on {\it GALEX} data but cross-correlated with optical
data
from 
SDSS, \citet{Haberzettl09} obtained a shallower sample of LBGs than
that of \citet{Burgarella06,Burgarella07} at $z\sim 1$ with the sky
coverage of 1 $\rm deg^2$.

Between these two regimes, i.e., $1.5\la z \la 2.5$
named as the ``redshift desert" (\citealt{Steidel04}), many efforts
have been devoted in recent years.
\citet{Genzel06,Genzel08,Genzel10,Genzel11}, for instance, studied
star forming galaxies at $z\sim 2$ according to the Spectroscopic
Imaging survey in the Near-infrared with SINFONI (SINS) at ESO's
Very Large Telescope (VLT). They found that disks appear at $z\sim
2$ with star forming regions being very clumpy. \citet{Huang09}
established
a sample of Ultra Luminous IR galaxies (ULIRGs)\footnote{ULIRGs
defined by their IR luminosities $10^{12} L_{\sun} < L_{\rm IR} <
10^{13} L_{\sun}$}
at $z\sim 1.9$ by the selection of IRAC color criterion and claimed
that ULIRGs at that redshift display variable morphologies in their
rest-frame UV band. \citet{Conselice11} found that massive galaxies
at $1.7<z<2.9$  show the bimodality in the color-magnitude space
according to deep NIR surveys utilizing the NICMOS on {\it HST}.
With the Photodetector Array Camera \& Spectrometer (PACS)
(\citealt{Poglitsch10}) onboard   newly launched {\it Herschel}
Space Telescope (\citealt{Pilbratt10}), \citet{Nordon10}
investigated star formation rates of massive galaxies at $1.5 < z <
2.5$ in the field of GOODS-N.
 By analyzing both their SEDs from rest-frame UV to
mid-IR and direct 160 $\mu$m FIR observations by PACS,
\citet{Oteo12} found that spectroscopically selected star-forming
Ly$\alpha$ emitting galaxies at $2.0 \la z \la 3.5$ show ages mostly
below 100 Myr with a wide variety of dust attenuations, SFRs, and
stellar masses, and their morphology is suggested to range from
bulge-like galaxies to highly clumpy systems.
According to the {\it Herschel} observations,
\citet{Magdis10b} investigated FIR properties of mid-IR-selected
ULIRGs at $z\sim 2$ and \citet{Bongiovanni10} studied FIR
counterparts of selected Ly$\alpha$ emitters at $z \sim 2.2$. By the
$NUV$ dropout from {\it GALEX} data, \citet{Burgarella11}
established a sample of LBGs at $z\sim 2$ and analyzed their FIR
luminosities.
\citet{Nilsson11b} studied Ly$\alpha$ emitters at $z \sim 2.3$ in
the COSMOS field and found that their stellar masses are higher than
those at higher redshift.
Taking the similar method, \citet{Haberzettl12} established a sample
of 73 LBG candidates at $z\sim 2$ recently, with the median stellar
mass of $\sim 10^{10}\Msun$ and most of them showing disk-like
structures.
 \citet{Basu-Zych11} studied LBGs at $0.5\la z \la2$
based on observations in the Chandra Deep Field South (CDF-S) by the
UV/Optical Telescope (UVOT) instrument on board Swift
(\citealt{Gehrels04}) together with multi-wavelength data. They
found that their LBGs have stellar masses slightly lower than those
at $z\sim 3$ and slightly higher than $z \sim 1$ CDF-S galaxies.

So far, one of the key questions in galaxy formation and evolution,
i.e., how galaxies evolve from high $z$ to present day,
arises after so many distant galaxies have been found. Most of the
current studies
focus on photometric properties,
such as color, spectrum features, size, mass and spatial clustering,
to understand the connections between high-$z$ and local galaxies.
For instance, spectra of some bright LBGs (\citealt{Pettini00};
\citealt{Teplitz00}) are remarkably similar to those of local
starbursts with their UV luminosities directly relating to the
number of short-lived, massive stars. Sub-mm observations reveal a
population of FIR-bright galaxies that might be similar to local
(Ultra) Luminous IR galaxies ((U)LIRGs)\footnote{LIRGs defined by
their IR luminosities $10^{11} L_{\sun} < L_{\rm IR} < 10^{12}
L_{\sun}$} (\citealt{Blain99}). Based on the correlation length of
clustering, \citet{Shu01} suggested that LBGs and SCUBA galaxies at
$z\sim 3$ correspond to present-day bulges of disk galaxies (see
also \citealt{Giavalisco96}) and giant ellipticals, respectively.
\citet{Heckman05} and \citet{Hoopes07} suggested that local
($z<0.3$) Lyman break analogs have similar UV surface brightness to
high redshift LBGs, different from \citet{Scarpa07}.

%
Although most galaxies at high redshift must somehow be the
progenitors of modern galaxies, that connecting specific high
redshift populations to low-redshift counterparts remains a
complicated issue (\citealt{Delgado-Serrano10}). Yet, we still have
very little knowledge of when and how the modern Hubble sequence
came into place. The direct way is to investigate morphologies of
galaxies at different redshifts and compare with local galaxies. It
is not an easy task since it strongly depends on the resolutions and
sensitivities of the observing facilities. Some observations have
shown that morphologies of galaxies at $z>2.5$ appear very
irregular, very different from local either ellipticals or disks
(\citealt{Giavalisco96}). Familiar Hubble types are observed with
perhaps similar comoving volume densities at $z\sim 1$ to those at
$z\sim 0$ (\citealt{Bergh00}). Galaxies at high-$z$ are
characterized observationally by compact structures with high
surface brightness in their central regions and are harder to detect
their outer regions with lower surface brightness.
So it is difficult to analyze their morphological properties. As
pointed out by \citet{Burgarella01}, only compact star forming
regions could be easily detected in deep {\it HST} observations.

%


In the present paper, we study the photometric and morphological
properties of LBGs at $z\sim 1$ in detail, adopting broad
multiwavelength data from UV, optical to NIR together with high
quality {\it HST} images from the Galaxy Evolution from Morphologies
and SEDs (GEMS) survey and the Great Observatories Origins Deep
Survey (GOODS). The preliminary LBG candidate sample is taken from
\citet{Burgarella07}.
The structure of this paper is as follows. The description of the
worked sample is set in Sect. \ref{sec:sample}.
Detailed studies on photometric properties of LBGs
are arranged in Sect. \ref{sec:pho}. {\it HST} images obtained from
the GEMS and GOODS surveys and visual classifications are described
in Sect. \ref{sec:HST}. Morphological properties are studied in
Sect. \ref{sec:mor}. Correlations between photometric and
morphological properties of LBGs are in Sect. \ref{sec:cor} with
conclusions in Sect. \ref{sec:con}.

Throughout the paper the ``concordance" cosmology with $\Omega_{0} =
0.3$ and $\Omega_{\Lambda} = 0.7$ is taken. The Hubble constant is
$H_0 = 100\,h^{-1}\kms\mpc^{-1}$ with $h=0.7$ if needed. Under this
cosmology, the luminosity distance $D_L$ and the angular diameter
distance $D_A$ are $6.607\rm Gpc$ and $1.652\rm Gpc$, respectively
at $z=1$, with the size scale being $8.01\kpc ~\rm arcsec^{-1}$ and
the age of the universe being 5.75Gyr. Moreover, magnitudes are in
the AB system (\citealt{Oke83}) and colors are given in the
rest-frame unless otherwise stated.

\section{The LBG Sample}
\label{sec:sample}

\subsection{Preliminary sample and photometric data}
\label{subsec:pre-sample}

The preliminary  LBG candidate sample at redshift $z\sim 1$ used in
the present paper is described by \citet{Burgarella07} (hereafter
B07), and belongs to CDF-S. It can be briefly summarized as follows.
%
%
Based on the deep (76 444s) near ultraviolet ({\it NUV}) image
available in {\it GALEX} Release 2 (GR2) centered at $\alpha =
03^{\rm h}32^{\rm m}30^{\rm s}.7$ and $\delta =
-27^{\circ}52'16''.9$ (J2000.0), and the deep (44 668s) image survey
(DIS) in far ultraviolet ({\it FUV}) also by {\it GALEX}
\citep{Martin05},
B07 cross-identified  {\it GALEX} sources within a radius of 2
arcsecs to their optical counterparts
 in the field of 0.263deg$^2$ overlapped with COMBO 17
(\citealt{Wolf04}).
%
LBG candidates are selected among sources with $NUV \la 26.2$,
taking the criteria
$FUV - NUV >2$ and the photometric redshift data from COMBO 17 into
account.
%
%
Discarding objects classified in COMBO 17 with the flags as `Star',
`WD', `QSO(gal)' and `Strange Object', 420 LBG candidates, with the
completeness of $\sim$80\%
down to $NUV \sim 24.8$,
are
selected with their photometric redshifts from COMBO 17 between 0.9
and 1.3. Among them, two candidates have quadruple counterparts
found in COMBO 17, 40 have pair counterparts and the other 378 have
single counterparts individually. Moreover, 62 among 420 LBG
candidates have been detected in the 24$\mu$m MIPS image by {\it
Spitzer}.
Based on SED studies, B07 obtained a median age of 250Myr and a
median star formation rate of $\sim 30\my$ for their LBG candidates.
A typical dust attenuation in $FUV$ of $\rm A_{FUV} \sim 2.5$ is
estimated for candidates with 24$\mu {\rm m}$ detections, while the
other candidates without 24$\mu {\rm m}$ detections display less
with $\rm A_{FUV} \sim 1.8$. Details can be found in B07.
%

\begin{figure}
\resizebox{\hsize}{!}{\includegraphics{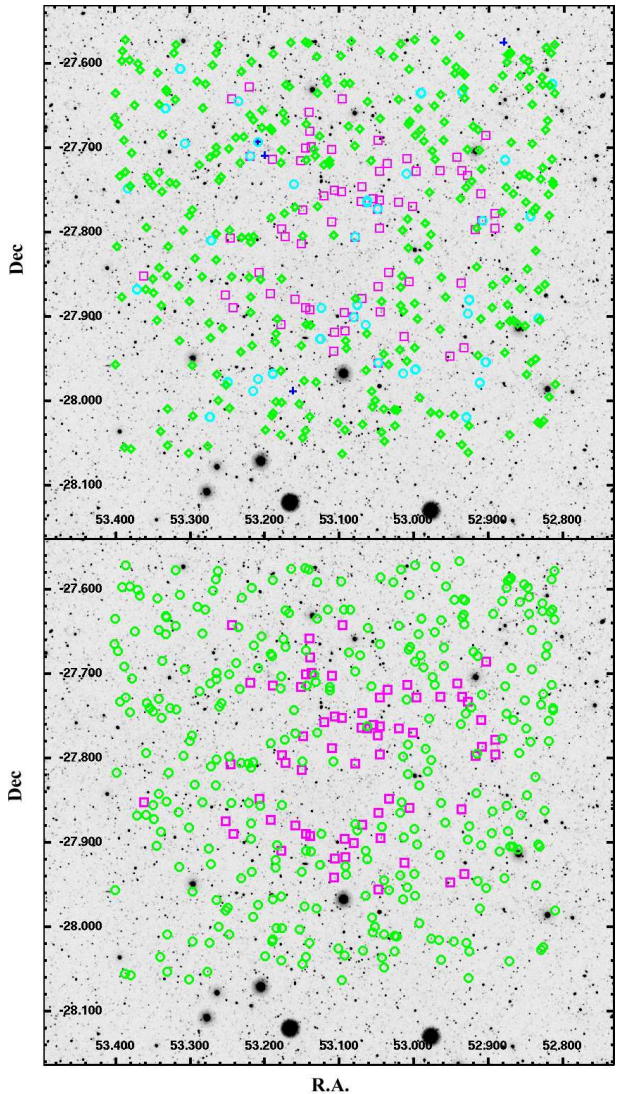}} \caption{(Upper
panel) LBG candidates of B07 in the field of the CDF-S with diamonds
``$\diamond$" and circles ``$\circ$", squares ``$\square$" and
crosses ``+" denoting candidates with single counterparts and
multiple counterparts in COMBO 17, spectral data available from the
MUSYC catalog and other previous studies, and AGNs respectively.
(Lower panel) Same as the upper panel but for the refined LBG sample
(Sect. \ref{sec:re-sam}) except circles ``$\circ$" denoting LBGs
with no spectral data available (see text for details).}
\label{fig:pos}
\end{figure}

Note that the MUSYC catalog (\citealt{Cardamone10}, see below in
this subsection for details), which is available more recently than
COMBO 17, can be also adopted to select LBG candidates. If we
replace the COMBO 17 catalog by the MUSYC catalog and take the same
processes as B07, 442 LBG candidates are found with more than 85\%
objects common to B07. The difference is not significant for the
statistical studies on the LBG population at $z\sim 1$.
%
%
Moreover, taking the updated catalog of COMBO 17 (\citealt{Wolf08})
into account, we find that more than $95\%$ LBG candidates in B07
can still be selected as LBG candidates and the mean offset of the
photometric redshifts is only 0.015.
Although the MUSYC catalog covers more bands, the COMBO 17 catalog
displays higher survey sensitivity and spectral resolution. We
prefer to keeping the selection of B07 for the following studies.
%
 It should be pointed out that the sample of B07 is
unbiased, much deeper and more complete than their previous ones
(\citealt{Burgarella06}).
For more clarity, we plot the distribution of B07 LBG candidates in
the field of the CDF-S in the upper panel of Fig. \ref{fig:pos},
with diamonds ``$\diamond$" and circles ``$\circ$", squares
``$\square$" and crosses ``+" denoting candidates with single
counterparts and multiple counterparts in COMBO 17, spectral data
available from the MUSYC catalog and other previous studies, and
AGNs respectively (see this section below).

Image resolutions of UV and IR observations are too low, e.g.,
$4.5''$ in {\it FUV} and $6''$ in {\it NUV} for $\it GALEX$, for us
to investigate morphological properties of LBGs at $z\sim 1$.
Optical images by {\it HST} are adopted for the morphological
studies (see Sect. \ref{sec:mor} for details). So, identifications
of optical counterparts for individual LBG candidates must be
carefully performed. We start to re-do the cross-identifications of
optical counterparts in the catalog of COMBO 17 for LBG candidates,
taking the same processes proposed by B07.
%
%
Since the space resolution of {\it GALEX} is 2.55 arcsecs (B07) and
the intrinsic astrometric precision of COMBO 17 is about 0.5 arcsecs
(\citealt{Wolf04}), together with the astrometric precision of
smaller than 0.2 arcsecs between the catalogs of COMBO 17 and MUSYC
(\citealt{Cardamone10}, see below), an appropriate radius of 3
arcsecs is taken in the present paper for cross-identifications.

Slightly different from B07, two LBG candidates in B07 with
quadruple optical counterparts are found as one with four
counterparts and the other one with two counterparts in COMBO 17.
One candidate with pair counterparts in B07 is found to have a
single counterpart. For the other LBG candidates, we have the same
results of optical counterparts found in COMBO 17 as B07. To
summarize, we find among 420
LBG candidates that 40 have pair, 1 has four and the other 379 (378
in B07) have single optical counterparts in COMBO 17, respectively.
In total, 463 optical counterparts are found.

It is important to check the contamination of Active
Galactic Nuclei (AGNs) in our LBG candidate sample.
According to \citet{Silverman10}, based on optical spectra of X-ray
sources, four confirmed AGNs are found among LBG candidates in B07,
with UV IDs 81300, 39096, 68162 and 69870. Their corresponding
spectral redshifts are 0.688, 0.652, 0.979, and 0.972, respectively.
These four AGNs
are classified as LBG candidates with single counterparts above.
Note that only three AGNs, which are among the above four, can be
found in the $Chandra$ 2 Ms CDF-S catalog (\citealt{Luo08, Luo10}).
It means that the AGN fraction in our LBG preliminary sample is
$\sim 1\%$, which is a bit less than those of LBGs at $z\sim 1.5$
(\citealt{Basu-Zych11}) and $z\sim 3$ (\citealt{Steidel03,
Lehmer08}).
Since we focus on photometric and morphological studies for LBGs,
they are excluded in the following part of the paper. So the number
of LBG candidates becomes 416 with 375 among them found to have
single optical counterparts in COMBO 17.

Since cross-identifications are done within a radius of 3 arcsecs
centered on the coordinates of individual LBG candidates, it can be
reasonably assumed that counterparts of those 375 LBG candidates
with only single counterparts are their ``true" optical counterparts
individually. For those candidates with more than one counterparts
in COMBO 17, more accurate estimates of their redshifts are needed
for further identifications.
To study in detail the photometric properties of LBGs at $z\sim 1$,
{\it FUV, NUV} and more wavelength observations must be considered
to re-estimate
their photometric redshifts
rather than simply taking the results from the optical catalog of
COMBO 17 only (see subsection below).

Thanks to the Multiwavelength Survey by Yale-Chile (MUSYC) {\it BVR}
selected catalog which covers the field of the present study with
multi-wavelength data available from optical to NIR together with
several tens of spectroscopic data from other observations
(\citealt{Cardamone10}).  The main features of the MUSYC {\it BVR}
selected catalog are as follows. In the catalog, {\it UBVRI} data
are from ESO archive collected and calibrated as part of GaBoDS (the
Garching-Bonn Deep survey, \citealt{Hildebrandt06}) and $z$-band
data are from the Mosaic-II camera on the CTIO 4m Blanco telescope
(\citealt{Gawiser06}; \citealt{ Taylor09}). $H$-band data are
obtained from SofI on the ESO NTT3.6 telescope (\citealt{Moy03}) and
$JK$-band data are from the ISPI camera also on the CTIO Blanco 4m
telescope (\citealt{Taylor09}). The IRAC data in four channels
($3.6\mu$m, 4.5$\mu$m, 5.8$\mu$m and 8.0$\mu$m) are from the {\it
Spitzer} IRAC/MUSYC public Legacy in the ECDF-S (SIMPLE) project
(\citealt{Damen11}),
respectively. Photometric redshifts of individual objects in MUSYC
are also listed. Note that the magnitude limit of the MUSYC {\it
BVR} catalog is $R~\la 25.3$, a bit shallower than COMBO 17 which is
$R~\la 26$ (\citealt{Wolf04,Faber07}). Details of the MUSYC {\it
BVR} catalog are referred to \citet{Cardamone10}.

\begin{table}
\caption{The cross-identification results of optical counterparts in
the COMBO 17 catalog and the correspondences in the MUSYC {\it BVR}
catalog for 416 LBG candidates (see text for details). The first and
second columns denote the number of counterparts of an LBG candidate
and the numbers of LBG candidates found in COMBO 17. The third
column denotes the results of cross-identifications from the COMBO
17 to MUSYC {\it BVR} catalogs and the fourth column shows the
number with the spectroscopical data available, respectively.}
\begin{tabular}{cclr}
\hline counterpart(s) & COMBO & MUSYC & $z_{\rm spe}$\\
\hline \hline single & 375 & 351 & 60 \\
\hline pair  & 40 & 36 single  & 9\\
&  & 2 pair  & 1 \\
\hline four  & 1  & 1 single  & /\\
\hline \hline total & 459 & 392 & 70\\
\hline
\end{tabular}
\textrm{} \label{tab:counter}
\end{table}

Aiming to extend the observational wavelength coverage of COMBO 17
for the photometric redshift re-estimates of LBG candidates,
we perform cross-identifications for the optical counterparts of LBG
candidates found in COMBO 17 to those in the MUSYC {\it BVR}
catalog. The astrometric precision less than 0.2 arcsecs between two
catalogs enables us to do cross-identifications without any
multi-counterpart problems from the COMBO 17 to MUSYC {\it BVR}
catalogs.
Since the magnitude limit is a bit shallower than that of COMBO 17,
less number of counterparts are found in the MUSYC  {\it BVR}
catalog. Among 375 LBG candidates with single counterparts found in
COMBO 17, 351 candidates are found with single counterparts and the
rest 24 candidates are found without any counterparts in the MUSYC
{\it BVR} catalog.
%
%
For 40 LBG candidates with pair counterparts in COMBO 17, 2 and 36
candidates are found with pairs and single counterparts available,
respectively,
while the rest 2 candidates are found with none
counterparts.
For the one candidate with 4 counterparts in COMBO 17, only one
counterpart is found in  the MUSYC {\it BVR} catalog.
Totally, 392 counterparts are found in the MUSYC {\it BVR} catalog
among 459 counterparts in COMBO 17 for 416 LBG candidates in B07.
There are 70 and 66 among these 459 and 392 counterparts in the
COMBO 17 and MUSYC {\it BVR} catalogs, respectively, with spectral
data available from the MUSYC {\it BVR} catalog, the K-selected
catalog of \citet{Taylor09} and the recent work done by
\citet{Nilsson11a}.

Detailed results of the cross-identifications are summarized in
Table \ref{tab:counter}. The first and second columns in Table
\ref{tab:counter} denote the number of counterparts of an LBG
candidate and the corresponding numbers of LBG candidates found in
COMBO 17. The third column denotes the results of
cross-identifications from the COMBO 17 to MUSYC {\it BVR} catalogs
and the fourth column shows the number with the spectroscopical data
available, respectively.

\subsection{Photometric redshifts of the preliminary sample}\label{subsec:fitting}

\begin{figure}
\resizebox{\hsize}{!} {\includegraphics{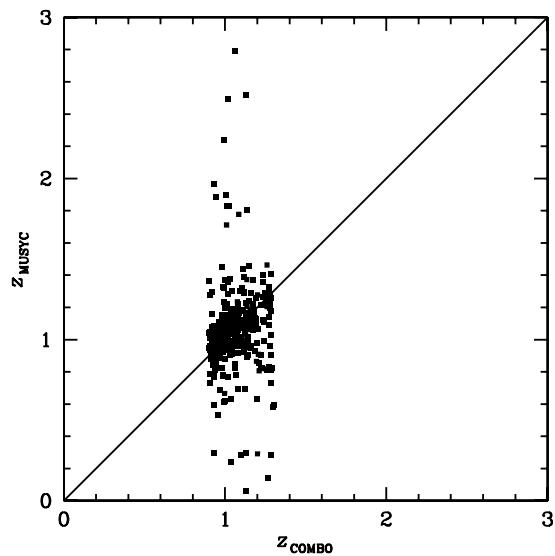}}
\caption{Comparison of photometric redshifts between $z_{\rm COMBO}$
from COMBO 17 and $z_{\rm MUSYC}$ from MUSYC for optical
counterparts of LBG candidates commonly available in two catalogs,
with the solid diagonal indicating where they would agree.}
\label{fig:zc-zm}
\end{figure}

As in the previous subsection, the preliminary LBG sample is
selected in UV together with
photometric redshifts from their optical counterparts in COMBO 17.
Most of the counterparts are also available in the MUSYC {\it BVR}
catalog with NIR and IR data.
The MUSYC {\it BVR} catalog covers observed spectral energy
distributions (SEDs) with more bands  while the COMBO 17 catalog
shows higher sensitivity and spectral coverage in general. Both of
them provide good estimates of photometric redshifts globally.
It is interesting to compare in Fig. \ref{fig:zc-zm} the photometric
redshifts of the 392 counterparts available in both the COMBO 17 and
the MUSYC {\it BVR} catalogs. Although the global accuracies of
photometric redshifts for both catalogs are high, it can be found
from the figure that many counterparts of the LBG candidates show
significantly different photometric redshifts between two catalogs.

It must be pointed out that  {\it FUV} and {\it NUV} data which are
adopted to select LBG candidates at $z\sim 1$ are not taken into
account for redshift estimates in either the COMBO 17 or the MUSYC
{\it BVR} catalog.
So,
%
it is worthy of re-estimating photometric redshifts for optical
counterparts of LBG candidates individually, combining {\it FUV} and
{\it NUV} data from B07 and data from MUSYC/COMBO 17 together.

We re-estimate photometric redshifts of individual counterparts for
LBG candidates by fitting their observed SEDs to
templates by the code
Hyperz\footnote{http://webast.ast.obs-mip.fr/hyperz/}
(\citealt{Bolzonella00}), taking the extinction law as
\citet{Calzetti00}. The templates adopted are
the BC03 (\citealt{BC03}) templates with the assumptions of an
exponentially decaying star formation history and a star formation
time scales  $\tau$ for different types (\citealt{Bolzonella00}),
the mean SEDs of local galaxies (\citealt{Coleman80}, hereafter CWW)
and the observed starburst templates of Arp220 types
(\citealt{Huang09}).
The spectral types
(with the star formation time scales $\tau$ if available)
and their corresponding SpT numbers of different templates are
listed in Table \ref{tab:spt} respectively.

 Considering survey sensitivity (\citealt{Wolf04,
Wolf08}), broad band data from COMBO 17 are preferred in the
following studies although intermediate-band data would have favored
a more accurate estimate of photometric redshifts. Moreover, the
calibration update of the COMBO 17 (\citealt{Wolf08}) has been taken
into account.
For the 392 counterparts of LBG candidates available in the both
COMBO 17 and MUSYC {\it BVR} catalogs (see Table \ref{tab:counter}),
we take data of 13 bands, i.e., {\it FUV, NUV, U, B, V, R, I, z, J,
H, K}, $3.6\mu$m and 4.5$\mu$m (the latter two from IRAC), to fit
their SEDs. For the rest 67 counterparts available only in the COMBO
17 catalog, data of 7 bands, i.e., {\it FUV, NUV, U, B, V, R} and
{\it I}, are taken. Since dust emission feathers are not taken into
account for the BC03 templates, we have ignored 5.8$\mu$m and
8.0$\mu$m data of IRAC for the specific case of galaxies at $z \sim
1$.

\begin{table}
\caption{The spectral types  (with the star formation time scales
$\tau$ if available) and their corresponding SpT numbers of
different templates adopted for SED fittings, with BC03 denoting
templates from \citet{BC03}, CWW from \citet{Coleman80}, Arp220 and
Arp220-modifying from \citet{Huang09}, respectively.}
\begin{center}

\begin{tabular}{rrr}
\hline
Spectral type (SpT) & Templates & $\tau$ (Gyr)\\
\hline \hline
 E (1) & BC03 & 1\\
   E (2) &  CWW & --\\
   S0 (3) & BC03 & 2\\
   Sa (4) & BC03 & 3\\
   Sb (5) & BC03 & 5\\
   Sbc (6) & CWW & --\\
   Sc (7) & BC03 & 15\\
   Scd (8) &  CWW & --\\
  Sd  (9) & BC03 & 30\\
  Irr (10) &  BC03 & $\infty ^*$\\
  Irr (11) &  CWW & --\\
  dusty (12) &  CWW & --\\
  Burst (13) & BC03 & $0 ^*$\\
  Arp220-modifying (14) & Arp220-rebuilt & --\\
  Arp220 (15) & Arp220 & --\\
\hline
\end{tabular}


\end{center}
\textrm{$^*$ Note that $\tau = \infty$ and 0 correspond to a
constant star formation rate and a single burst, respectively.}
\label{tab:spt}
\end{table}

\begin{figure}
\resizebox{\hsize}{!} {\includegraphics{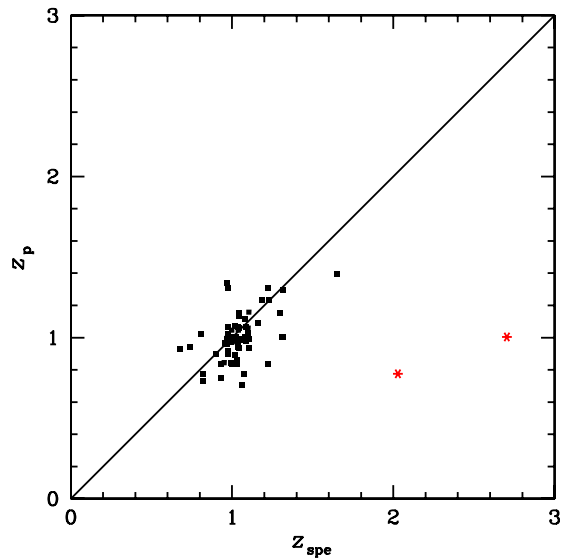}}
\caption{Comparison between re-estimated photometric redshifts
$z_{\rm p}$ and spectroscopic redshifts $z_{\rm spe}$ for 70 LBG
counterparts with available $z_{\rm spe}$. The solid diagonal
indicates where they would agree.
Unreliable $z_{\rm spe}$ with their UV IDs of 67958 and 59121 (IDs
56523 and 42593 in MUSYC) are labeled as ``$\ast$" due to their bad
quality in spectroscopic observations.} \label{fig:zpsec}
\end{figure}

As can be seen in Table 1, there are 70 among 459 optical
counterparts of LBG candidates with known spectroscopic data. We
compare their re-estimated photometric redshifts with their
spectroscopic redshifts in Fig. \ref{fig:zpsec}.
%
%
Two counterparts with their UV IDs 67958 and 59121 in B07 (IDs 56523
and 42593 in MUSYC) are labeled in the figure as ``$\ast$" because
of their unreliable spectroscopic redshifts due to very bad
spectroscopic observations.
To quantitatively clarify, we list uncertainties of our re-estimated
photometric redshifts for LBGs candidates as a function of their
apparent magnitudes in $R$-band in Table 3.
The corresponding uncertainties for the specific case of LBGs
candidates in the COMBO 17 and MUSYC {\it BVR} catalogs are also
listed in the table.
%
Four AGNs and two counterparts with bad quality in spectral
observations are excluded in the table.
%
Here $\Delta z$
and $\delta z$ in the table are defined respectively as the average
of $(z_{\rm p} - z_{\rm spe})/(1+z_{\rm spe})$, i.e., the mean
offset of $z_{\rm p}$ to $z_{\rm spe}$ in a given magnitude range,
and its corresponding standard deviation.
The first column in the table lists the magnitude ranges in
$R$-band. Columns 2-4 are
results of $\Delta z$ and $\delta z$
%
%
for the present paper,
the COMBO 17 (C) and the MUSYC (M) catalogs, respectively.

\begin{figure}
\resizebox{\hsize}{!} {\includegraphics{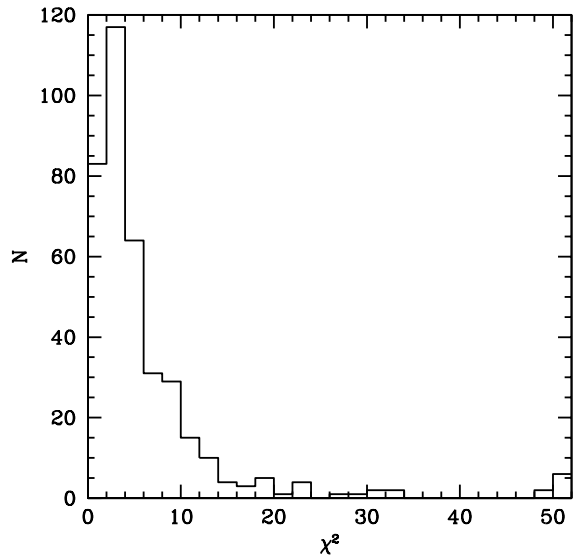}} \caption{The
reduced $\chi^2$  distribution of SED fittings.} \label{fig:chi2}
\end{figure}

%
It can be seen from the figure that our re-estimated photometric
redshifts of counterparts agree well with their corresponding
spectral data if available, except two with unreliable spectroscopic
redshifts.
It can be also found from Table 3 that
%
the averaged offset overall is -0.017 with its deviation of 0.063
for re-estimated photometric redshifts, comparable to those from the
COMBO 17 and the MUSYC catalogs. For counterparts with $R > 23$,
which contains half of spectral data available, the accuracies of
our re-estimated photometric redshifts are better than those from
the COMBO 17 catalog and a bit worse than those from the MUSYC
catalog, respectively.
%
%
To further demonstrate the goodness of SED fittings for LBG
counterparts, the distribution of reduced $\chi^2$ (per degree)
is shown in Fig. \ref{fig:chi2}. It can be found that most of SED
fittings are good ($\chi^2 \la 4$) with a median value of 1.8 for
$\chi^2$.
Together with Fig. \ref{fig:zpsec} and Table 3, it can be concluded
that our photometric redshift re-estimates are fairly good.
Note that there are 15 counterparts with large $\chi^2 ~ (> 10)$,
because most of them are very bright with small magnitude error
bars, especially in NIR bands. Discussions of reliability for the
SED fittings are in next subsection.



\begin{table}
\begin{center}
\caption{Photometric redshift accuracies as a function of apparent
magnitudes in $R$-band for LBG optical counterparts. The first
column lists the magnitude ranges. Columns 2-4 are the averaged
offsets of $z_{\rm p}$ to $z_{\rm spe}$ and their corresponding
standard deviations $\delta z$ for the present paper, the COMBO 17
(C) and the MUSYC (M) catalogs, respectively (see text for
details).}
%

\begin{tabular}{r|c|c|c}
\hline & $\Delta{z}$, $\delta z$  & $\Delta{z}$, $\delta z$ (C)
& $\Delta{z}$, $\delta z$ (M)\\
\hline \hline
$R<22$ &  -0.021, 0.061 & 0.014, 0.021 & 0.002, 0.014 \\
$22<R<23$  & -0.020, 0.063 & -0.014, 0.041 & -0.004, 0.024 \\
$23<R<24$ &  0.002, 0.061 & 0.015, 0.089 & 0.027, 0.074 \\
$24<R $ &  -0.067, 0.062 & 0.082, 0.097 & 0.017, 0.045 \\
\hline
\hline total &  -0.017, 0.063 & 0.009, 0.072 & 0.010, 0.050 \\

\hline \hline
\end{tabular}
\end{center}
\textrm{} \label{tab:acc}
\end{table}

\begin{figure}
\resizebox{\hsize}{!} {\includegraphics{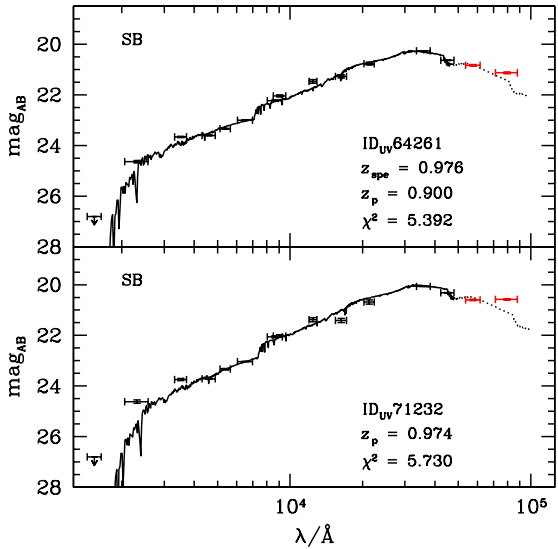}]} \caption{SED
fittings in solid lines for two LBG candidates having 24$\mu$m
detections (Red-LBGs) with $\rm ID_{UV}$, $z_{\rm p}$, $z_{\rm
spe}$, arrows and dashed lines denoting their IDs in UV from B07,
re-estimated photometric redshifts, spectroscopic redshifts, upper
limits of the flux detections and the extrapolations of the SED
fittings. The upper and lower panels are for candidates with and
without spectroscopic redshifts, respectively (see text for
details).}

\label{fig:sed24}
\end{figure}

\begin{figure}
\resizebox{\hsize}{!} {\includegraphics{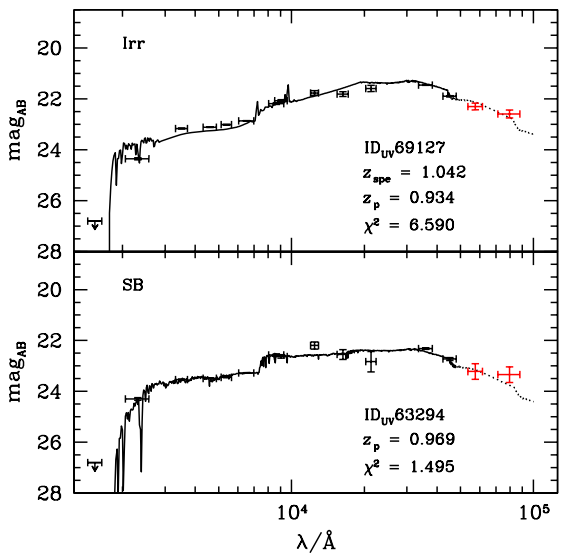}} \caption{Same
as Fig. \ref{fig:sed24} but for two LBG candidates without 24 $\mu$m
detections (Blue-LBGs, see text for details).}\label{fig:sedno24}
\end{figure}

As examples, Figs. \ref{fig:sed24} and
\ref{fig:sedno24} show individually SED fittings of two LBG
candidates without/with 24$\mu$m detections from B07, i.e.,
Blue-/Red-LBG candidates (see next subsection), respectively,
%
available in both the COMBO 17 and the MUSYC catalogs. Their
corresponding IDs in UV from B07 are labeled in the figures.
In each figure one LBG candidate with (upper panel) and the other
one without (lower panel) spectroscopic redshifts are selected for
clear comparison with their reduced $\chi^2$ (not small),
re-estimated photometric $z_{\rm p}$ and spectroscopic redshifts
$z_{\rm spe}$ being labeled. They are classified as three starburst
(SB) galaxies and an irregular (Irr) galaxy.
%
%
%
The arrows in the figures are upper limits of the flux detections.
The dashed lines are the extrapolations of template SEDs since
5.8$\mu$m and 8$\mu$m data of IRAC are not taken into account during
SED fittings. Together with Table 3 and Figs. \ref{fig:zpsec} \&
\ref{fig:chi2}, it further shows that our results of SED fittings
for LBG candidates are good.

\begin{figure}
\resizebox{\hsize}{!} {\includegraphics{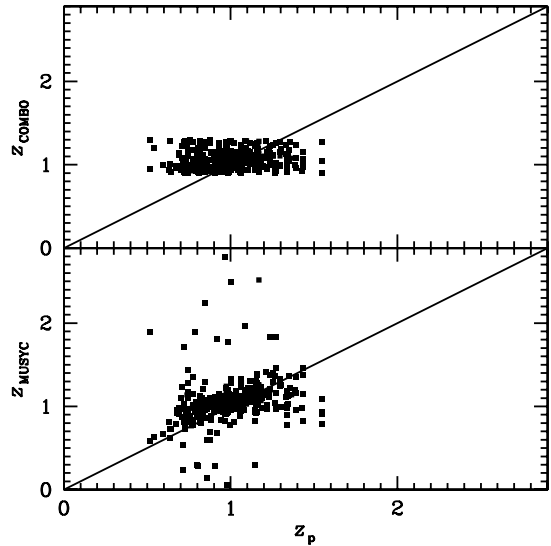}}
\caption{Comparisons of photometric redshifts (replace by $z_{\rm
spe}$ if available) between our re-estimated results with those in
the COMBO 17 and available in the MUSYC catalogs, respectively, with
the solid diagonal indicating where they would agree.}
\label{fig:z_zc_zm}
\end{figure}

Comparisons of photometric redshifts (replaced by $z_{\rm spe}$ if
available) between our re-estimated results with those in the COMBO
17 and available in the MUSYC catalogs are interesting and shown in
Fig. \ref{fig:z_zc_zm}. It can be clearly found that there exist
significant differences between our results and COMBO 17.
Since we use broader coverage in wavelength, especially UV and NIR
data, than COMBO 17 to estimate photometric redshifts, the quality
of our re-estimated $z_{\rm p}$ for the specific case of LBGs
candidates at $z\sim 1$ should be better (see also Table 3).
Moreover, the agreement between our results and MUSYC is better
mainly because NIR data have been included in MUSYC.
Note that there still are some outliers of LBG candidates between
our re-estimated photometric redshifts and those from the MUSYC
catalog (see the lower panel of Fig. \ref{fig:z_zc_zm}). After
checking carefully, two sources (MUSYC IDs of 56523 and 42593) with
their spectral redshifts $z_{\rm spe}> 2$ in the MUSYC catalog have
very bad quality in their spectroscopic observations (see also Fig.
\ref{fig:zpsec}). For the other outliers with photometric redshifts
higher than 1.6 or lower than 0.5 in the MUSYC catalog, they are
either very bright or very faint in NIR bands compared to their
optical magnitudes. This leads to over- or under-estimating their
photometric redshifts if {\it FUV} and {\it NUV} are not considered
as in the MUSYC catalog.
We conclude that the consideration of UV data really plays an
important role in the determination of photometric redshifts for LBG
candidates.

\subsection{Sample refinement} \label{sec:re-sam}

The processes of selecting  the ``true" optical counterpart of an
LBG candidate are as follows. There are 70 counterparts with
spectroscopic redshifts available from previous studies.
We replace the photometric redshifts of 68 among these 70
counterparts by their observed spectroscopic redshifts except two
with very bad spectroscopic observations (see previous subsections).
The counterparts of 375 LBG candidates with only single counterparts
are assumed to be their corresponding counterparts as discussed
above. For a candidate with pair or four counterparts, their
re-estimated photometric/spectroscopic redshifts firstly and the
reduced $\chi^2$  during SED fittings secondly are considered for
its ``true" counterpart selection. If only one counterpart with
photometric redshifts/spectroscopic is within the range $0.9\la
z_{\rm p}\la 1.3$, we assume this one to be the ``true" optical
counterpart of the candidate firstly, since the selection criterion
of $FUV-NUV>2$ strongly suggests that it biases to locate at $z \sim
1$. If more than 2 counterparts of a candidate with their
re-estimated photometric redshifts or none of them are within the
redshift range, the counterpart with smaller $\chi^2$ is selected as
the ``true" counterpart.


\begin{figure}
\resizebox{\hsize}{!} {\includegraphics{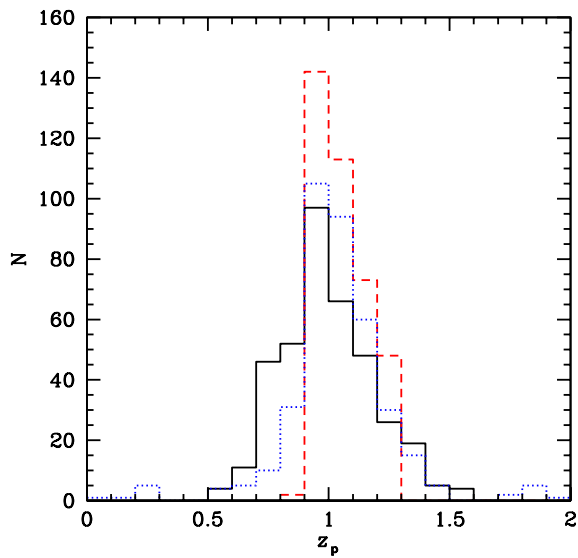}}
\caption{The distributions in histograms of the re-estimated (solid)
and the original photometric redshifts from COMBO 17 (dashed) for
the LBG candidates with the dotted histogram denoting results from
MUSYC.}\label{fig:zhis}
\end{figure}

Now we have finished the determination of the ``true" optical
counterparts of all 416 LBG candidates selected by $FUV-NUV>2$ from
the preliminary LBG sample of B07 with 4 AGNs being excluded. The
distribution of their re-estimated photometric (spectroscopic if
available) redshifts $z_{\rm p}$ is plotted as the solid histogram
in Fig. \ref{fig:zhis}. For comparison, the distributions of the
original photometric redshifts from COMBO 17 and the results of 394
candidates from MUSYC are also plotted as the dashed and dotted
histograms in the figure.
It can be found that the re-estimated photometric (spectroscopic if
available) redshifts of all LBG candidates
are between 0.5 and 1.6, because of the selection criterion
$FUV-NUV>2$ as expected. Almost two thirds of LBG candidates (268 of
416) are within the redshift range of $0.9\la z_{\rm p}\la 1.3$ as
B07. The numbers of candidates are 339 and 387
within the redshift ranges of $0.8 \la z_{\rm p}\la 1.4$ and $0.7
\la z_{\rm p}\la 1.4$, respectively.
%

\begin{figure}
\resizebox{\hsize}{!} {\includegraphics{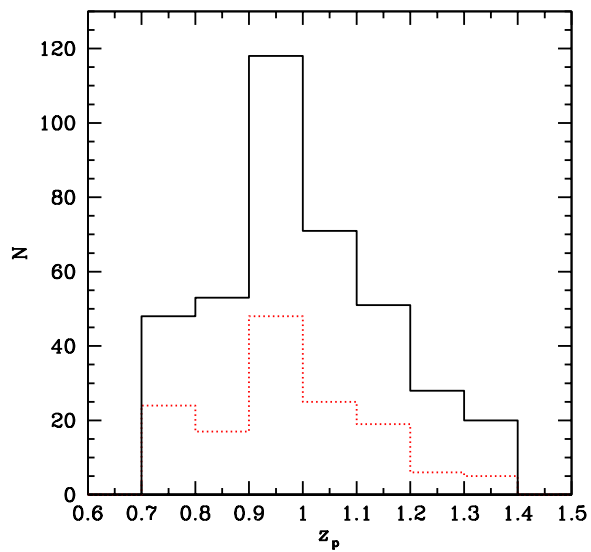}}
\caption{The redshift distributions of all LBGs and LBGs for
morphological studies (see Sect. \ref{sec:mor}) as solid and dashed
histograms, respectively.}
\label{fig:zphot_144_all}
\end{figure}


Although the goodness of SED fitting is parameterized by $\chi^2$,
we need to check the reliability of individual LBG candidates one by
one, especially for 15 objects with $\chi^2
>10$.
%
%
We find that 399 among 401 LBG candidates with $\chi^2<10$ display
reliable SED fittings except two candidates (UV IDs of 64312 and
75583) with their resulted $z_{\rm p}$ of 0.883 and 0.802
respectively, showing very uncertain photometric data.
%
A candidate of UV ID 47588 with its resulted $z_{\rm p}=0.930$ and
$\chi^2>10$ displays a very strange SED with a shape jump from
$FUV$, $NUV$ to $U$ ($FUV - NUV \sim 4$ and $NUV - U \sim -2.5$).
Since the error bars of its photometric data is $\sim 0.25$, it is
hard to find a template to fit it reliably. Moreover, a candidate
with its UV ID of 54502, $z_{\rm p}= 0.906$ and $\chi^2 =11.2$ shows
its SED with a bump between $I$ and $H$ bands and then drops very
steeply from $H$ band to $4.5\mu$m. Together with its magnitudes in
$5.8\mu$m and $8.0\mu$m which are much brighter than its magnitude
in $4.5\mu$m, we suggest that its SED fitting is unconvincing. For
the rest 13 candidates with $\chi^2>10$, we find that their SED
fittings are reasonable, together with the consideration of the
extrapolations of the SED fittings to $5.8\mu$m and $8.0\mu$m. Their
large $\chi^2$ values are because of small photometric error bars in
individual bands.
%
In total, 4 candidates with unreliable SED fittings are discarded in
the following studies.
%
%

\begin{figure}
\resizebox{\hsize}{!} {\includegraphics{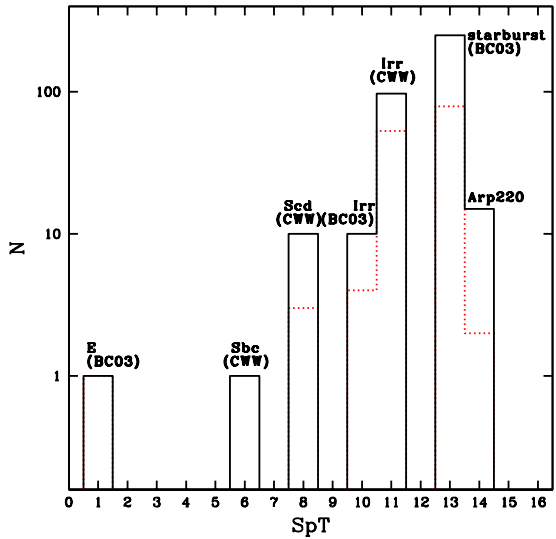}}
\caption{The spectral type distribution of LBGs obtained through
their SED fittings with the corresponding templates, i.e., SpT
numbers, being listed in Table \ref{tab:spt}. The solid and dashed
histograms denote all LBGs and LBG for morphological studies (see
Sect. \ref{sec:mor}), respectively.} \label{fig:spt}
\end{figure}

We define 383 LBG candidates with reliable SED fittings at $0.7 \la
z_{\rm p}\la 1.4$ to be our LBG sample at $z\sim 1$ for the further
studies below because of the following three reasons: (1) the Lyman
limit, i.e., 912${\AA}$, moves to 1550${\AA}$ at redshift 0.7, just
away from the effective wavelength of {\it FUV}, 1530${\AA}$; (2)
the Lyman limit leaves from {\it NUV} if redshift larger than 1.4;
and (3)
the uncertainty $\delta z$ of re-estimated photometric redshifts
obtained above is about 0.1 at $z\sim 1$.
%
To clarify and easily compare, we plot the distribution of
%
%
LBGs in the field of the CDF-S in the lower panel of \ref{fig:pos}
with circles ``$\circ$" and squares ``$\square$" denoting LBGs
without and with spectral data available from the MUSYC catalog and
other previous studies.
%
%
Their redshift distribution as the solid histogram and the
corresponding distribution of 142 LBGs for morphological studies
(see Sect. \ref{sec:mor} for details) as the dashed histogram are
plotted in Fig. \ref{fig:zphot_144_all}, respectively.

Note that two AGNs would be included in the LBG sample if we only
consider the color selection criterion and the redshift range.
Comparing with \citet{Steidel03} and \citet{Lehmer08}, the AGN
contamination ($\sim 0.5\%$) is smaller for LBGs at $z\sim 1$ than
LBGs at $z\sim 3$ ($\sim 3\%$). It may imply that the connection
between star formation activities and AGNs is stronger at higher
redshifts for color selected samples.
Moreover, we take the same criterion of with/without 24$\mu$m
detections to classify Red-/Blue-LBGs as suggested by B07 in the
present paper.
%
For our LBG sample at $z\sim 1$, there are 324 (85\%) and 59 (15\%)
Blue- and Red- LBGs, among total 383 LBGs, respectively.

\section{Photometric properties of LBGs}\label{sec:pho}

\subsection{Spectral types}


To investigate the spectral type (SpT) of a galaxy through its SED
fitting, we have to bear in mind that its resulted $z_{\rm p}$ and
SpT during SED fitting are degenerated with each other. To check the
reliability, we consider results of the second solutions of
photometric redshifts $z_{\rm p2}$ for individual LBGs.
%
%
As expected, the reduced residuals of SED fittings for $z_{\rm p2}$,
named as $\chi^2_2$, of individual LBGs are always larger than
$\chi^2$.  More than 70\% of LBGs show $\chi^2_2$ larger than 7.8.
The median value of $\chi^2_2$ for LBGs is 16.7 while that of
$\chi^2$ is 1.8. There are more than 80\% and 70\% of LBGs with
$\chi^2$ smaller than 4 and 3, while about 10\% and 5\% of them show
$\chi^2_2$ smaller than 4 and 3, respectively. This means that the
contamination of the degeneracy between $z_{\rm p}$ and SpT for LBGs
is less than 5\% in a ``$1-\sigma$" level.
%
%
We further check spectral types of LBGs with $\chi^2_2 < 3$ when the
second solutions are adopted. It is found that more than half of
them still display the same types as obtained for the first
solutions.
%
We conclude that spectral types of LBGs obtained through SED
fittings are reliable for the further studies below.


The distribution of spectral types for LBGs is shown in Fig.
\ref{fig:spt} as the solid histogram with the corresponding
templates, i.e., SpT numbers, being listed in Table \ref{tab:spt}.
It can be clearly found from the figure that LBGs in our sample are
dominated by starburst galaxies, with 249 and 15 LBGs being
classified as BC03 and Arp 220 types, respectively. There are 97 and
10 LBGs being classified as CWW and BC03 type irregulars.
Moreover, 10 and 1 LBGs are classified as CWW type Scd galaxies and
a CWW type Sbc galaxy respectively.
The spectral type distribution is consistent with the selection of
LBGs which must be star forming galaxies at $z\sim 1$. And it is
also in consistence with that most of LBGs for morphological studies
are late type galaxies with their $\sersic$ indexes smaller than 2
in Sect. \ref{sec:mor} .

Note that only one LBGs in our sample with its UV ID of 74754 is
classified as a BC03 type elliptical galaxy at  $z_{\rm p}=0.834$
with the age of $\sim 1$Gyr.
%
%
The reduced $\chi^2$ of its SED fitting is 7.68, relatively larger
than most LBGs but still acceptable. Its star formation rate and
stellar mass are $22.9\my$ and $1.4\times 10^{11}\msun$ (see below).
%
Note that its second solution of the SED fitting is
not reliable because of very large $\chi^2_2 \sim 900$ with $z_{\rm
p2}=1.75$.
It is detected in both F606W and F850LP bands with its morphology
being classified visually as a ``bulge" LBG (see Sect.
\ref{sec:HST}).
Its $\sersic$ indexes are of 0.7 and 2.2 with the similar sizes of
$\sim 5\kpc$ in two bands respectively (see Sect. \ref{sec:mor}). It
implies that the distribution of its star formation activity is
extended while that of stars is compact.
%
Since LBGs are selected according to their rest-frame UV
luminosities (star formation activities), early type galaxies with
star formation rates higher than the selection criterion can be
selected as LBGs, although they are rare such as only one in our
sample.

To investigate statistical properties of different spectral types in
the following studies, we always divide LBGs into the ``starburst"
(SB) group for starburst type LBGs, the ``irregular" (Irr) group for
irregular type LBGs and the ``S+E" (EST) group for spiral and
elliptical type LBGs. Note that spectral types are later to earlier
from the SB, Irr to EST groups. The numbers of LBGs  are 264 (227
blue and 37 red), 107 (92 blue and 15 red) and 12 (5 blue and 7 red)
for the SB, Irr and EST groups respectively. The fraction of
Blue-/Red-LBGs decreases/increases from the SB, Irr to EST groups,
in consistence with spectral types obtained. Since the number of
LBGs in the EST group is too small to have statistical significance,
we mainly focus on the SB and Irr groups for statistical discussions
and take the EST group for only illustration in the following
studies.

\begin{figure}
\resizebox{\hsize}{!} {\includegraphics{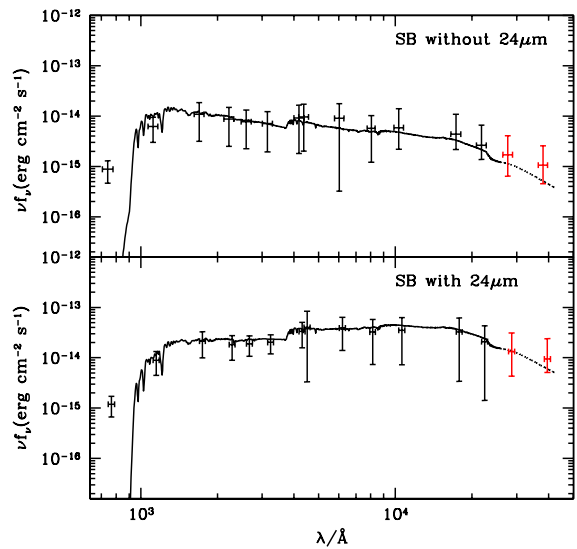}} \caption{The
averaged rest-frame SEDs of LBGs in the SB group with the up and
lower panels denoting results without (Blue-LBGs) and with
(Red-LBGs) detection in $24\mu$m by MIPS, respectively.}
\label{fig:SB}
\end{figure}

\begin{figure}
\resizebox{\hsize}{!} {\includegraphics{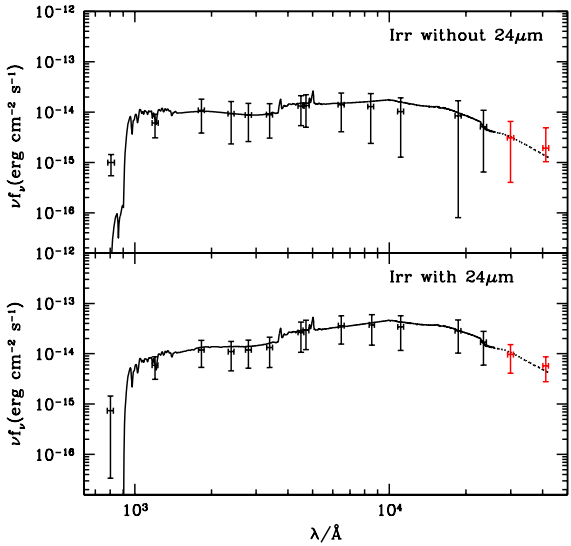}} \caption{Same
as Fig. \ref{fig:SB} but for LBGs in the Irr group.} \label{fig:Irr}
\end{figure}

\subsection{SEDs and UV luminosity function}

Because of the limitation of the paper space, we do not show SED
fitting results of LBGs individually. Instead, we compile figures of
SED fitting results for LBGs with images detected both available in
F606W and F850LP bands
in Table 6, together with their IDs, morphological parameters and
images (see Sect. \ref{sec:mor} for details).

The ages of LBGs spread from several Myr to 1.5Gyr
with a median value of $\sim 50$Myr, comparable to
\citet{Haberzettl12} for LBGs at $z\sim 2$ and \citet{Papovich01}
for LBGs at $z\sim 3$. Similar to \citet{Shapley01}, we find
$\sim$10\% of LBGs, classified as SB types, are with ages younger
than 10Myr which is different from \citet{Haberzettl12} for LBGs at
$z\sim 2$.
To be more specific, the median ages for LBGs in the SB and Irr
groups are $\sim$30Myr and $\sim$800Myr, respectively. It is
consistent with that ages of galaxies are younger generally for
later types. Similar to B07, the median ages of Red- and Blue-LBGs
are $\sim$400Myr and $\sim$250Myr, respectively.
Moreover intrinsic reddening of individual LBGs can be simply
obtained through the adopted dust extinction law of
\citet{Calzetti00}.  We find that the dust extinctions in their
rest-frame $FUV$ for Blue- and Red-LBGs show narrow distributions
with the median values $\rm A_{FUV}$ to be 1.7 and 2.5 magnitudes
respectively, which is consistent well with B07.

For easy comparison with observations, we plot the averaged SEDs in
the rest-frame for Red- and Blue-LBGs, i.e., with and without
24$\mu$m detection, in the SB and Irr group at $z\sim 1$ in Figs.
\ref{fig:SB} and \ref{fig:Irr}, respectively. The mean observed data
with error bars are also shown in the figures. Same as the previous
section, data of the two reddest bands by IRAC, i.e., 5.8$\mu$m and
8$\mu$m, are plotted to compare with the extrapolations of SEDs as
dashed lines since the templates adopted do not consider the dust
emission features.

\begin{figure}
\resizebox{\hsize}{!} {\includegraphics{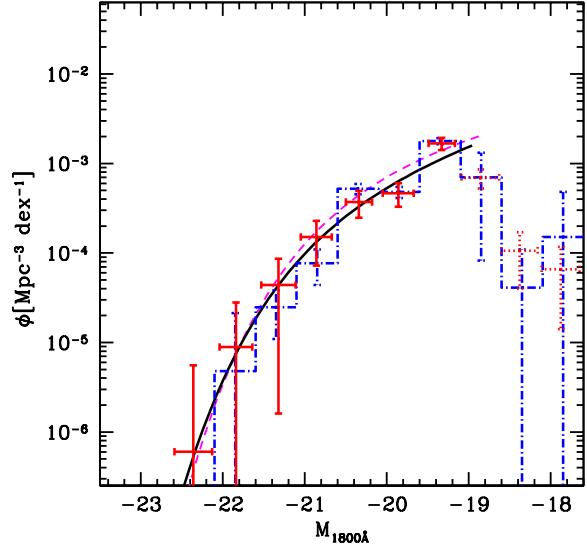}} \caption{The
rest-frame FUV luminosity function of LBGs at $z\sim 1$ as points
with the solid and dotted error bars denoting the complete and
incomplete parts, respectively. The best-fit Schechter function is
plotted as the solid line. Result of B07 is as the dash-dotted
histogram with error bars and the best-fit Schechter function of NUV
selected galaxies at $z\sim 1$ by \citet{Arnouts05} is as the dashed
line (see text for details).} \label{fig:LF}
\end{figure}

\begin{figure}
\resizebox{\hsize}{!} {\includegraphics{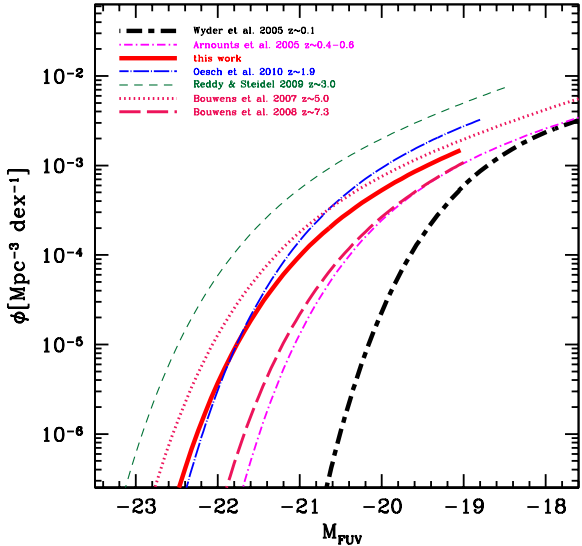}} \caption{The
best-fit Schechter functions of FUV LFs at different redshifts
listed in Table 4, except that of $0.8\la z \la 1.2$
(\citealt{Arnouts05}) shown in Fig. \ref{fig:LF}.}
\label{fig:LF_all}
\end{figure}

It can be found from the figures that LBGs in the SB group on
average are brighter in their rest-frame UV than LBGs in the Irr
group.
This is consistent with that starburst galaxies
are on average more active in star formation than irregular
galaxies.
Within each group, LBGs with $24\mu$m detections, i.e., Red-LBGs,
are on average brighter in their rest-frame NIR,
i.e., on average more massive in their stellar masses, than LBGs
without $24\mu$m detections, i.e., Blue-LBGs (see also next
subsection).
%


Luminosities of individual LBGs in the rest-frame FUV can be
estimated through their SEDs.
As pointed out by B07, the completeness of the LBG sample is 83\%
down to $NUV = 24.5$.
%
To be more safe, we adopt $NUV = 24.2$ as the completion of the LBG
sample (see also \citealt{Burgarella08}), which corresponds to
%
the absolute magnitude $M_{\rm 1800\AA}\sim -19$ at $z\sim 0.7$.
Based on the widely adopted $1/V_{\rm max}$ method suggested by
\citet{Schmidt68}
 and considering uncertainties of individual luminosity
bins due to uncertainties of photometric redshifts,
%
we obtain the rest-frame FUV ($1800\AA$) luminosity function (LF) of
LBGs at $z\sim 1$, which is shown
as points in Fig. \ref{fig:LF}, simply taking the error bars as
Poisson statistics. Note that the points with solid and dashed
error bars denote the complete and incomplete parts of LBG sample
respectively. Adopting the Levenberg-Marquardt method (see
\citealt{Zaninetti08}), we fit the rest-frame FUV LF of LBGs by a
Schechter function as the solid line in Fig. \ref{fig:LF}, without
considering its incomplete part. We get $\alpha = -1.61\pm 0.40 $,
$\rm M^* = -20.40 \pm 0.22$ and $\phi^* = (0.89\pm 0.30)\times
10^{-3}\mpc^{-3}{\rm dex^{-1}}$, respectively.
%
%
For easy comparison, the rest-frame FUV LF for LBG candidates in B07
and NUV selected galaxies at $z\sim 1$ by \citet{Arnouts05} are also
plotted as the dash-dotted histogram with error bars and the dashed
line in the figure, respectively.
We find that our FUV LF agrees well with that of B07 for $\rm
M_{1800\AA} < -18.5$. It decreases faster for $\rm M_{1800\AA} >
-18.5$ since the number of LBGs in our sample is 10\% less and the
redshift range is wider in the present paper.
%
%
We also find that the rest-frame FUV LF of LBGs at the bright end
well matches the LF of NUV selected galaxies at $z\sim 1$
(\citealt{Arnouts05}). And it is lower by a factor less than 2 even
at the complete magnitude of $M_{1800\AA} = -19$. It implies that
most of NUV selected galaxies at $z\sim 1$ are selected as LBGs, in
consistence with the results obtained by B07.


For further comparison, previous results of LFs in the rest-frame
FUV ($\sim1500{\AA}$) at different redshifts, together with ours,
are listed in Table 4. We also plot in Fig. \ref{fig:LF_all} the
Schechter functions fitted to FUV LFs at different redshifts listed
in Table 4 individually, except that of $0.8< z < 1.2$
(\citealt{Arnouts05}) shown in Fig. \ref{fig:LF}.
As can be seen, our result of $\rm M^*$ locates in the trend of
monotonic fading of $\rm M^*$ from $z\sim 3$ to $z\sim 0$, in
consistence with \citet{Reddy09} and \citet{Oesch10}. For the
faint-end slope $\alpha$, our result is similar to that of
\citet{Arnouts05} for NUV selected galaxies at $z\sim 1$ and along
the trend that $\alpha$ increases with the decreasing of redshift
$z$ (\citealt{Reddy09}). For $\phi^{\ast}$, our result also agrees
well with \citet{Arnouts05} within a factor less than 2 (see above).
Detailed discussions on the evolution of FUV LFs can be seen in
\citet{Reddy09}.



Same as B07, we can also estimate the contribution of luminosity
density in FUV for the Blue- and Red-LBGs in our sample to be
 \beq
 \rho_{\rm FUV} (1800\AA) = (2.6 \pm 0.6) \times 10^{7} h (\rm
 L_{\odot} Mpc^{-3}), \nonumber
 \eeq
 and
\beq
 \rho_{\rm FUV} (1800\AA) = (2.3 \pm 0.8) \times 10^{6} h (\rm
 L_{\odot} Mpc^{-3}), \nonumber
 \eeq
respectively. Comparing with B07, the contributions of our LBG
sample contains 75\% FUV luminosity densities in 1800$\AA$  of the
total NUV-selected galaxies at $z\sim 1$.

\begin{table}\label{tab:LF}
\caption{Best-fit Schechter parameters of FUV LFs for galaxies at
different redshift $z$ including the results of present paper for
LBGs at $z\sim 1$. Refs 1-6 denote the reference of \citet{Wyder05},
\citet{Arnouts05}, \citet{Oesch10}, \citet{Reddy09},
\citet{Bouwens07} and \citet{Bouwens08}, respectively.}
{\small
\begin{center}
\begin{tabular}{c|c|c|c|c}
\hline  z& $\alpha$ & $\phi^{\ast}(10^{-3}{\rm Mpc^{-3}dex^{-1}})$ & $\rm M^{\ast}$   & refs\\
\hline \hline
0-0.1               & $-1.16\pm0.07$     &    $5.50\pm0.14      $     &    $-18.23\pm0.11     $   &    1\\
0.4-0.6             & $-1.55\pm0.21$     &    $1.69\pm0.88      $     &    $-19.49\pm0.25     $    &    2\\
0.8-1.2 & $-1.63\pm0.45$ & $1.14\pm0.76$ & $-20.11\pm0.45$ & 2\\
0.7-1.4             & $-1.61\pm0.40$     &    $0.89\pm0.30      $     &    $-20.40\pm0.22     $   &    this work\\
1.7-2.1             & $-1.60\pm0.51$     &    $2.19\pm0.83      $     &    $-20.16\pm0.52     $   &    3\\
2.7-3.4             & $-1.73\pm0.13$     &    $1.71\pm0.53      $     &    $-20.97\pm0.14     $   &    4\\
3.8                 & $-1.73\pm0.05$     &    $1.30\pm0.20      $     &    $-20.98\pm0.10     $   &    5\\
5.0                 & $-1.69\pm0.09$     &    $0.9^{+0.3}_{-0.2}$     &    $-20.69\pm0.13 $   &    5\\
7.3                 & $-1.74       $     &    $1.1^{+1.7}_{-0.7}$     &    $-19.8\pm0.4   $   &    6\\
\hline \hline
\end{tabular}
\end{center}
}
\end{table}

\begin{figure}
\resizebox{\hsize}{!}
 {\includegraphics{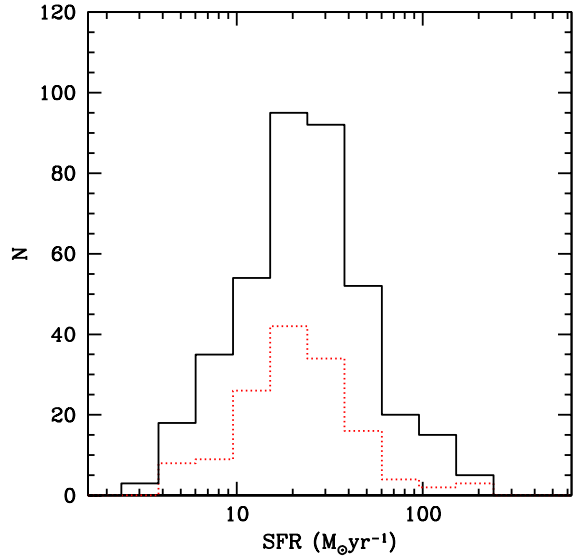}}
\caption{The dust-corrected SFR distribution of LBGs with the solid
and dashed histograms denoting all LBGs and LBGs for morphological
studies in Sect. \ref{sec:mor}, respectively.} \label{fig:sfr}
\end{figure}

\begin{figure}
\resizebox{\hsize}{!}
 {\includegraphics{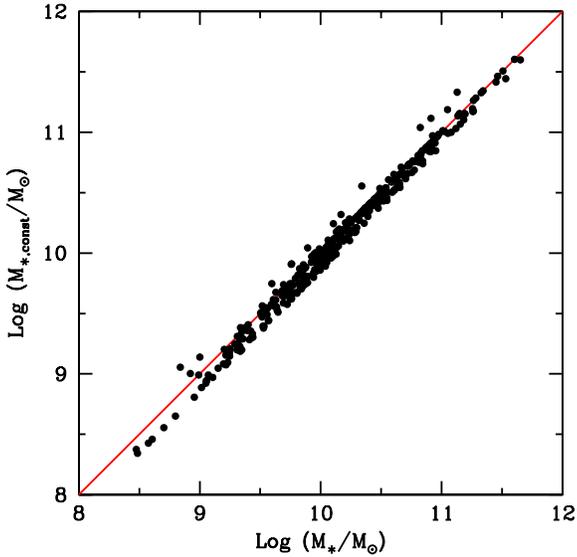}}
\caption{Comparison between stellar masses of LBGs obtained in the
present study ($M_*$) and ($M_{*,\rm const}$) by taking a constant
$M_*/L_K = 0.5M_\odot/L_{\odot,K}$ with the solid diagonal
indicating where they would agree.} \label{fig:mass}
\end{figure}

\begin{figure}
\resizebox{\hsize}{!} {\includegraphics{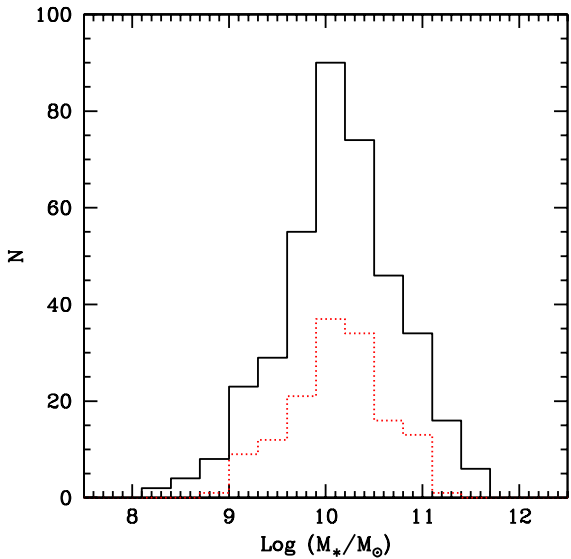}} \caption{Same
as Fig. \ref{fig:sfr} but for the stellar mass $M_*$ distribution of
LBGs.} \label{fig:star}
\end{figure}

\subsection{SFRs and stellar masses}
\label{sub:SFRvsM}

We can estimate the star formation rates (SFRs) of individual LBGs
according to their rest-frame {\it NUV} flux as suggested
 by \citet{K98}), i.e.,
\beq
 {\frac{\rm SFR}{\my}} = 1.4 \times 10^{-28}\;\ {\frac{L_{\rm NUV}}{\rm
 ergs^{-1}Hz^{-1}}},
\eeq
which is obtained by converting the calibration of \citet{Madau98}
to a \citet{Salpeter55} initial mass function with the stellar mass
range from 0.1$\msun$ to $100\msun$.
Dust corrections of individual LBGs are taken into account according
to their SEDs and the extinction law adopted. As discussed in the
previous subsection, the dust corrections are on average similar to
the median values obtained by B07, respectively, for Red-/Blue-LBGs
with/without 24$\mu$m detections.

The dust-corrected SFR distribution of LBGs at $z \sim 1$ is plotted
in Fig. \ref{fig:sfr} as the solid histogram, with the dashed
histogram denoting the SFR distribution of LBG for morphological
studies in Sect. \ref{sec:mor}. It can be seen from the figure that
dust-corrected SFRs of LBGs range from $4\my$ to $220\my$ with a
median value of $\sim 25 \my$. The median SFR is similar to that of
LBGs at $z\sim 3$ (\citealt{Shapley01, Steidel03}), and a bit
smaller than (comparable to) those of LBGs at $z\sim 5$
(\citealt{Verma07}) and at $z\sim 2$ (\citealt{Haberzettl12}).
Moreover, the mean dust-corrected  SFR for LBGs in the SB group is
$35.9\my$, larger than that for LBGs, $21.3\my$, in the Irr group,
as expected in the previous subsection.


Stellar masses $M_{*}$ of individual LBGs are obtained as follows.
For an LBG with a given SpT type of the BC03 template, its star
formation history is assumed to be exponentially decaying with a
given timescale $\tau$ (see Sect. \ref{subsec:fitting} and Table
\ref{tab:spt}).
We can calculate its mass-to-light ratio, $M_*/L$, for any given
bands as a function of its age.
In the present study, $M_*/L$ of the rest-frame $K$-band is
calculated. So, stellar masses of LBGs classified as BC03 templates
can be obtained through their rest-frame
$K-$band luminosities according to their spectral types (star
formation histories) and ages.
For LBGs classified as Arp220 and CWW types, we calculate their
stellar masses according to their rest-frame $K$-band luminosities,
taking the $M_*/L_k$ suggested by \citet{Huang09} and a simple
constant $M_*/L_K = 0.5M_\odot/L_{\odot,K}$, respectively.
%

As pointed out by the previous studies (e.g.,
\citealt{Kauffmann98, Bell01, D04}), the $K$-band luminosity of a
galaxy is insensitive to its star formation history.
%
A simple way to estimate the stellar mass of a galaxy is to adopt a
constant mass-to-light ratio in $K$-band, e.g., $M_*/L_K =
0.5M_\odot/L_{\odot,K}$ (\citealt{Zheng07}), since $M_*/L_K$ varies
very slowly for young stellar populations and increases within a
factor of two for stellar populations older than $\sim$3Gyr
(\citealt{D04}).
Comparison between stellar masses of LBGs obtained in the present
paper ($M_*$) and ($M_{*,\rm const}$) by simply taking a constant
$M_*/L_K = 0.5M_\odot/L_{\odot,K}$ for all LBGs is plotted in Fig.
\ref{fig:mass}.
It can be seen from the figure that stellar masses obtained by these
two methods are in good agreement with each other within a factor of
2, since ages of LBGs in the present study are younger than 1.5Gyr.
This also further confirms that rest-frame $K$-band luminosity is a
good indicator to estimate stellar mass of a galaxy, especially when
its age younger than 3Gyr (\citealt{D04}).
%
%

%

%
The distribution of stellar masses $M_{*}$ for LBGs at $z \sim 1$ is
shown in Fig. \ref{fig:star} as the solid histogram.
%
Same as Fig. \ref{fig:sfr}, the corresponding
distribution of LBGs for morphological studies in Sect.
\ref{sec:mor} is also plotted as the dashed histogram.
We can see that stellar masses $M_{*}$ of LBGs in our sample
distribute from $2.3 \times 10^8 \msun$ to $7.7 \times 10^{11}
\msun$, with a median value of $\sim 10^{10} \msun$ which is a bit
smaller than those for LBGs at $z\sim 2$ (\citealt{Haberzettl12})
and at $z\sim 3$ (\citealt{Shapley01, Papovich01}), and about a
factor of 10 bigger than that of LBGs at $z\sim 5$
(\citealt{Verma07}).
%
Moreover, the mean stellar masses for Red-LBGs in the SB and Irr
groups are $6.6 \times 10^{10} \msun$ and $7.8 \times 10^{10}
\msun$, larger than those of $7.7 \times 10^9 \msun$ and $1.7 \times
10^{10} \msun$ for Blue-LBGs in the SB and Irr groups, respectively.
This is consistent with the discussions in the previous subsection
that Red-LBGs are on average brighter in their
rest-frame NIR, i.e., more massive in stellar masses, than
Blue-LBGs.

\begin{figure}
\resizebox{\hsize}{!} {\includegraphics{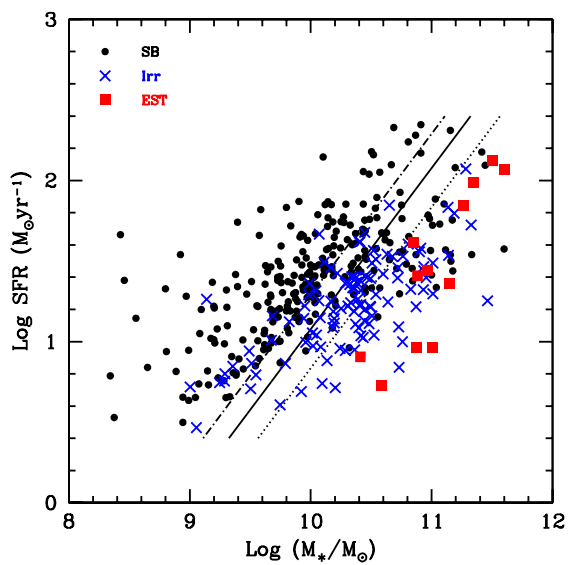}}
\caption{The correlation between the SFRs and $M_*$ for LBGs with
dots, crosses and squares denoting LBGs in the SB, Irr and EST
groups. The lower limits of the defined starbursts (Eq. (14) in
\citealt{Elbaz11}) at $z = 1.4$, 1, and 0.7 are plotted as
dash-dotted, solid and dotted lines for illustration, respectively
(see text for details).} \label{fig:mass_sfr}
\end{figure}

\begin{figure}
\resizebox{1.2\hsize}{!} {\includegraphics{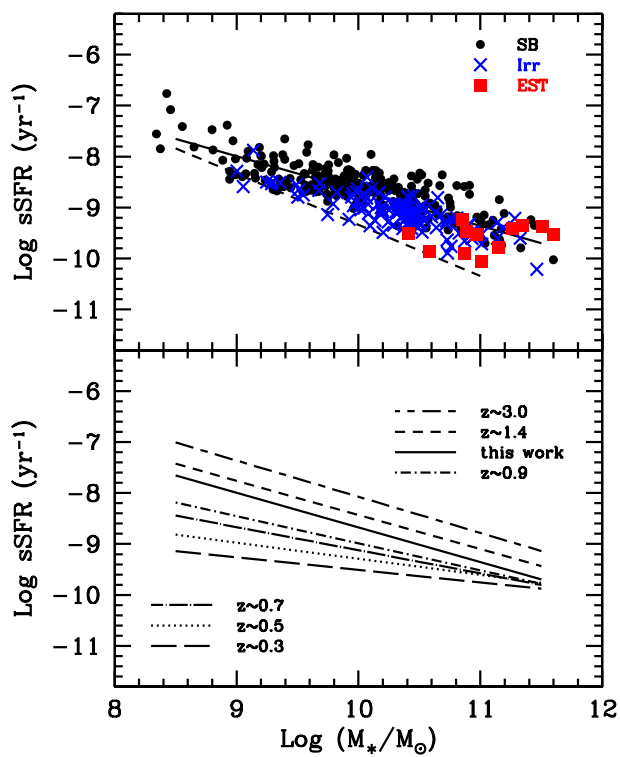}}
\caption{(Upper panel) Similar to Fig. \ref{fig:mass_sfr} but for
the correlation between sSFR and $M_*$ for LBGs with the dashed and
solid lines denoting the observational limit of $NUV \la 26^{\rm
m}.2$ and the simple least square fit to the data; (Lower panel)
Results of least square fit to data at different redshifts (see text
for details).} \label{fig:downsizing}
\end{figure}

\begin{figure}
\resizebox{\hsize}{!} {\includegraphics{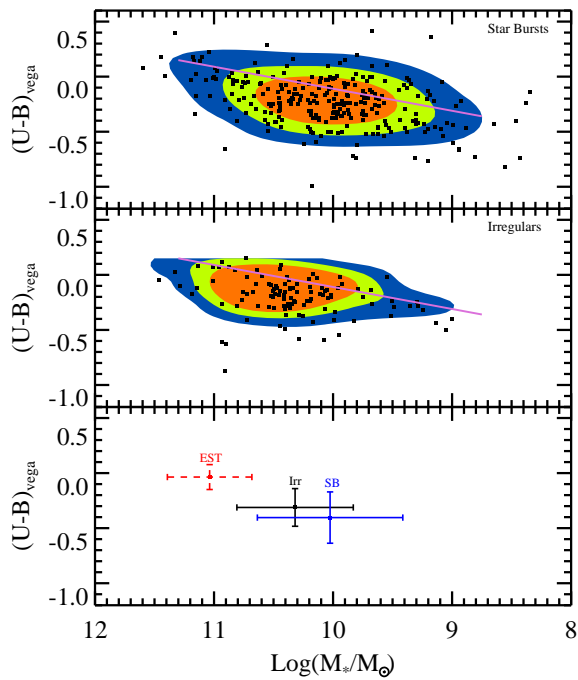}} \caption{
The distributions and the corresponding density contours for LBGs in
the SB (upper panel) and Irr (middle panel) groups, respectively, in
the diagram of rest-frame color $U-B$ vs stellar mass $M_*$.
Contours from outside to inside correspond to the mean number
densities of LBGs, and the over-dense contrasts of 2 and 3 times of
the mean densities, respectively.}
%
%
\label{fig:color_mass}
\end{figure}

The correlation between SFRs and $M_*$ for LBGs in our sample is
shown in Fig. \ref{fig:mass_sfr} with dots, crosses and squares
denoting LBGs in the SB, Irr and EST groups respectively. LBGs of
individual groups show clear trends that SFRs increase with the
increasing of $M_*$.
Studying IR SEDs of galaxies at $0<z<2.5$ in detail, \citet{Elbaz11}
established a prescription of the ``main sequence" for star forming
galaxies with the uncertainty of $\sim0.3$dex (i.e., a factor of 2)
 and defined starbursts at different redshifts.
Similar result is also obtained
by \citet{Sargent12}.
For comparison, we plot in Fig. \ref{fig:mass_sfr} the lower limits
of the defined starbursts (Eq. (14) in \citealt{Elbaz11}) at $z =
1.4$, 1, and 0.7 as dash-dotted, solid and dotted lines for
illustration, respectively. For a given line in the figure (i.e., a
given redshift), its upper-left part denotes the starburst region
and its lower-right region within a factor of 4 in SFR (y-axile) of
the line denotes the ``main sequence" of star forming galaxies.
%
%
%

\begin{figure}
\includegraphics[width=\figsmall,height=\figsmall]{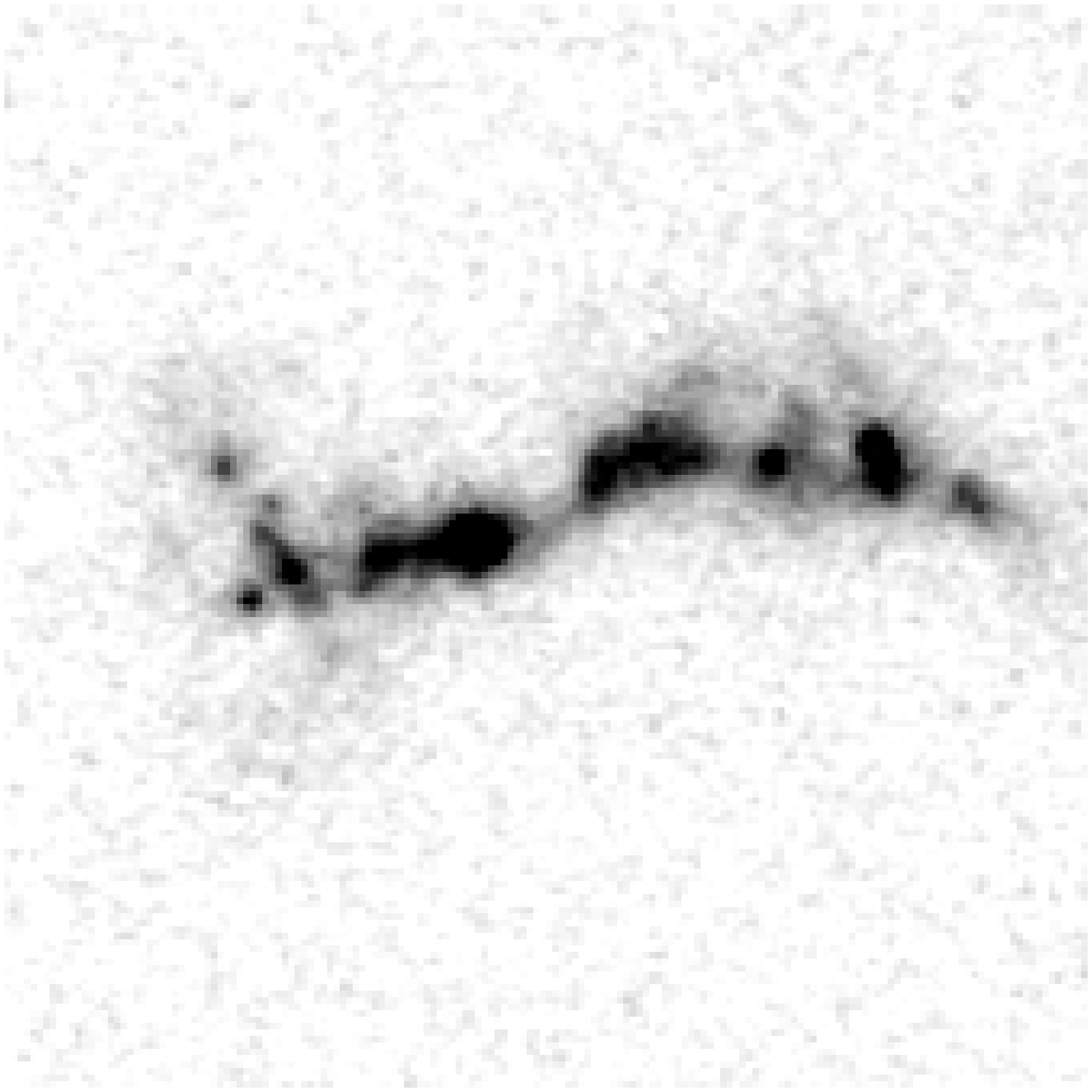}\quad
\includegraphics[width=\figsmall,height=\figsmall]{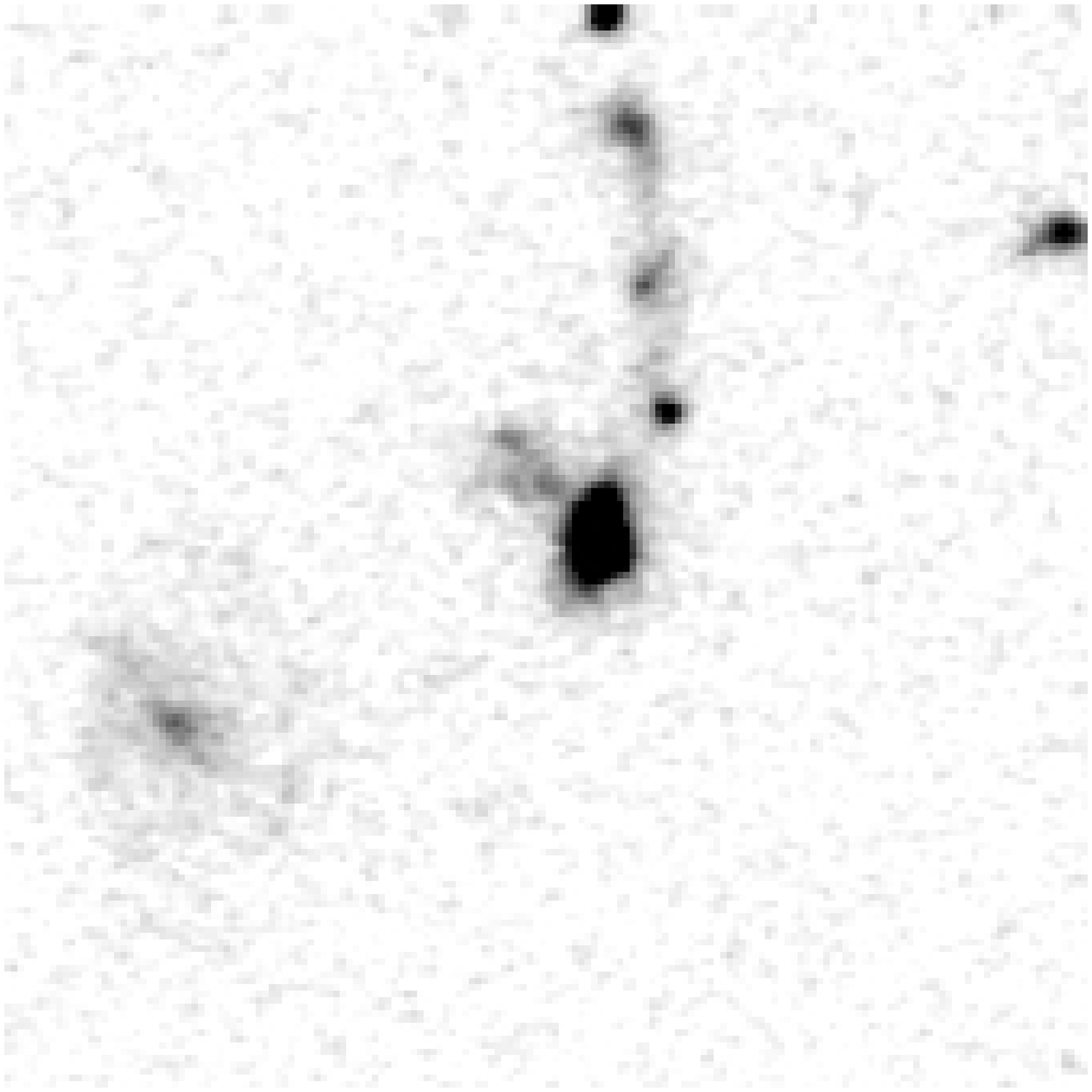}\quad
\includegraphics[width=\figsmall,height=\figsmall]{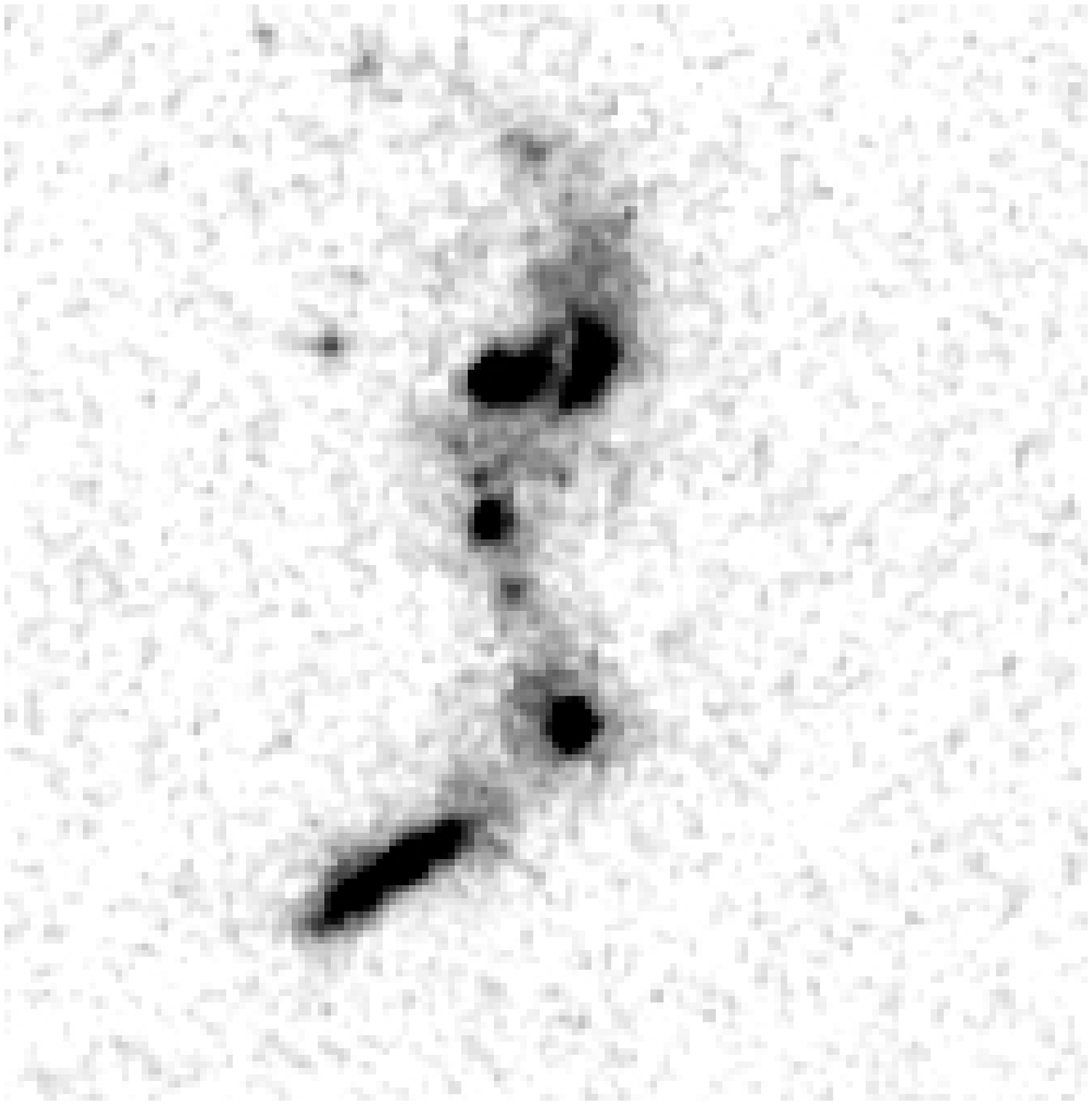}\\
\includegraphics[width=\figsmall,height=\figsmall]{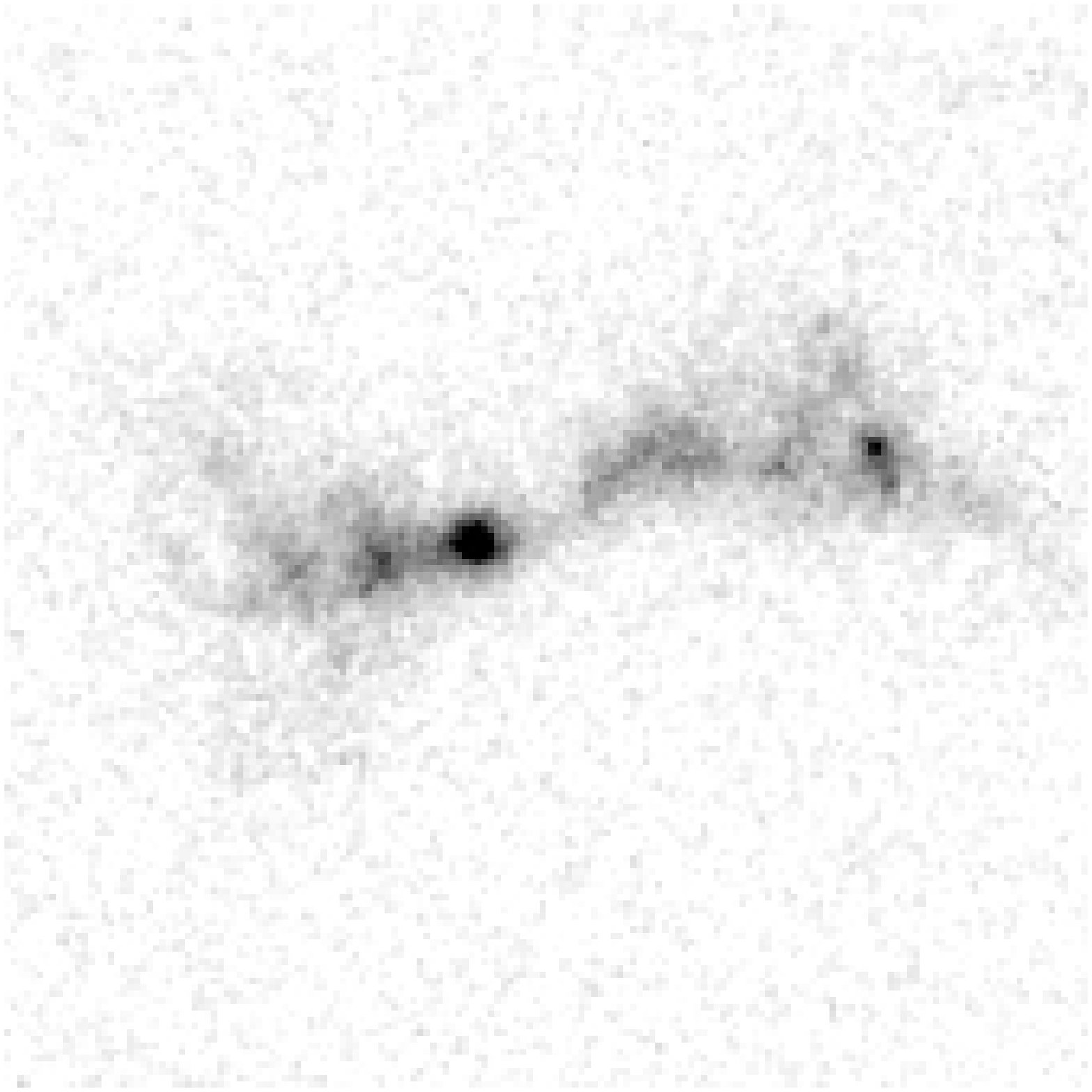}\quad
\includegraphics[width=\figsmall,height=\figsmall]{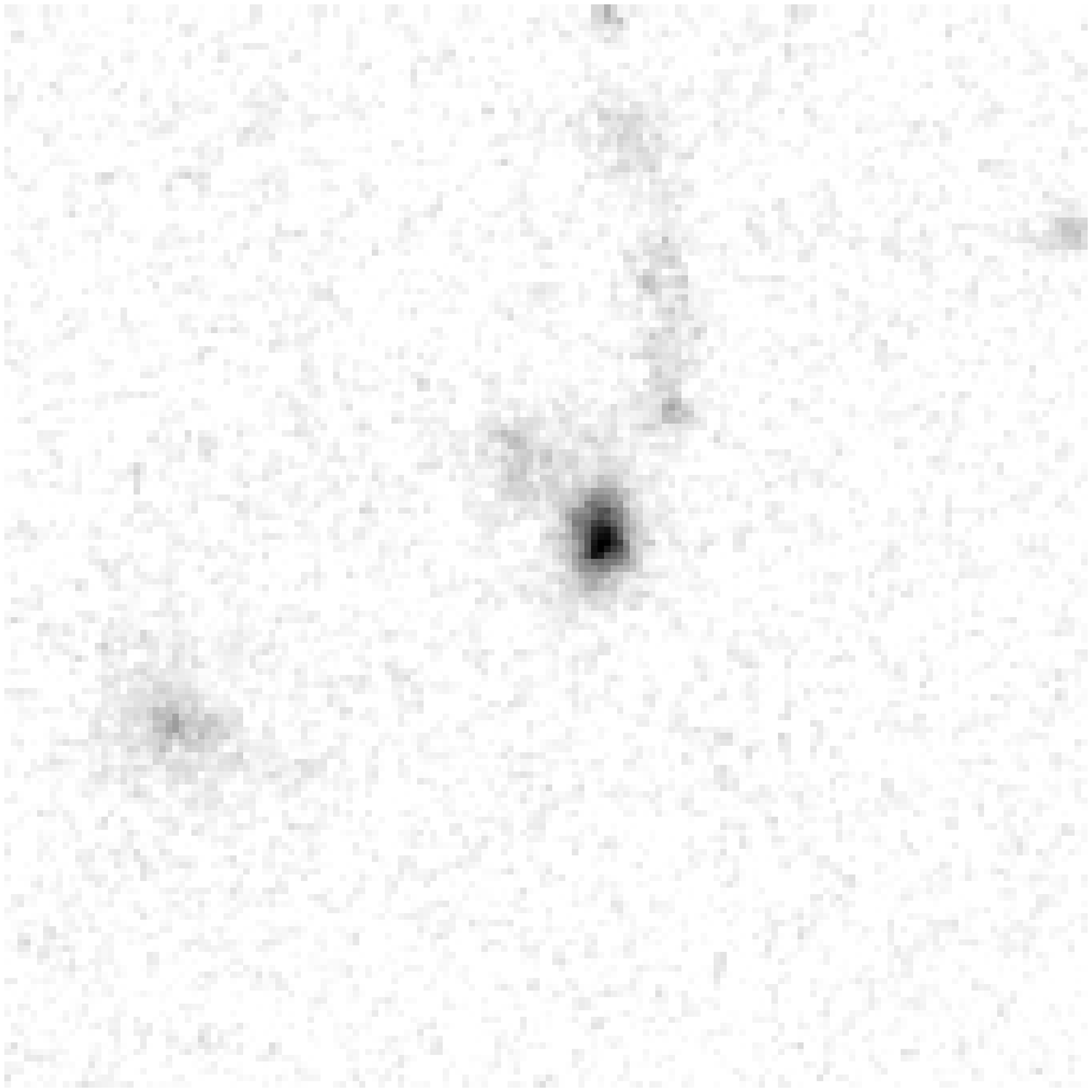}\quad
\includegraphics[width=\figsmall,height=\figsmall]{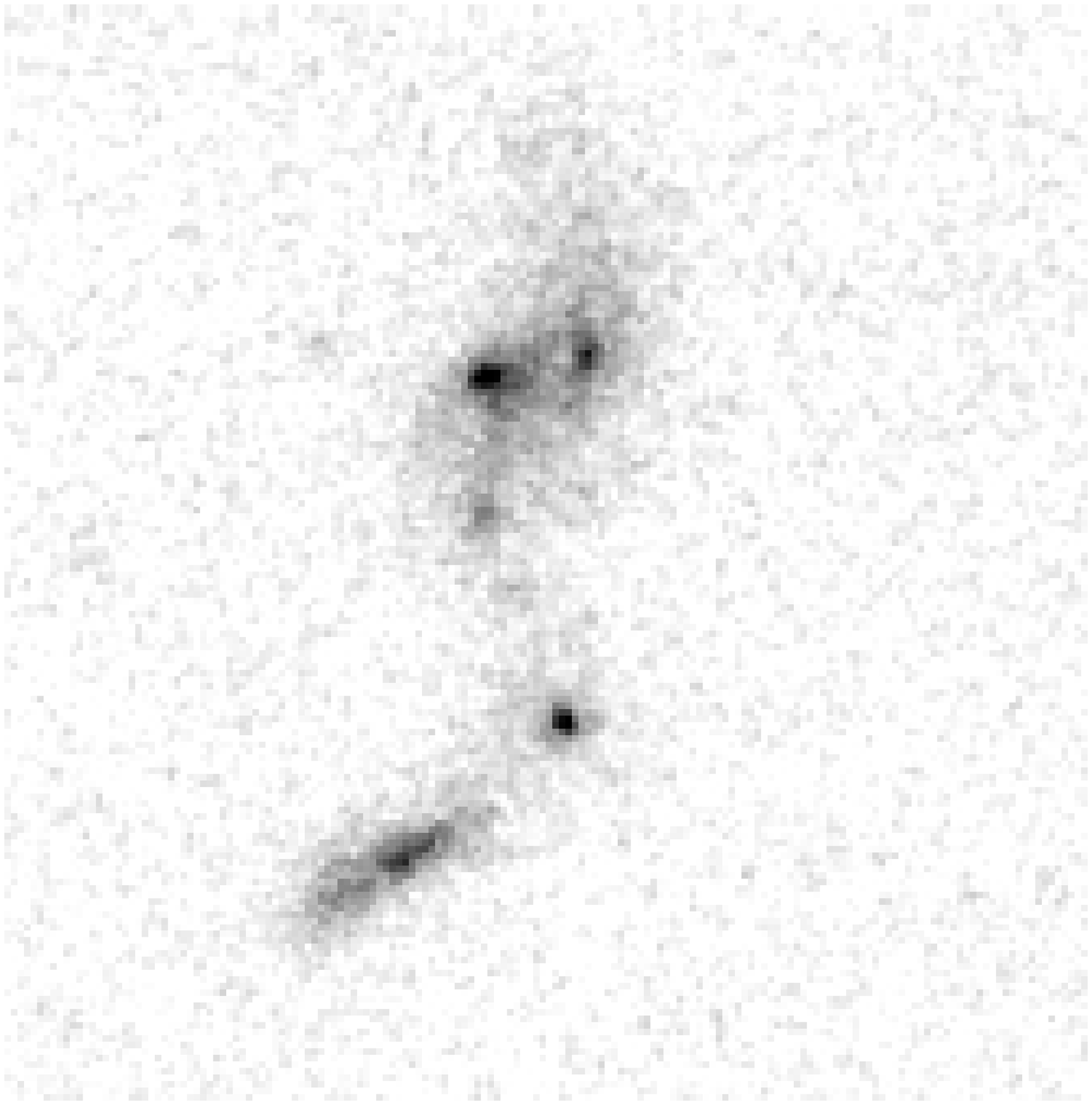}
\caption{Images of three ``chain" LBGs with their UV IDs of 60187,
75408 and 43975 in B07 from left to right. The upper and lower
panels denote the images in F606W and F850LP,
respectively.}\label{fig:figgalaxy1}
\end{figure}

\begin{figure}
\includegraphics[width=\figsmall,height=\figsmall]{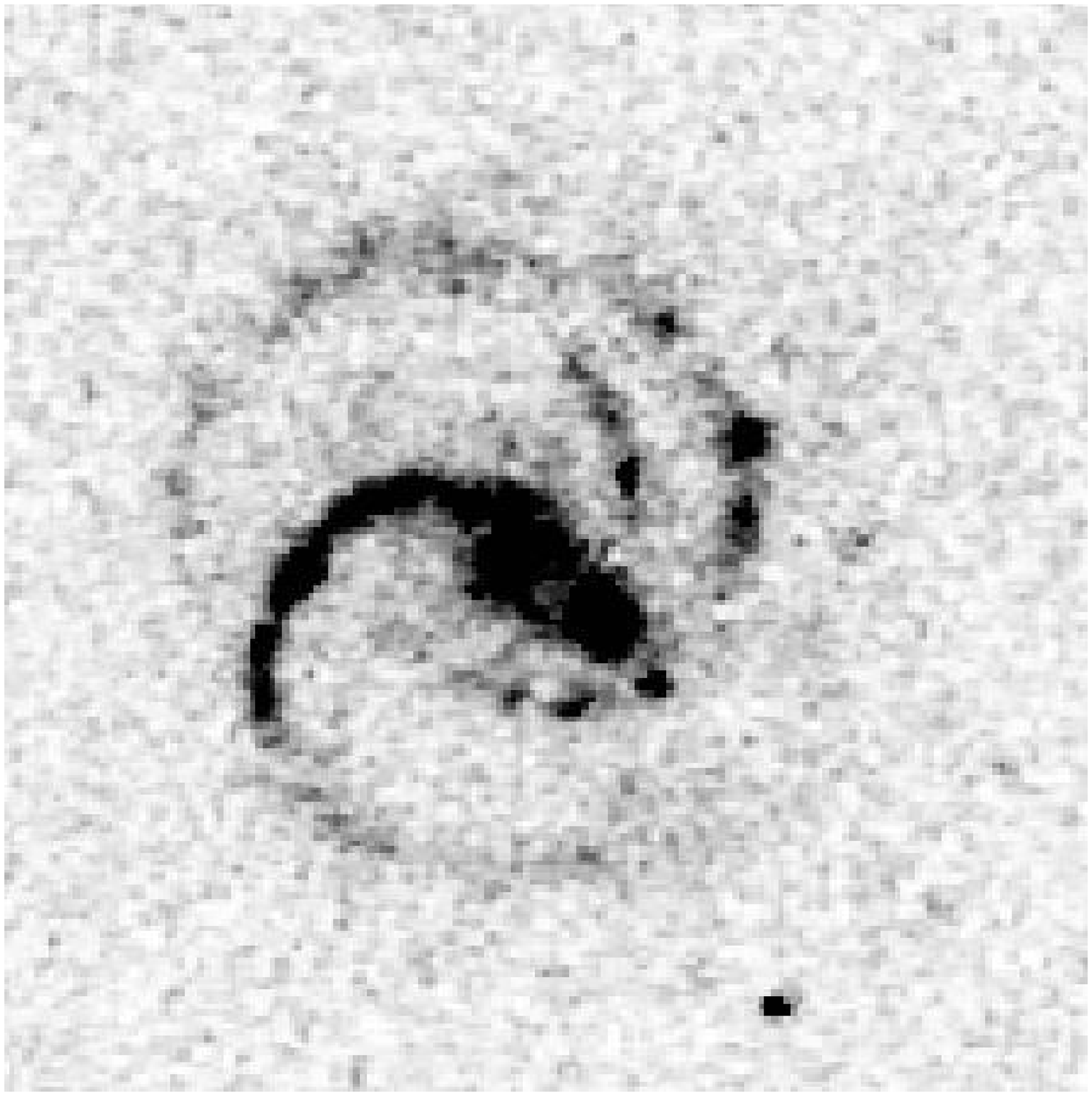}\quad
\includegraphics[width=\figsmall,height=\figsmall]{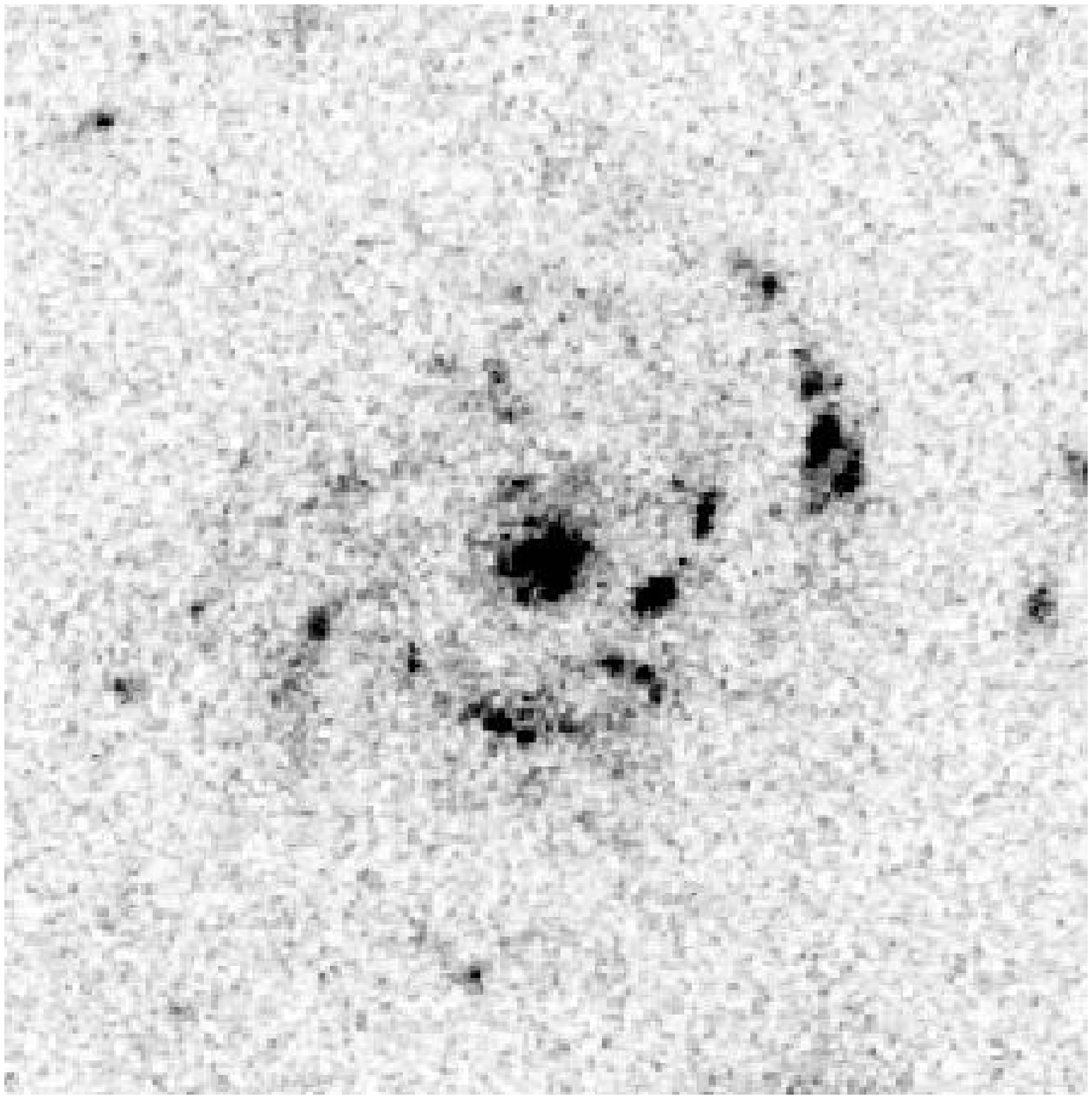}\quad
\includegraphics[width=\figsmall,height=\figsmall]{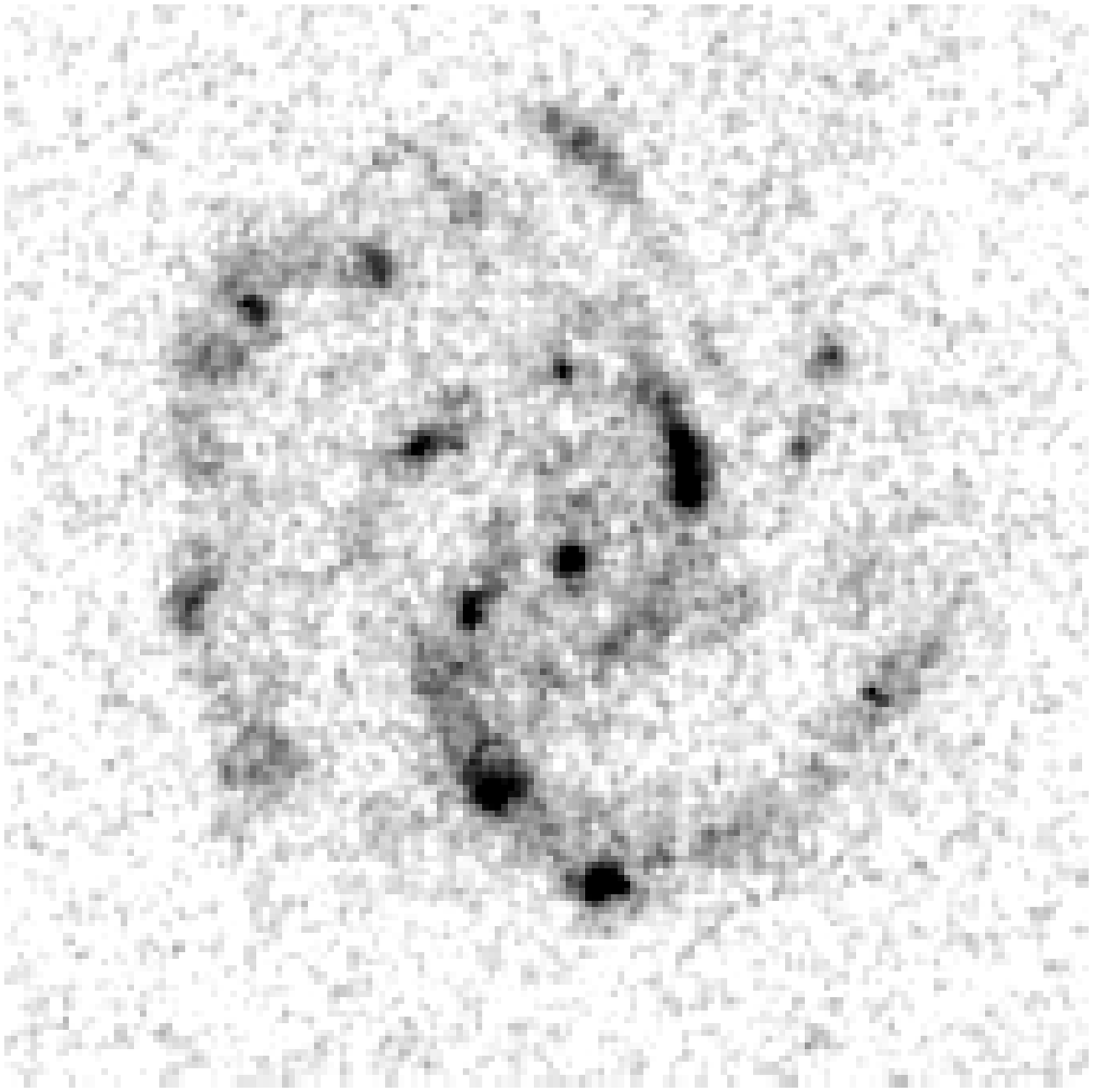}\\
\includegraphics[width=\figsmall,height=\figsmall]{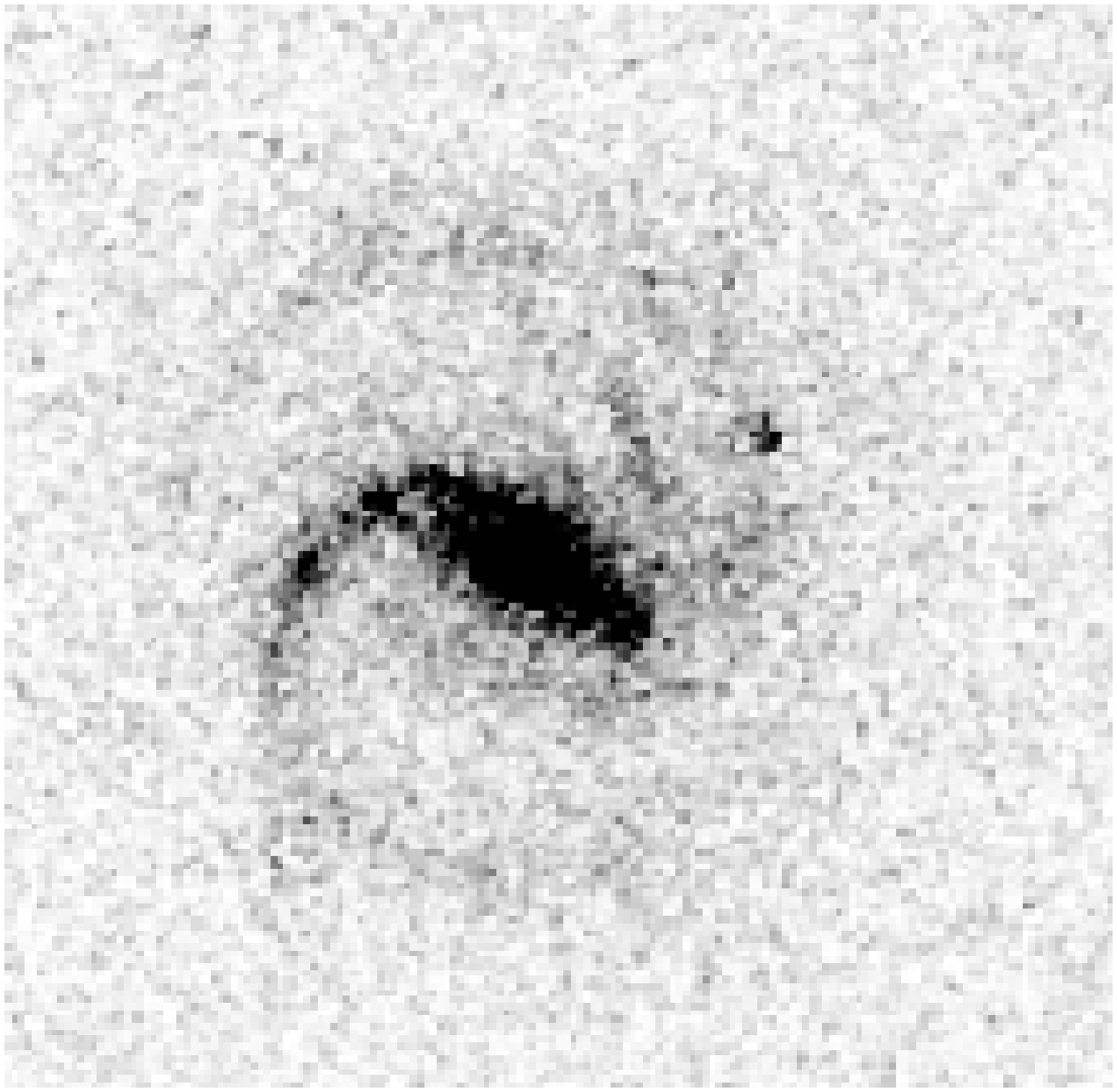}\quad
\includegraphics[width=\figsmall,height=\figsmall]{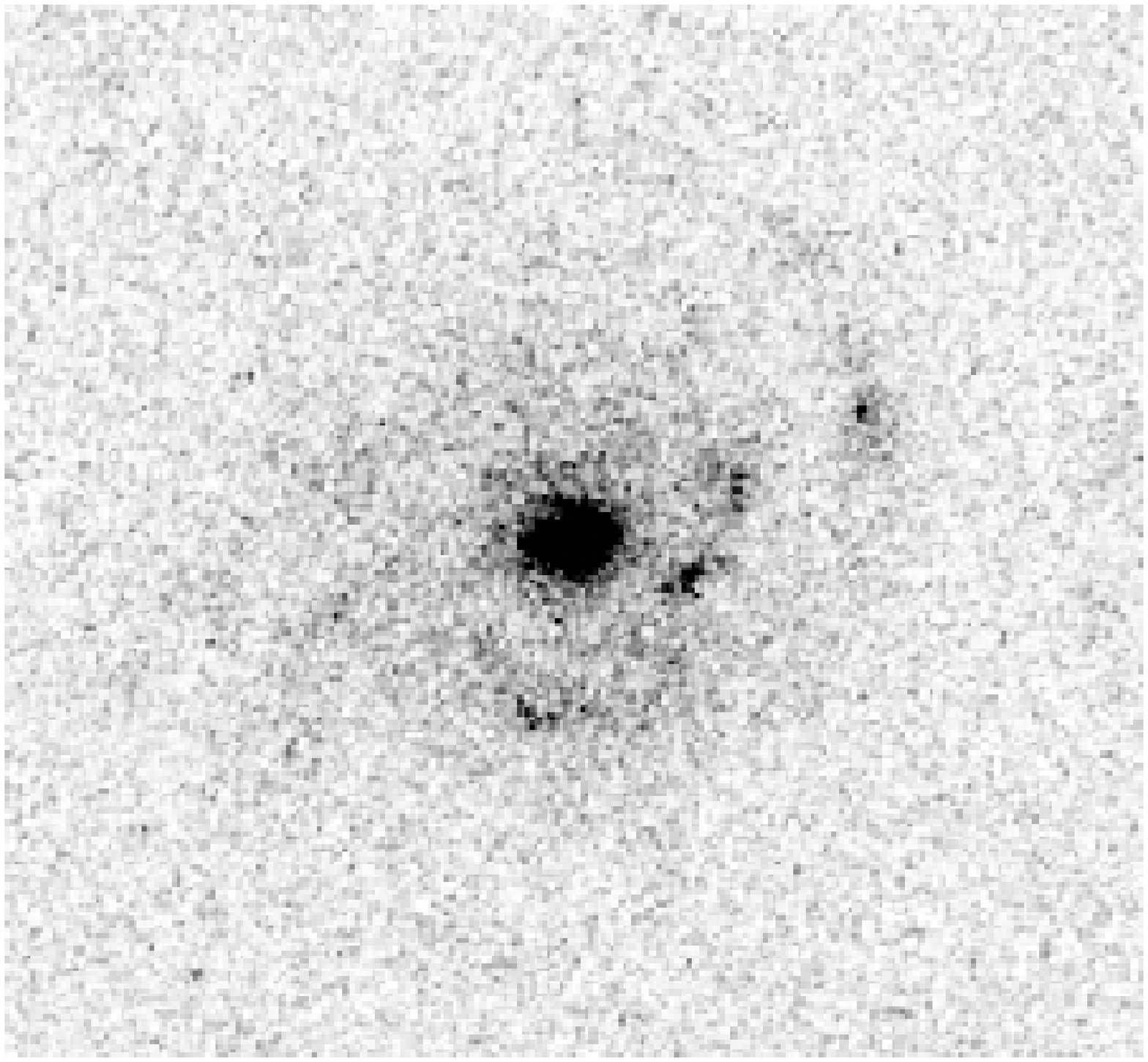}\quad
\includegraphics[width=\figsmall,height=\figsmall]{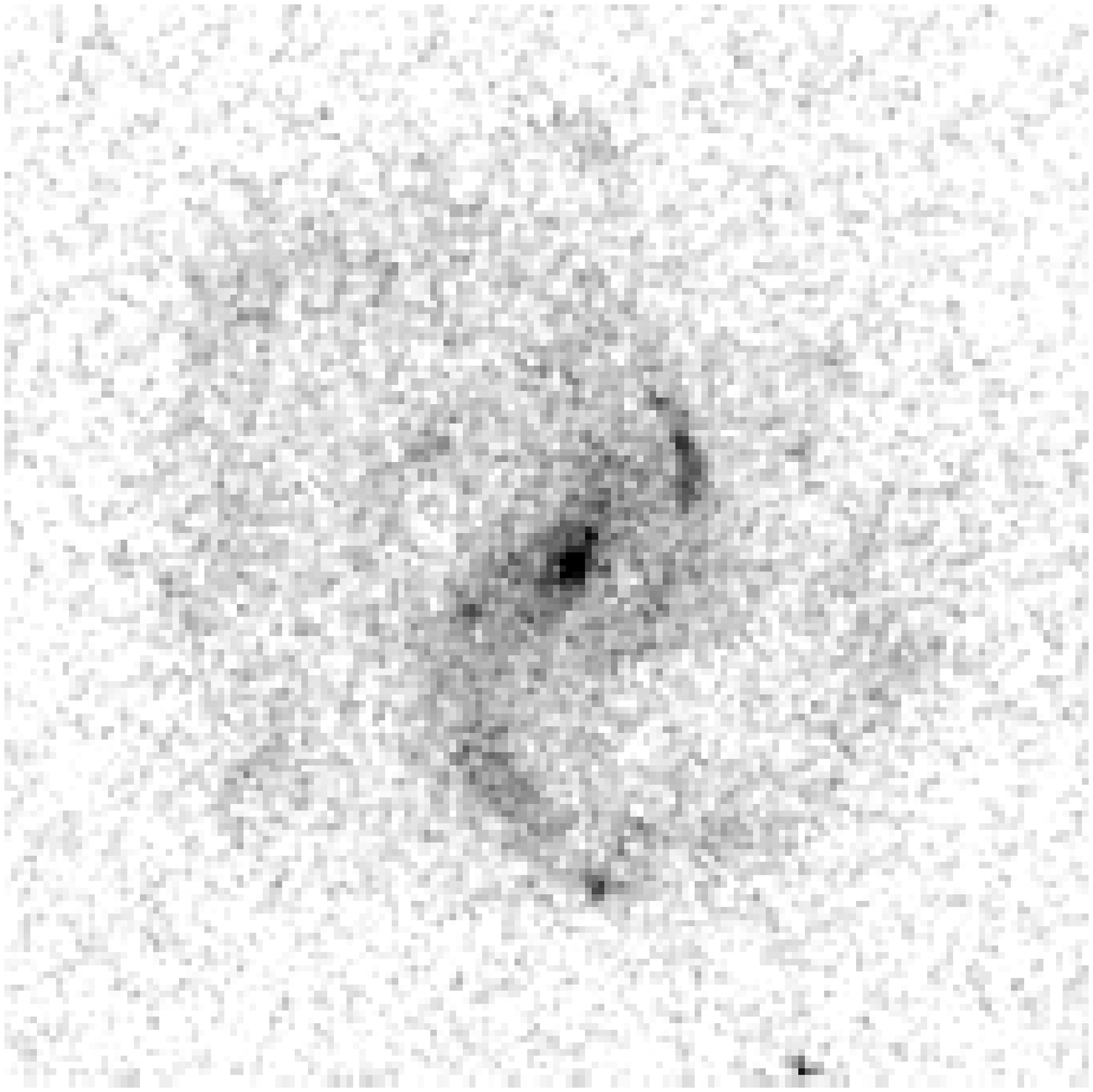}
\caption{Same as Fig. \ref{fig:figgalaxy1} but for three ``spiral"
LBGs with their UV IDs of 47798, 36485 and 59934 in B07 from left to
right, respectively.}\label{fig:figgalaxy2}
\end{figure}

\begin{figure}
\includegraphics[width=\figsmall,height=\figsmall]{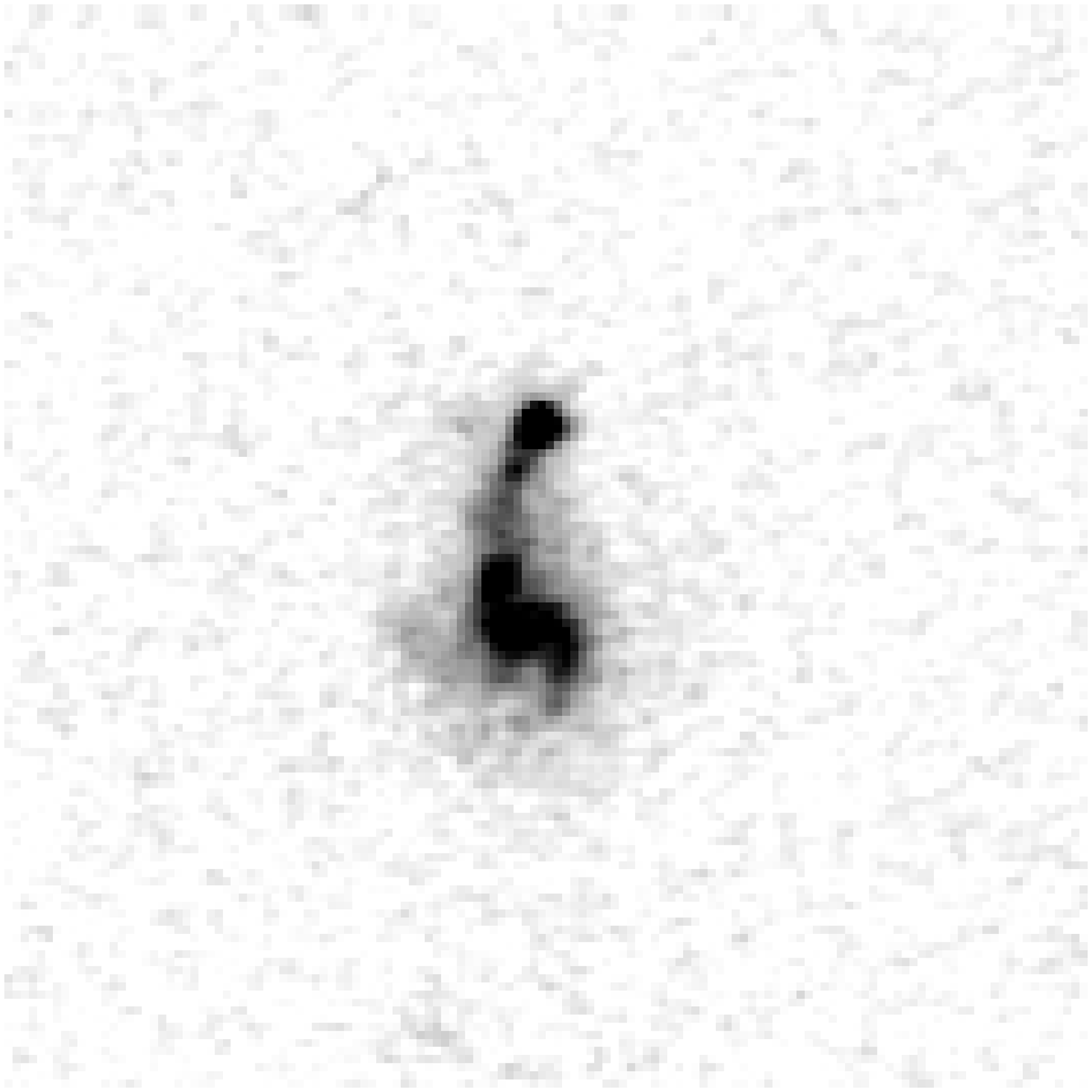}\quad
\includegraphics[width=\figsmall,height=\figsmall]{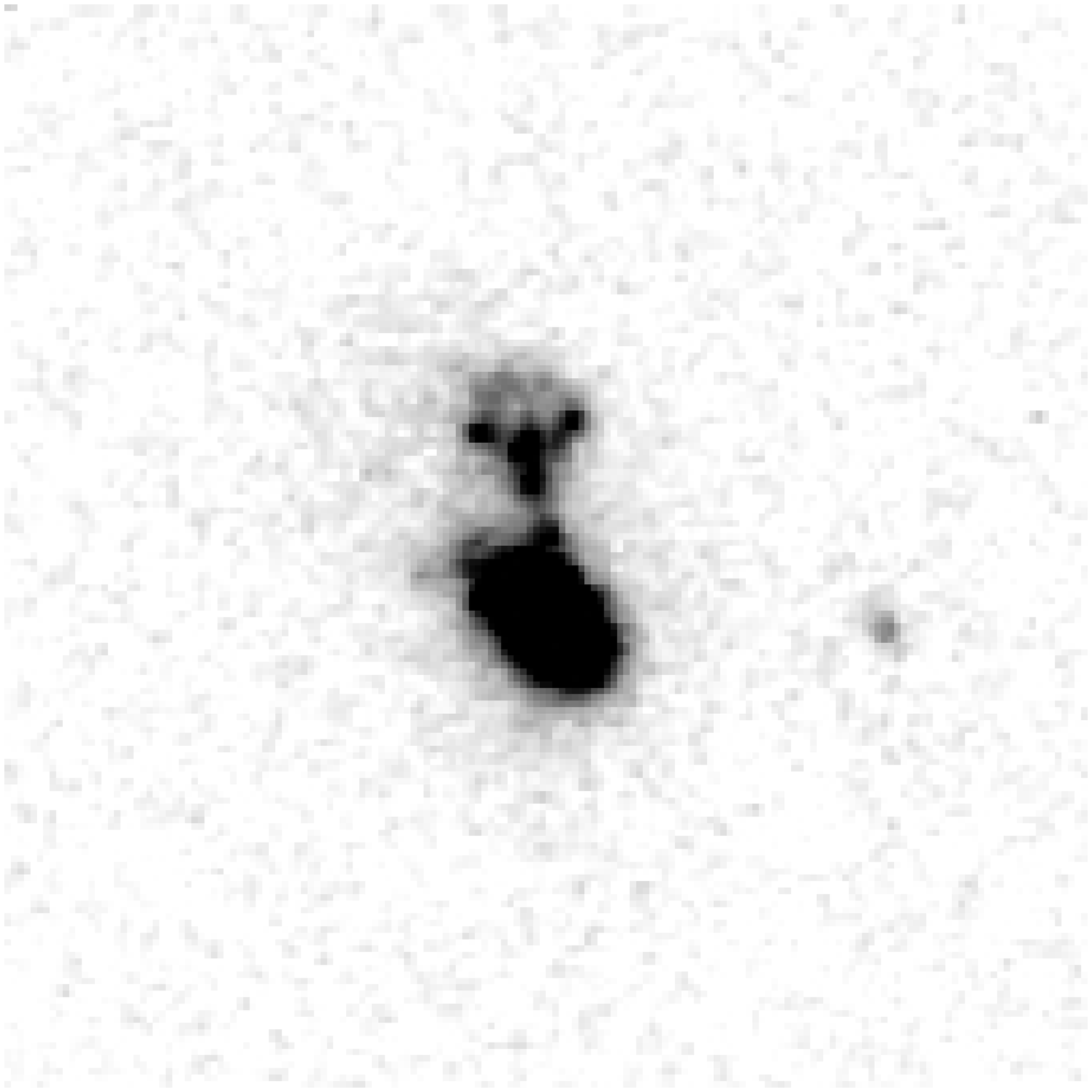}\quad
\includegraphics[width=\figsmall,height=\figsmall]{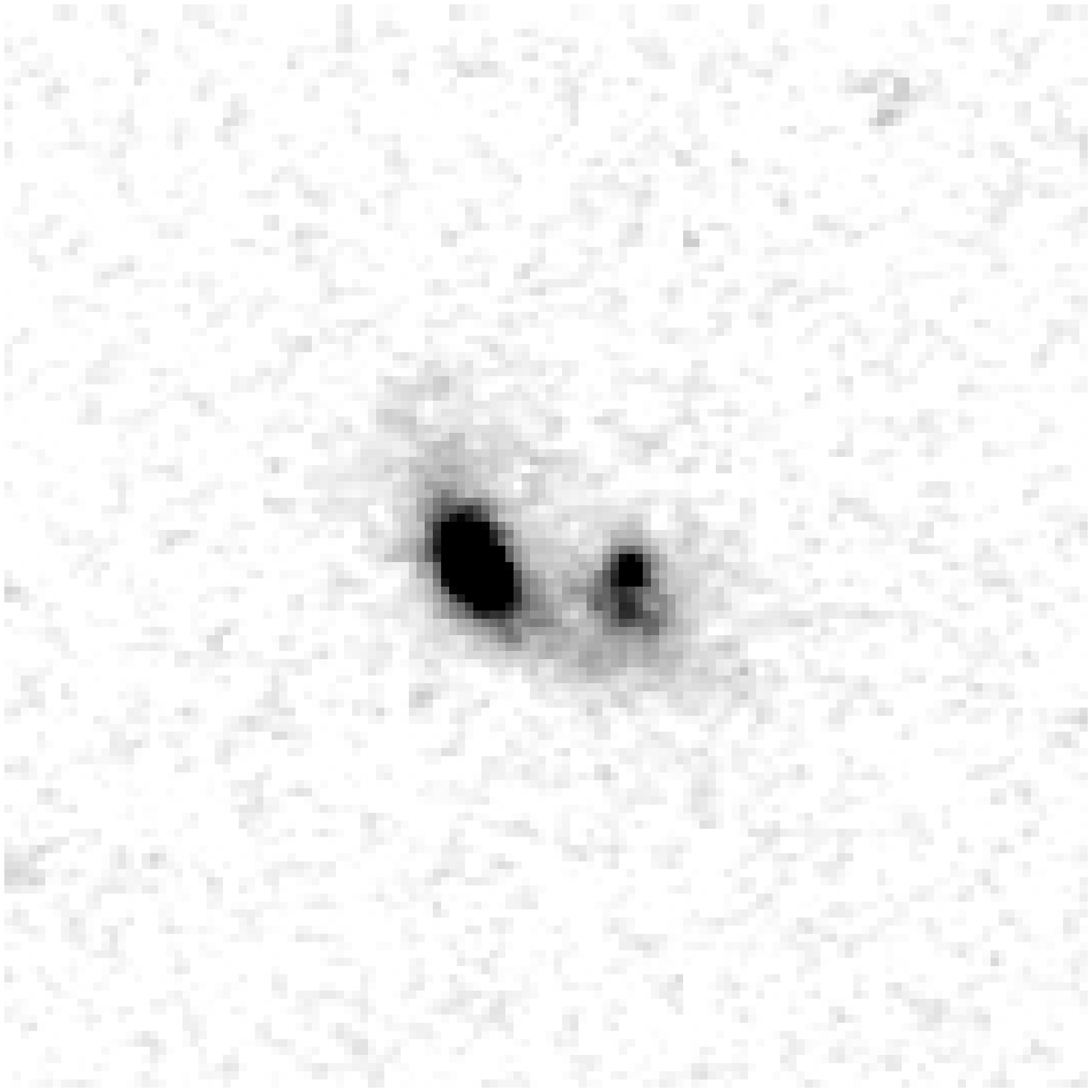}\\
\includegraphics[width=\figsmall,height=\figsmall]{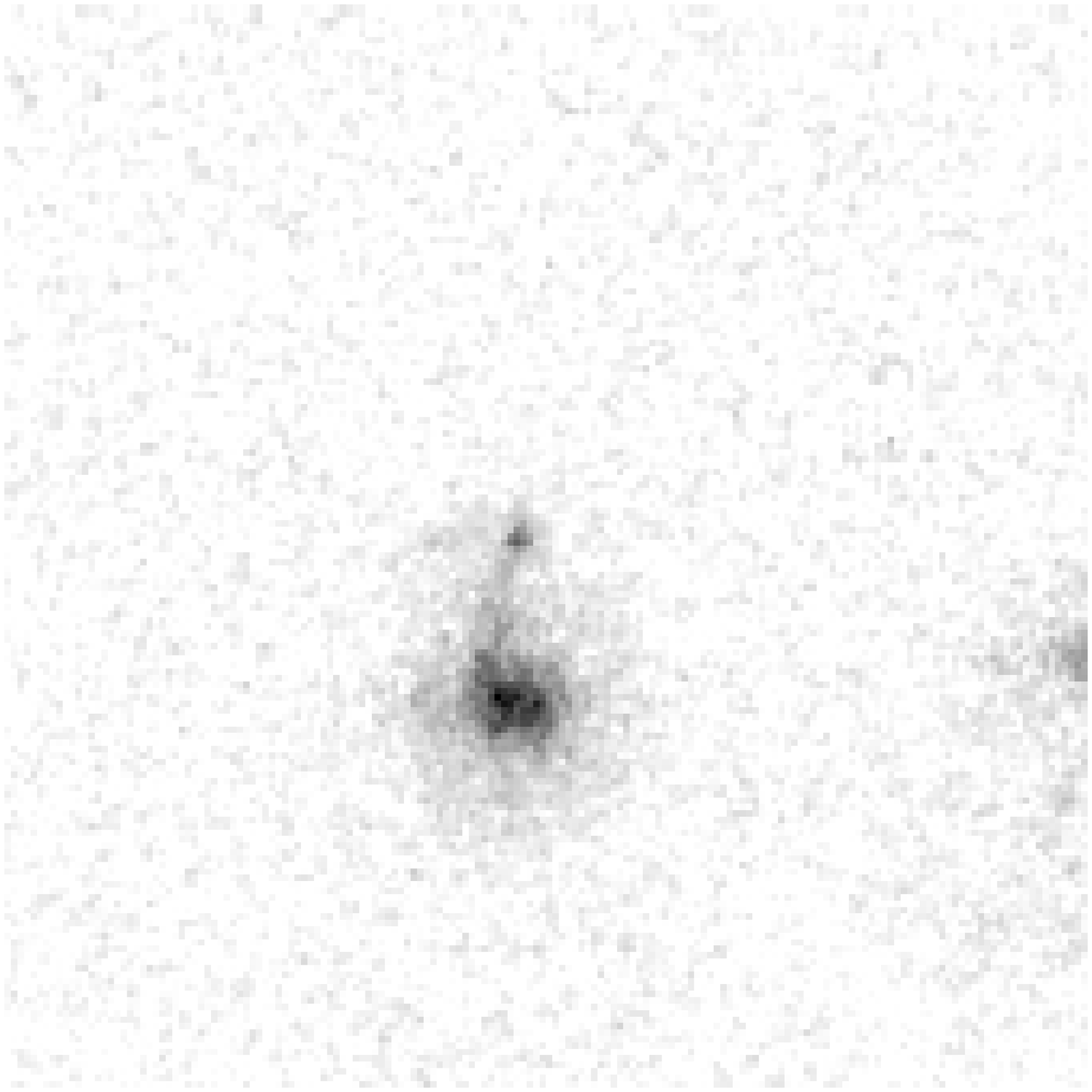}\quad
\includegraphics[width=\figsmall,height=\figsmall]{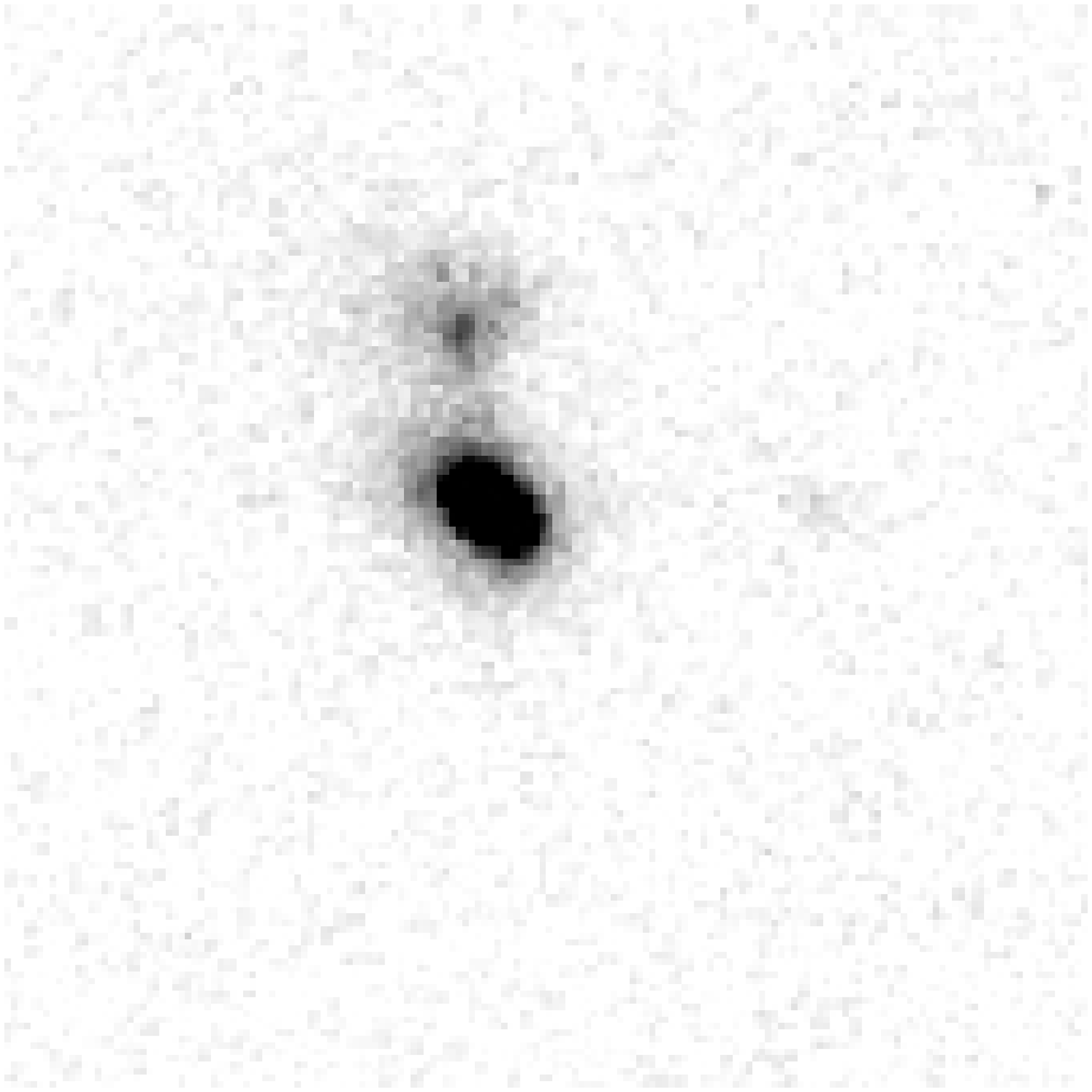}\quad
\includegraphics[width=\figsmall,height=\figsmall]{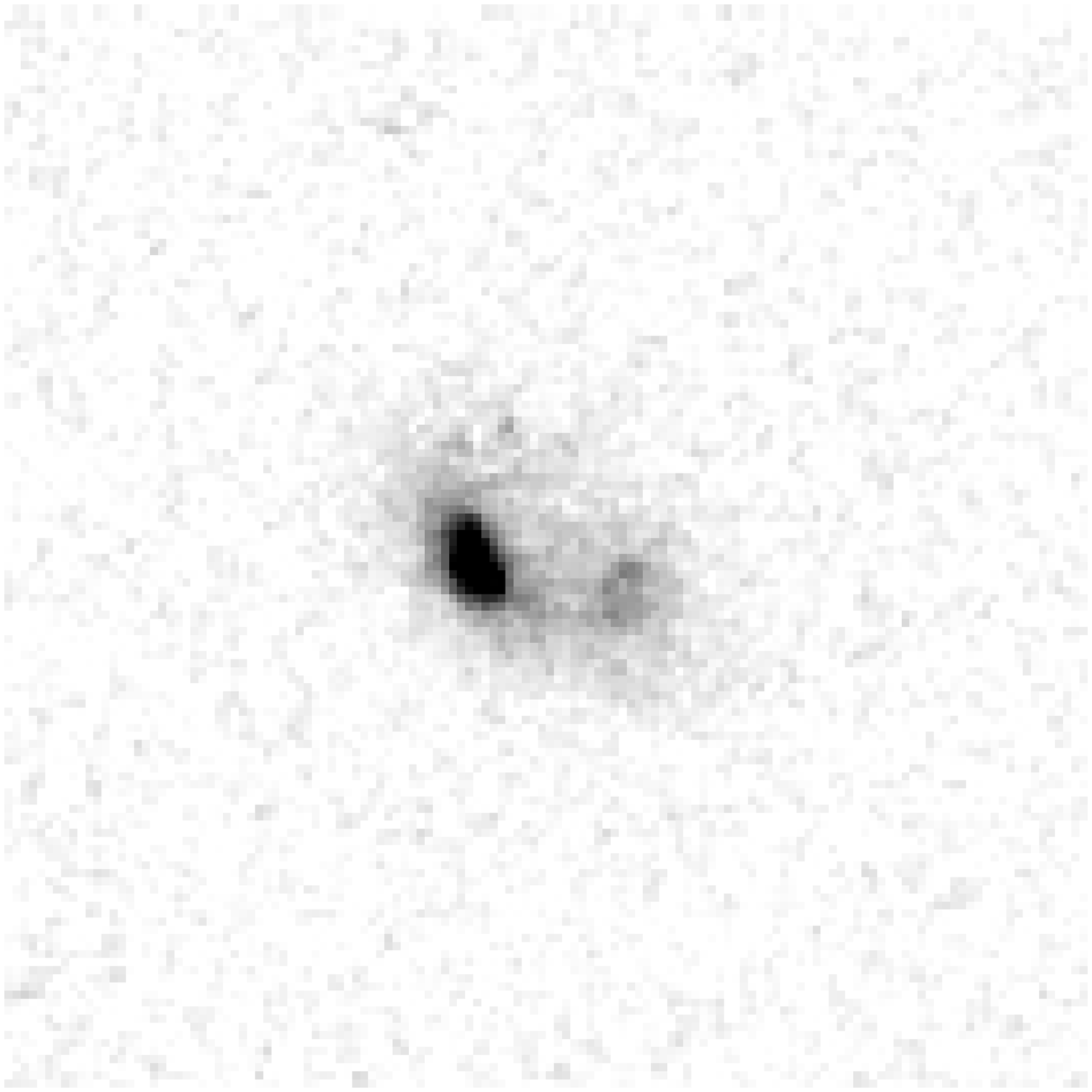}
\caption{Same as Fig. \ref{fig:figgalaxy1} but for three ``tadpole"
LBGs with their UV IDs of 79716, 81474 and 73352 in B07 from left to
right, respectively.}\label{fig:figgalaxy3}
\end{figure}

\begin{figure}
\includegraphics[width=\figsmall,height=\figsmall]{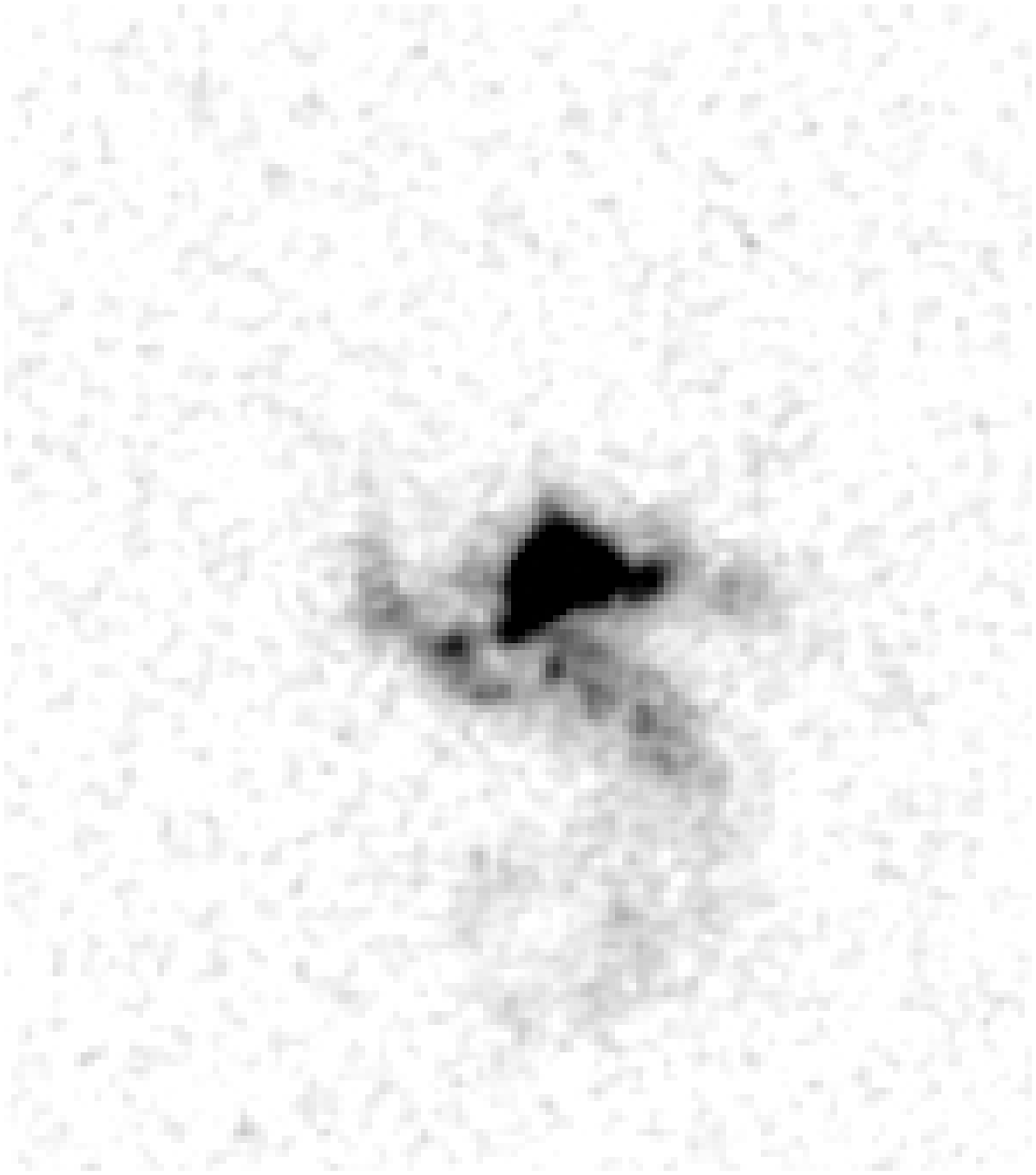}\quad
\includegraphics[width=\figsmall,height=\figsmall]{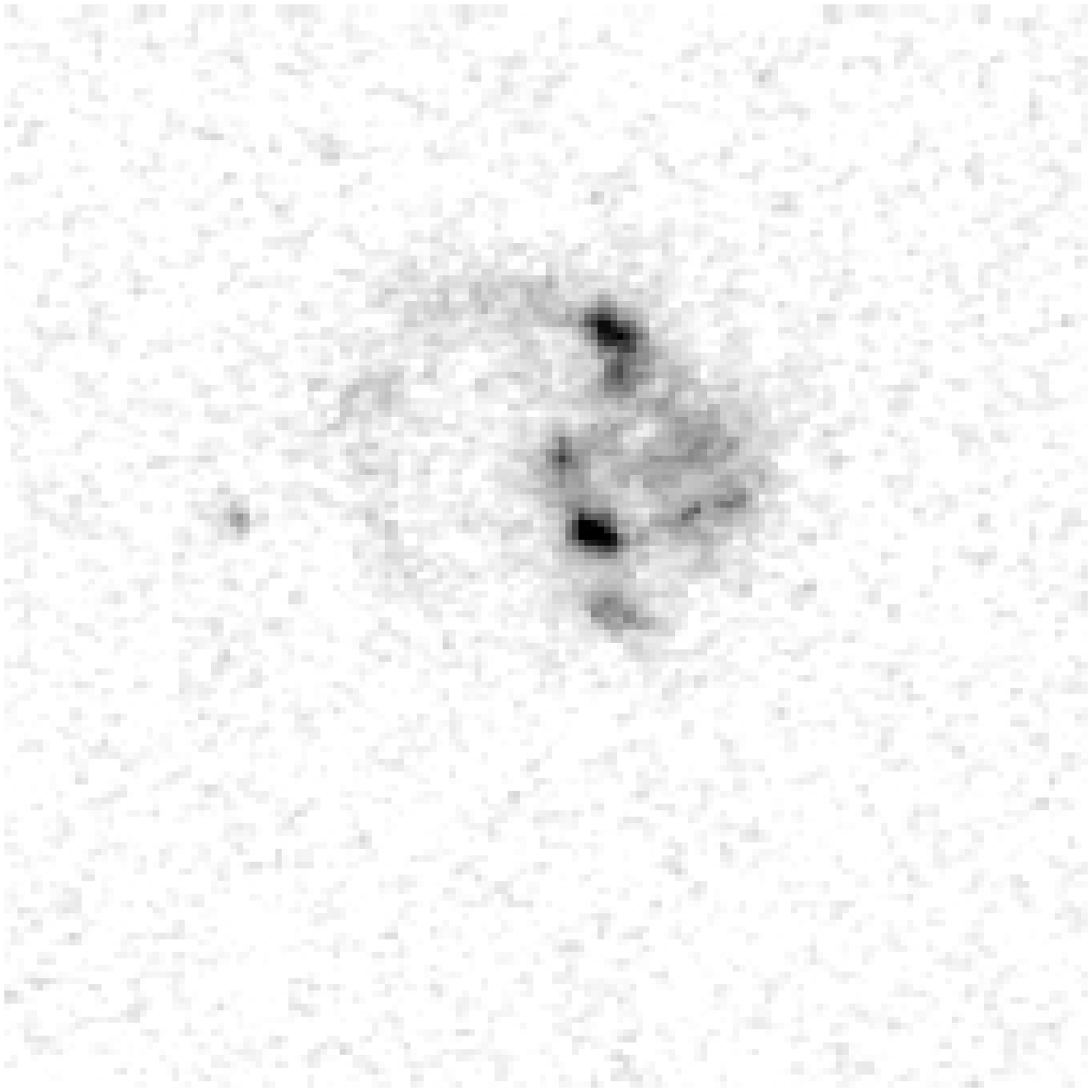}\quad
\includegraphics[width=\figsmall,height=\figsmall]{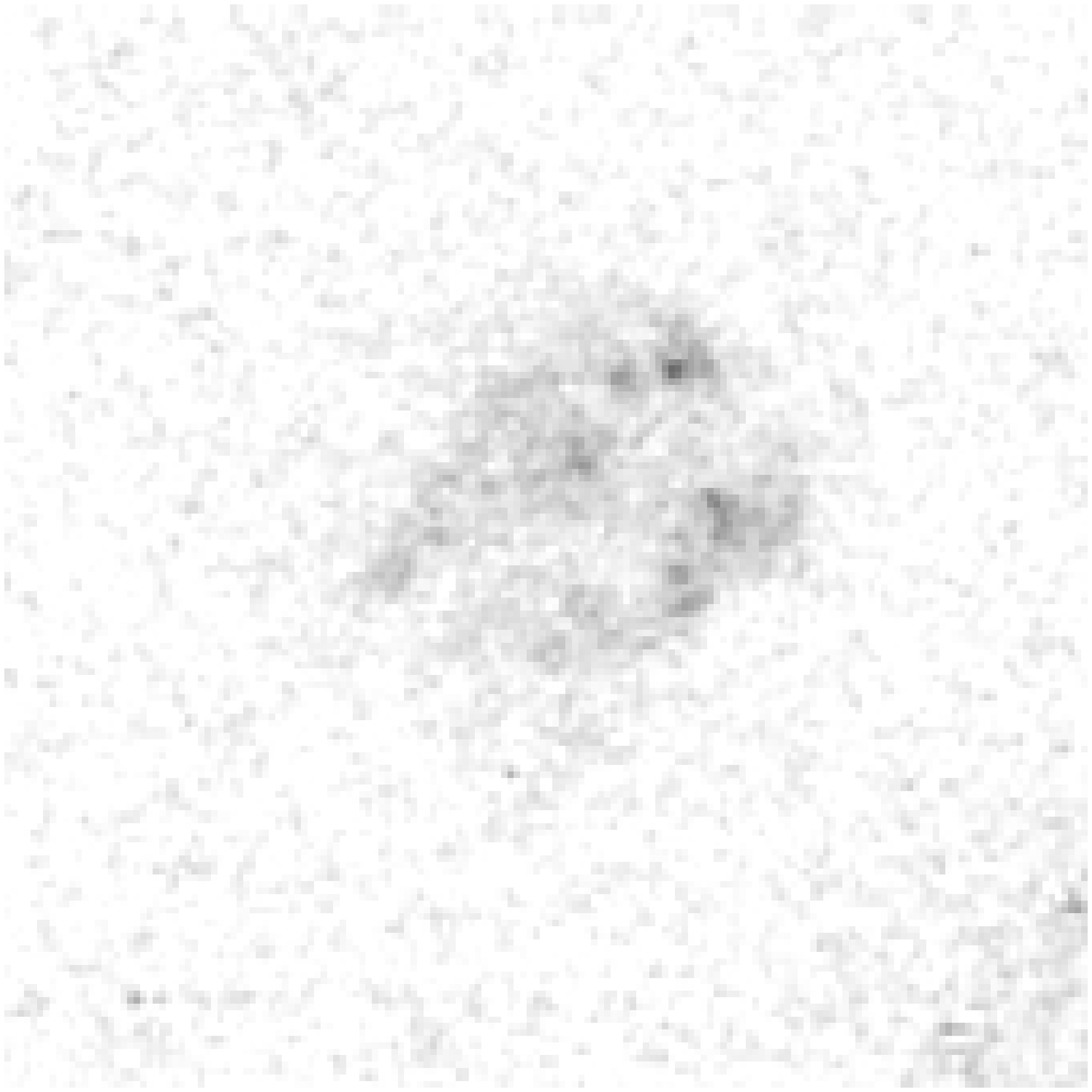}\\
\includegraphics[width=\figsmall,height=\figsmall]{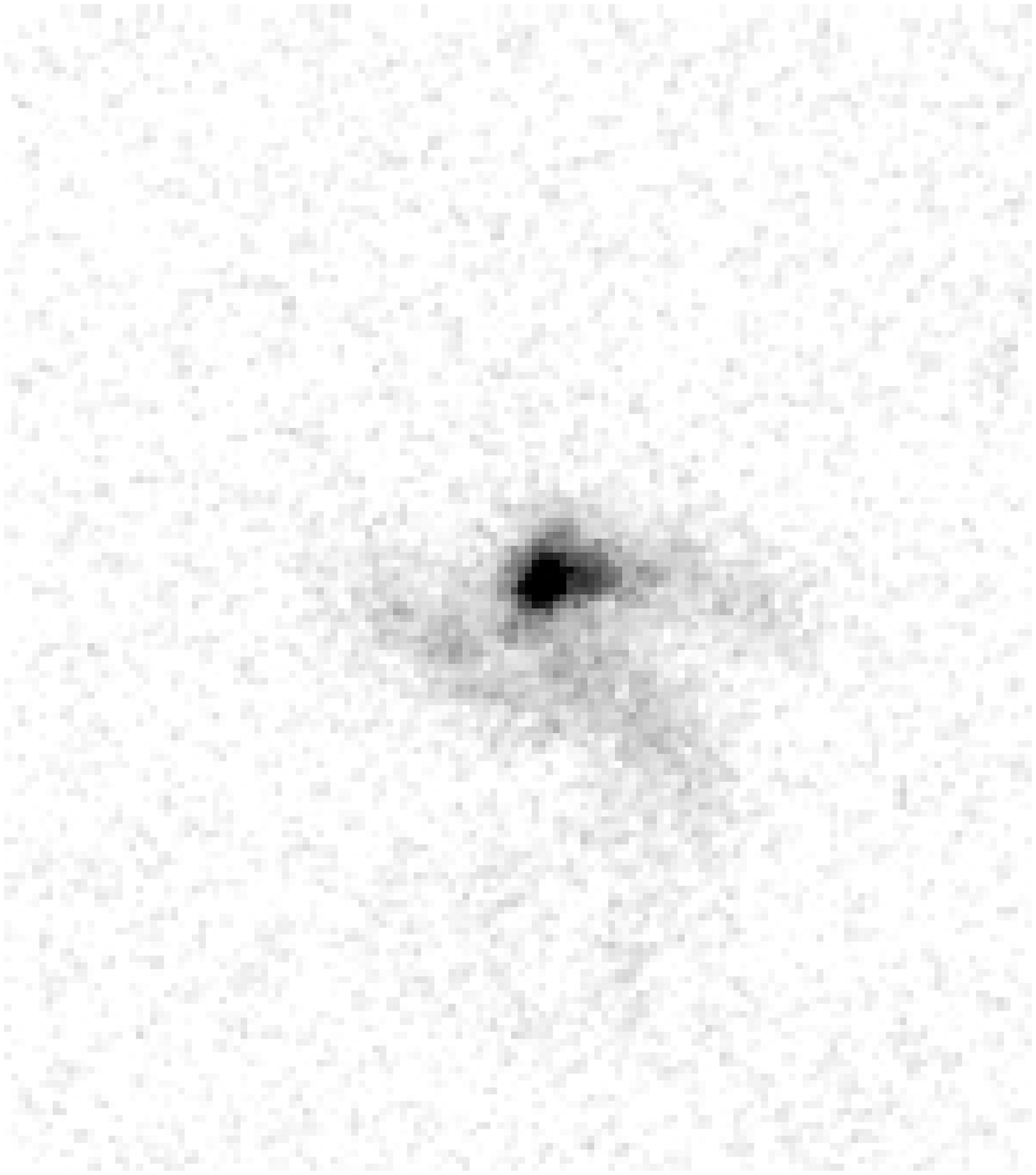}\quad
\includegraphics[width=\figsmall,height=\figsmall]{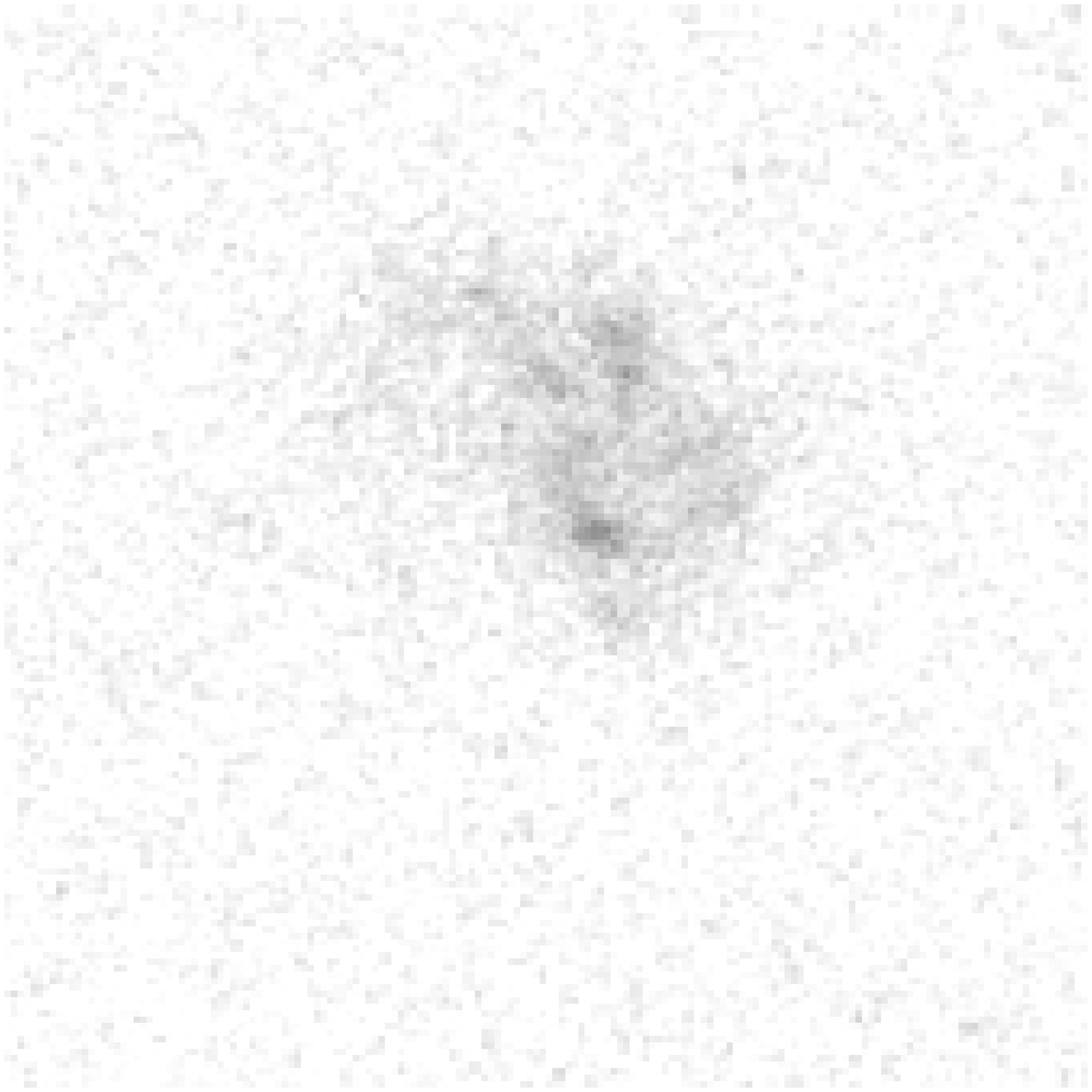}\quad
\includegraphics[width=\figsmall,height=\figsmall]{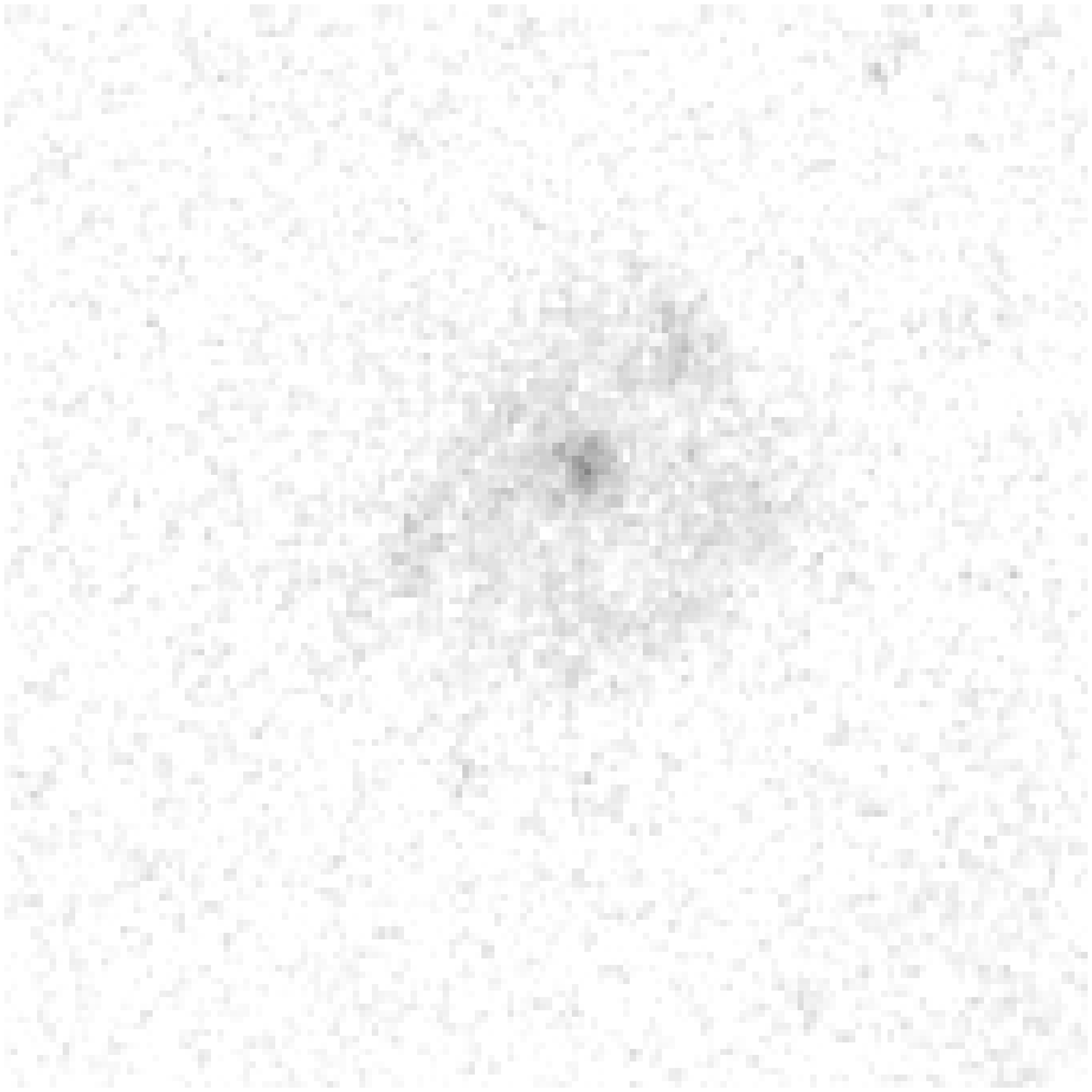}
\caption{Same as Fig. \ref{fig:figgalaxy1} but for three ``clump"
LBGs with their UV IDs of 78053, 36914, 58532 in B07,
respectively.}\label{fig:figgalaxy4}
\end{figure}

\begin{figure}
\includegraphics[width=\figsmall,height=\figsmall]{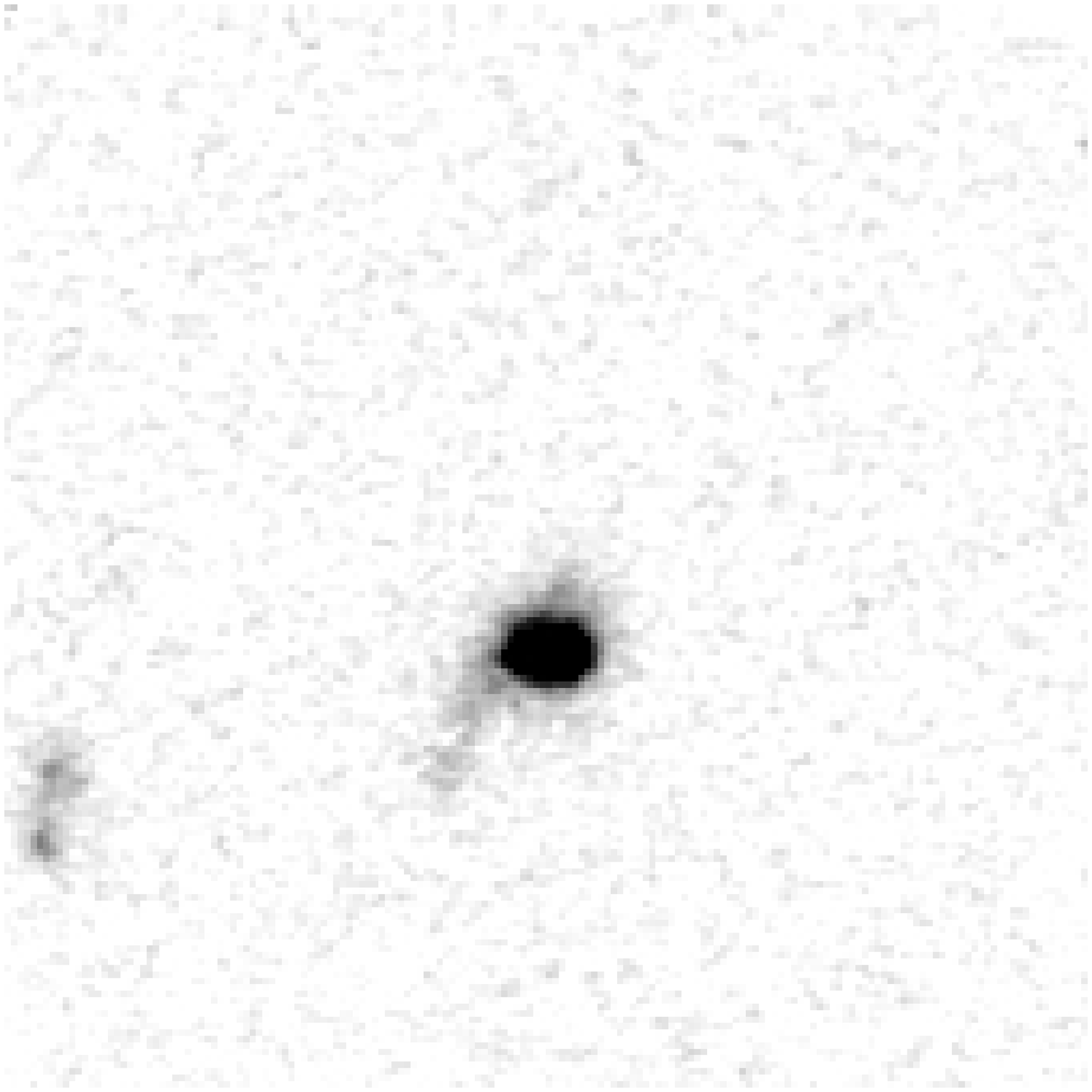}\quad
\includegraphics[width=\figsmall,height=\figsmall]{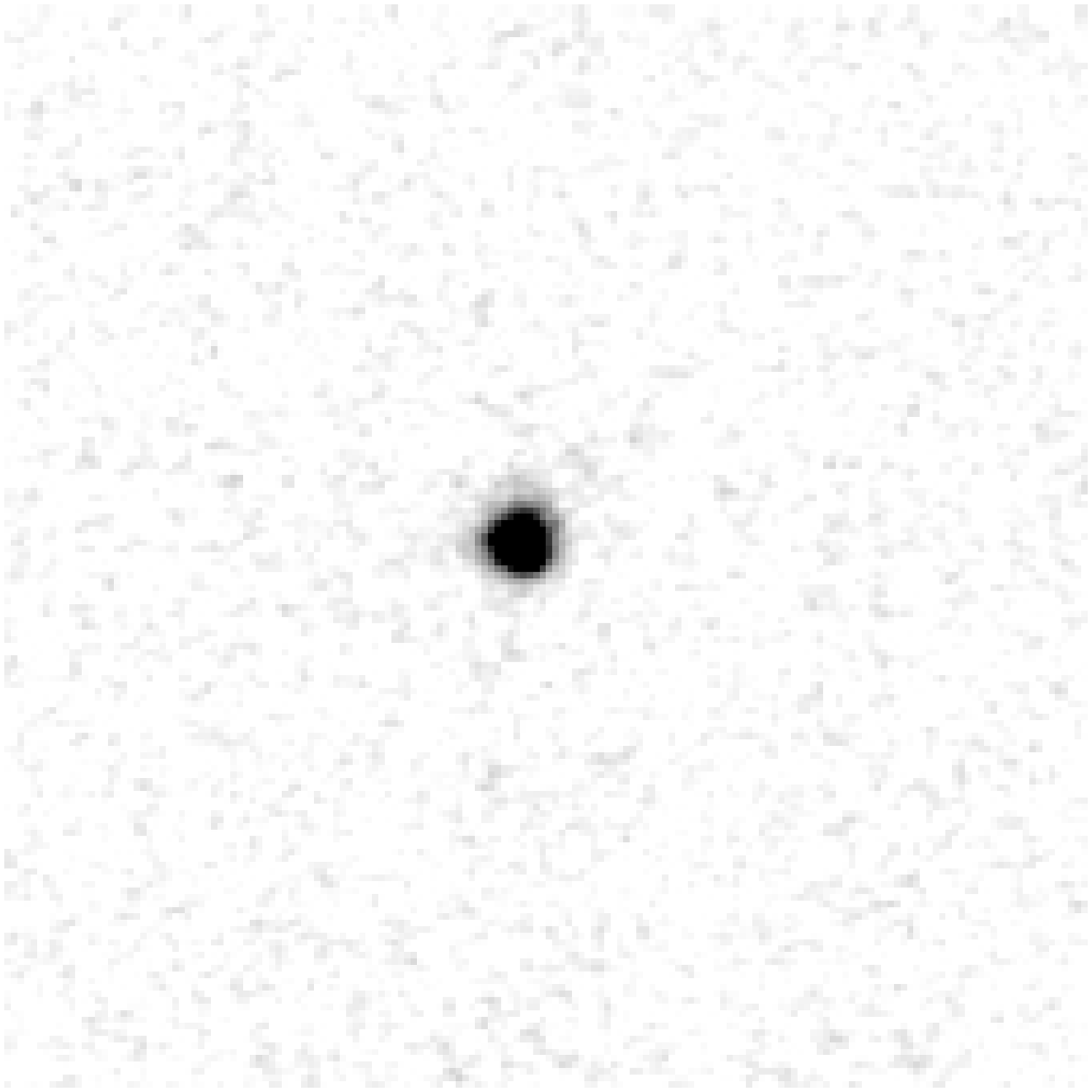}\quad
\includegraphics[width=\figsmall,height=\figsmall]{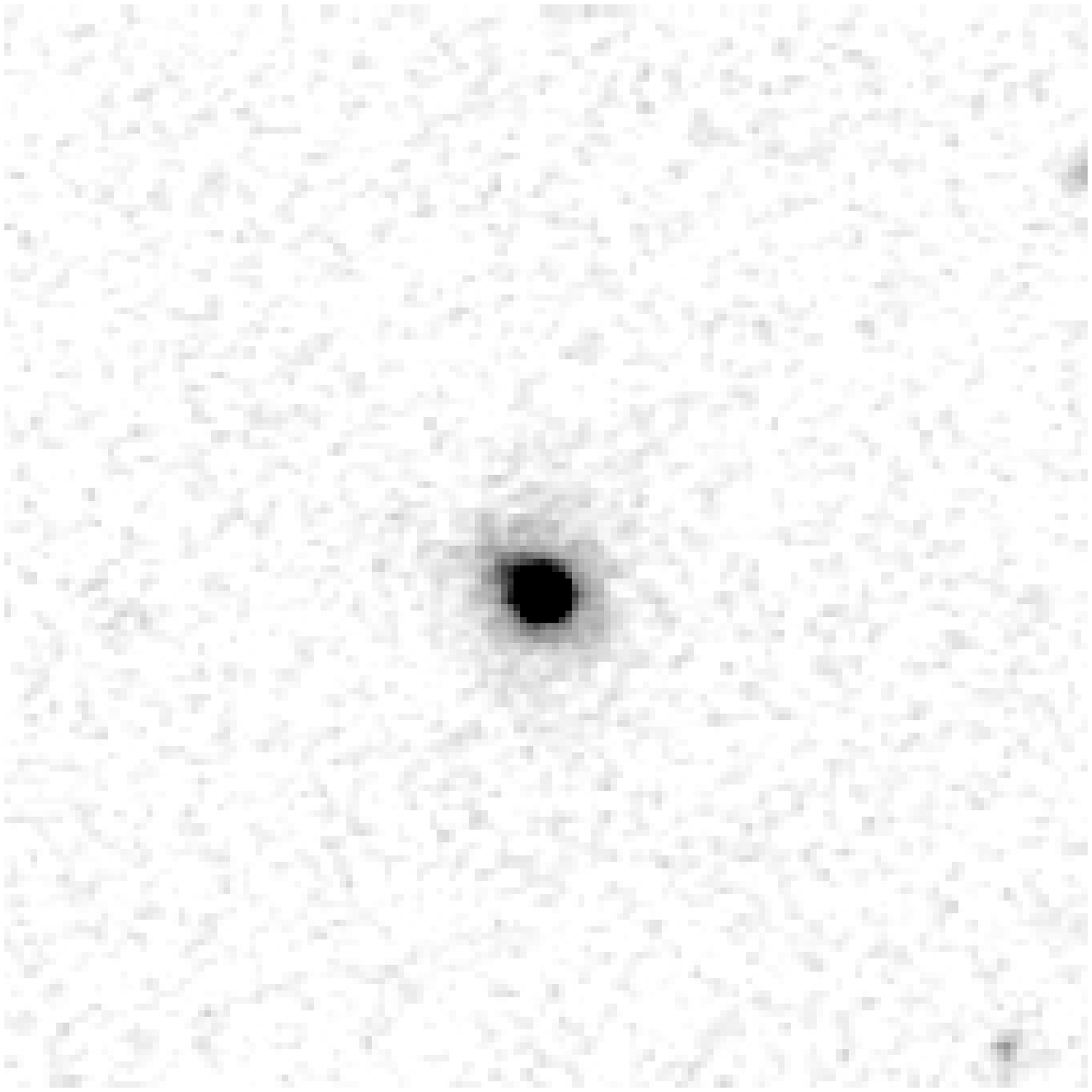}\\
\includegraphics[width=\figsmall,height=\figsmall]{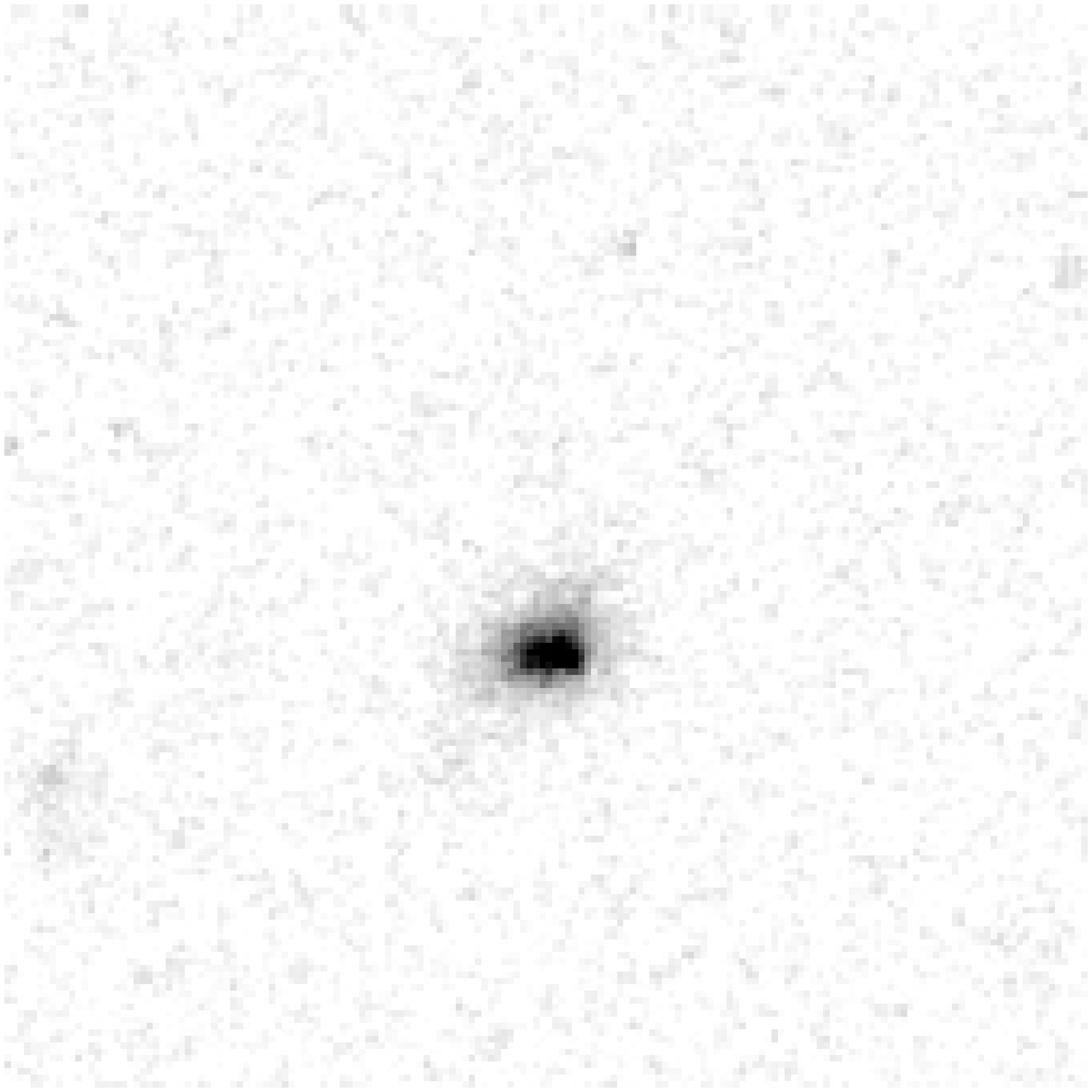}\quad
\includegraphics[width=\figsmall,height=\figsmall]{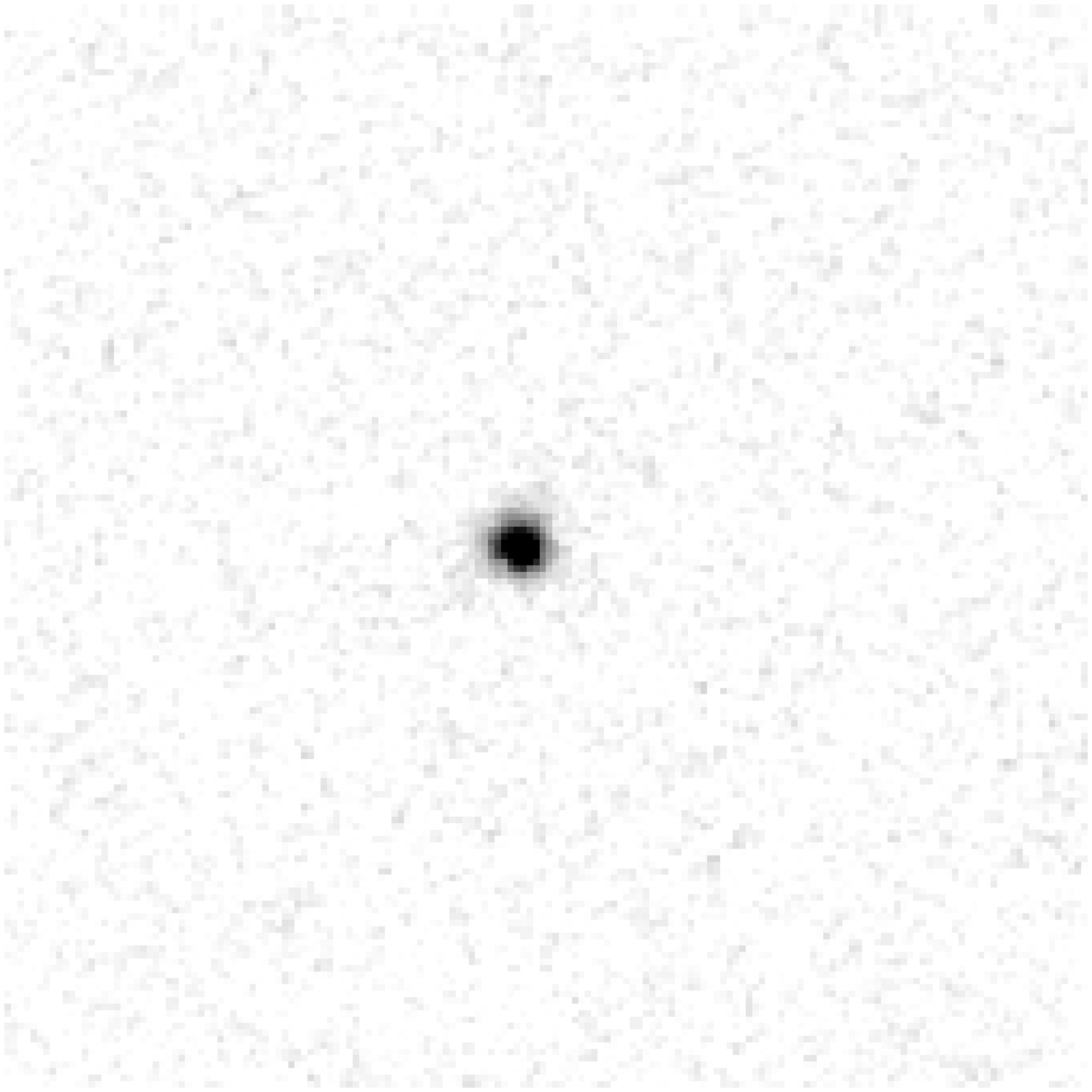}\quad
\includegraphics[width=\figsmall,height=\figsmall]{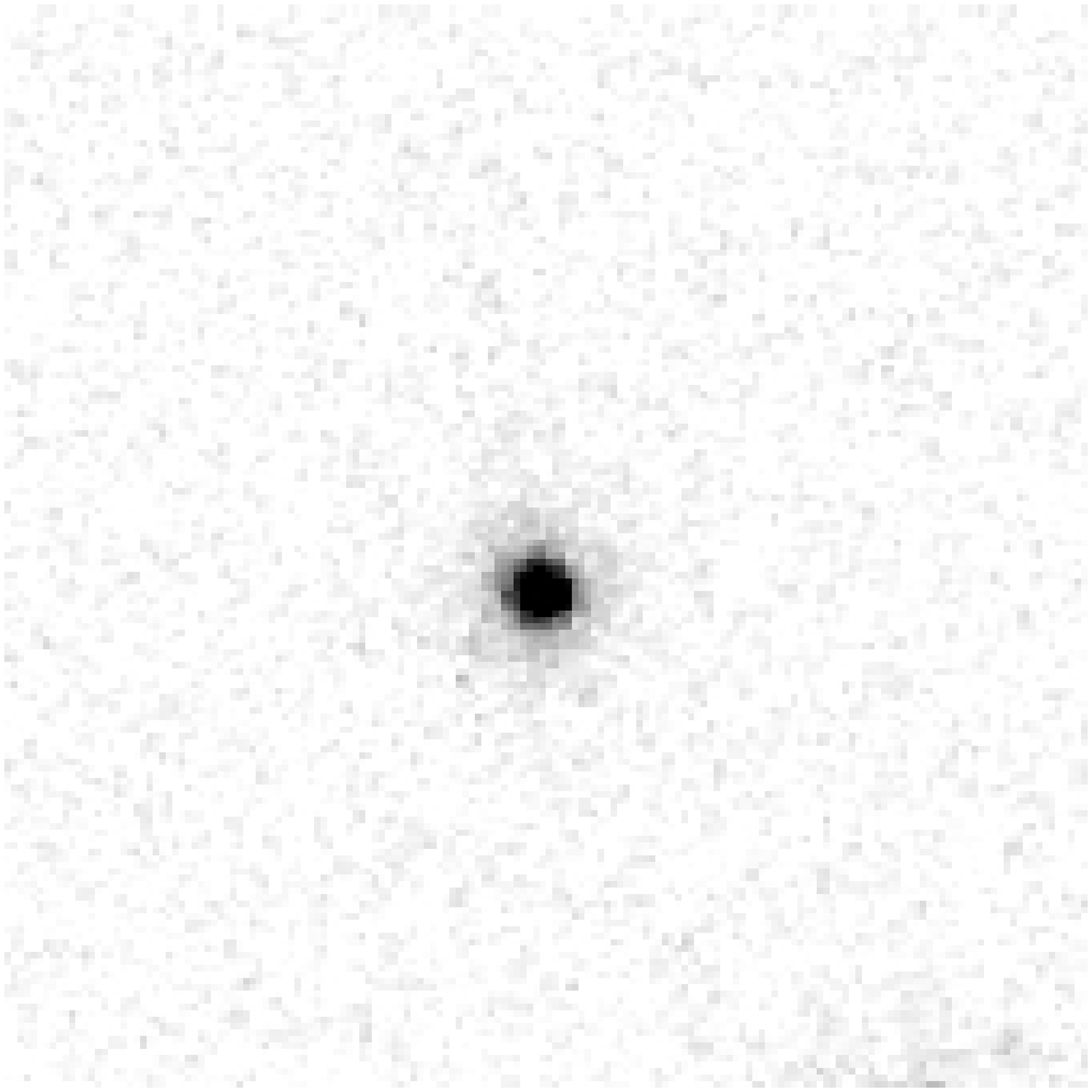}
\caption{Same as Fig. \ref{fig:figgalaxy1} but for three ``bulge"
LBGs with their UV IDs of 69074, 57264, 54502 in B07,
respectively.}\label{fig:figgalaxy5}
\end{figure}

It can be seen from the figure that LBGs in the EST
and Irr groups distributes along the ``main sequence" of star
forming galaxies suggested by \citet{Elbaz11}, with the
considerations of uncertainties and the redshift range. Most of LBGs
in the SB group locate in the ``starburst" region, away from the
``main sequence". It is in consistence with the spectral type
classifications of these LBGs during SED fittings.
%
%
Distances to the ``main sequence", i.e., the inverse way of star
forming timescales, increase on average for LBGs from the EST, Irr
to SB groups. Farer distances to the main sequence imply more
dramatic in star formation.
%
%
Because two thirds of LBGs are classified in the SB
group, it means that LBGs are dominated by active star forming
galaxies from either a spectral type or a star forming timescale
point of view.

It is interesting that median $M_*$ of LBGs increases very rapidly
from $z\sim 5$ to $z\sim 3$ and then varies slowly to $z\sim 1$
while median SFRs are similar for LBGs at different redshifts (see
above in this section).
If we simply extrapolate the prescription
suggested by \citet{Elbaz11} and \citet{Sargent12} to $z\ga 2.5$,
%
the ``main sequence" of star forming galaxies changes by a factor of
$\sim 3.5$ while the median SFR/$M_*$ changes by a factor of
$\sim10$ from $z\sim 5$ to $z\sim 3$. It  implies that the star
formation mode for LBGs at $z>3$  might be mainly starburst with the
significance increasing with $z$. And it  might evolve to be more
significant to the quenching mode after $z\sim 3$. This possible
picture is in consistence with that the epoch of major mergers ends
at $z\sim 2$ (\citealt{Conselice03,Conselice11}). We suggest that
fraction of starburst galaxies increases with redshift for LBGs at
$z>3$.

%


\subsection{The ``downsizing" effect}
\label{sub:down}

The correlation between specific star formation rates (sSFR), i.e.,
$\rm SFR/M_*$, and stellar masses of LBGs is plotted in the upper
panel of Fig. \ref{fig:downsizing} with the same notations of points
as Fig. \ref{fig:mass_sfr}. Dashed and solid lines, respectively,
denote the observational limit of $NUV \la 26^{\rm m}.2$ and the
simple least square fitting to the data
with the slope of $-0.68$.
A clear trend for LBGs can be seen from the figure, i.e., the
``downsizing" effect, that sSFR decreases with the increasing of
stellar mass $M_*$, except the observational limit in $NUV$. It can
be found also from the figure that the significance of this
``downsizing" effect decreases for LBGs from the SB, Irr to EST
groups,
%
It is in consistence with that distance to the ``main sequence"
decreases for LBGs from the SB, Irr to EST groups (see the previous
subsection).
%
The ``downsizing" effect indicates that LBGs with less massive
stellar masses build up their stellar blocks relatively faster. This
effect has been already found by \citet{Zheng07} and \citet{Buat08}
for star forming galaxies at $z \la 0.9$ (see also
\citealt{Elbaz07}), \citet{Rodighiero10} and \citet{Gruppioni10} for
galaxies at $z\sim 2$.


Usually, the ``downsizing" effect can be parameterized as ${\rm
sSFR} \propto M_*^{-\gamma}$, with $\gamma \sim 0 $ for the extreme
case of the quenching (quiescent) star formation. Together with the
observed correlation between sizes $R$ - stellar masses $M_*$ for
galaxies, the ``downsizing" effect in fact has many physical
implications for galaxy formation and evolution, especially closely
relating to structure parameters of galaxies, such as compactness
and surface densities of galaxies. More qualitatively discussions
are as follows.


The observed correlation of $R$-$M_*$ for galaxies can be described
as $R \propto M_*^{\beta}$. For active star formation galaxies, such
as merging galaxies, $\beta \sim 0.2$ at the redshift range of $0
\la z \la 3.5$ (\citealt{Shen03, Mosleh11, Ichikawa12}). For
quenching star formation by gas accretion such as local quiescent
galaxies, mean stellar surface densities in central regions of
galaxies display a small dynamic ranges, i.e., with $\beta \sim 0.5$
(\citealt{Shen03}, see also \citealt{Chang10}).

If we simply assume that sizes of stellar and star formation
distributions of a galaxy are similar (see also in Sect.
\ref{sec:mor}), we obtain $R \propto {\rm SFR}^{\beta/(1-\gamma)}$.
In the case of LBGs dominated by star forming galaxies, $R \sim {\rm
SFR}^{0.6}$ with $\gamma \sim 0.68$ and $\beta \sim 0.2$.
It implies that LBGs with larger SFRs are on average more compact
for their star forming regions, which is consistent with the
existence of a very weak correlation with scatters between $\sersic$
indices/sizes in the F606W images and SFRs for LBGs in Sect.
\ref{subsec:SFRvsMor} and \citet{Elbaz11}.
Furthermore, for the mean SFR surface density, i.e., the mean
res-frame UV brightness, in the central regions $\dot{\Sigma}_{*,c}$
of a galaxy, we have $\dot{\Sigma}_{*,c} \propto {\rm
SFR}^{1-2\beta/(1-\gamma)}$. For $\alpha ~\sim 0.68$ and $\beta \sim
0.2$ in our case for LBGs, $\dot{\Sigma_*} \propto {\rm
SFR}^{-0.2}$. It implies that mean rest-frame UV surface densities
in central regions of LBGs display a small dynamic range, i.e.,
insensitive to their total SFRs, which is consistent with the
previous studies on the surface brightness for LBGs at $z\ga 3$ and
star forming galaxies at $z\la 1.1$ (\citealt{Giavalisco96,
Barden05, Akiyama08}).
So, if we simply take $\dot{\Sigma}_{*,c}$ to be constant for LBGs,
their $R$ - $M_*$ relation at $z\sim 1$ can be directly resulted
from the ``downsizing" fitting
with the slope $\beta \sim 0.68/2 = 0.34$.  This is consistent with
the slopes of $\sim 0.36$ for the correlations obtained in Sect.
\ref{sec:cor} between sizes in F606W and F850LP and stellar masses
for LBGs, respectively (see Fig. \ref{fig:r_mass}).

%
%
%

It is interesting to compare the ``downsizing" effect of our LBG
sample with other previous studies on star forming galaxies at
different redshifts from $z\sim 0.3$ to $z\sim 3$, which is shown in
the lower panel of Fig. \ref{fig:downsizing}.
For simplicity, lines of least square fitting results to data are
presented in the figure.
%
From top to bottom, lines denote results of $z\sim$ 3, 1.4, 1, 0.9,
0.7, 0.5 and 0.3 with the slopes $\gamma$ of 0.68, 0.67, 0.68, 0.53,
0.45, 0.32 and 0.25, respectively. Data are from \citet{Zheng07} for
$z\sim$ 0.3, 0.5, 0.7 and 0.9, from \citet{Dunne09} for $z\sim 1.4$
and from \citet{Magdis10a} for $z \sim 3$, respectively.

As can be seen from the figure, the significance of the
``downsizing" effect for star forming galaxies increases, i.e.,
slope $\gamma$ increases, with redshift till $z\sim 1$.
And it is very interesting that  slope $\gamma$ remains unchanged
for $z\ga 1$. According to the analysis above, it implies that
distributions of star formation for galaxies are on average less and
less compact after $z\sim 1$ while they show similar compactness
before $z\sim 1$. Since star formation is dominated by galaxy
merging at high $z$, this also implies that galaxy mergers become
rare after $z\sim 1$. Although slope $\gamma$ decreases from 0.68 at
$z \ga 1$ to 0.25 at $z\sim 0.3$, the power index of the mean star
formation activities in the central region of a galaxy
$|{1-2\beta/(1-\gamma)}| < 0.5$. It implies that mean central
rest-frame UV surface brightness of a galaxy is not sensitive to its
total SFR.
%

\subsection{Colors and stellar masses}
\label{sub:UVvsM}


Based on the DEEP2 (\citealt{Davis03}) and the COMBO-17 surveys,
\citet{Faber07} studied in detail LFs of galaxies to $z\sim 1$.
Their sample contains 39,000 galaxies in total with 15,000 beyond $z
= 0.8$. They found that galaxies display clear bimodality in the
plane of rest-frame color $U-B$ vs stellar mass, i.e., the ``red"
sequence and the ``blue" cloud.
The number and total stellar mass of blue galaxies have been
substantially constant since $z\sim 1$, while the number densities
of red galaxies (near $L^*$) have significantly increased.
Some modes of how galaxies evolve from the ``blue" cloud to the
``red" sequence are suggested. Because the results of
\citet{Faber07} provide strong constraints on galaxy formation and
evolution, many efforts have been done since then, together with
discussions on the ``green valley" between the ``blue" cloud and the
``red" sequence (\citealt{Cattaneo08}, \citealt{Cortese09},
\citealt{Schawinski10}, \citealt{Skelton12}, \citealt{Lemaux12}).

%
We plot in  Fig. \ref{fig:color_mass} the
distributions and the corresponding density contours for LBGs in the
SB (upper panel) and Irr (middle panel) groups, respectively, in the
diagram of rest-frame color $U-B$ vs stellar mass $M_*$.
%
%
Contours from outside to inside in the figure correspond to the mean
number densities of LBGs, and the over-dense contrasts of 2  and 3
times of the mean densities, respectively.
%
The lower panel in the figure shows the mean values and the
corresponding standard deviations for LBGs in the SB and Irr groups
respectively. Results for LBGs in the EST groups are plotted in the
lower panel as dashes just for simple illustration. For easy
comparison with \citet{Faber07},
%
solid lines in the upper and middle panels of the
figure correspond exactly the position and direction of the major
axile of the ``blue" cloud in Fig. 10 of \citet{Faber07}.
%
%
The rest-frame color $U-B$ is in Vega system and the stellar mass in
the x-axile decreases from left to right.


We find from the figure that LBGs in our sample locate in the
``blue" cloud, in consistence with that they are star forming
galaxies selected by rest-frame UV luminosities.
%
LBGs in the SB group spread a wider range in stellar
masses as also seen in Fig. \ref{fig:mass_sfr}.
It is interesting that most of LBGs locate in the lower part if we
divide the  ``blue" cloud into two parts along the major axile. It
implies that LBGs on average display lower dust attenuation than
star forming (blue) galaxies selected by the other methods.  This is
in consistence with the analysis of dust properties for star forming
galaxies by \citet{Buat05,Buat12}, \citet{Reddy12}, and the
references therein.
%
Although the number of LBGs in the EST group is too
small to have statistical significance,
there shows a clear trend that LBGs of earlier types are on average
redder and more massive in stellar masses. As claimed by
\citet{Conselice03, Conselice11}, the epoch of major mergers ends at
$z\sim 2$. LBGs at $z\sim 1$, especially starburst types with low
stellar masses, seem have to maintain their star formation
activities for a relative long time so that they can evolve to have
masses as the local $L^*$ galaxies at present day. We suggest that
LBGs at $z\sim 1$ evolve along the ``blue" cloud and then arrive in
the ``red" sequence as red galaxies finally. Other mechanisms have
been also investigated, such as the re-merging processes in
particular at $z\sim 1$ by \citet{Firmani10}.

\subsection{Summary of the photometric properties}

We list detailed photometric properties of 383 LBGs in our sample at
the redshift range of $0.7 \la z_{\rm p}\la 1.4$  in Table
\ref{tab:LBGs}. The first to the third columns denote their
corresponding IDs of individual LBGs in UV from B07, and from the
COMBO 17 and MUSYC catalogs respectively. Re-estimated photometric
redshifts $z_{\rm p}$ of individual LBGs together with their
corresponding error estimates in 3-$\sigma$ are listed in Column 4.
Note that $z_{\rm p}$ are replaced by spectroscopic redshifts
$z_{\rm spe}$ if available except two with bad spectroscopic
observations (see Sect 2.1). SFRs and stellar masses are in Columns
5 and 6.  Detections of MIPS 24$\mu$m compiled by B07 are in Column
7 with number 99 indicating without detections. Goodness of SED
fittings (reduced $\chi^2$) are in Column 8 with resulted spectral
types (SpT numbers) in Column 9. Full table can be found at
http://202.121.53.133/ZhuChen/. Since figures of SED fittings for
individual LBGs are not easy to compile in the table, we show
SEDs of those LBGs for morphological studies in Table 6 together
with their $HST$ images (see below).

\begin{table*}
\caption{Detailed photometric properties of LBGs  at the redshift
range of $0.7 \la z_{\rm p}\la 1.4$. The first to the third columns
denote their corresponding IDs of individual LBGs in UV from B07,
and from the COMBO 17 and MUSYC catalogs respectively. Re-estimated
photometric redshifts $z_{\rm p}$ of individual LBGs together with
their corresponding error estimates in 3-$\sigma$ are listed in
Column 4. Note that $z_{\rm p}$ are replaced by spectroscopic
redshifts $z_{\rm spe}$ if available except two with bad
spectroscopic observations. SFRs and stellar masses are in Columns 5
and 6. Detections of MIPS 24$\mu$m compiled by B07 are in Column 7
with number 99 indicating without detections. Goodness of  SED
fittings (reduce $\chi^2$) are in Column 8 with resulted spectral
types (SpT numbers) in Column 9. Full table can be found at
http://202.121.53.133/ZhuChen/.}

\begin{tabular}{ r r r r r r r r r}
\hline  ID$_{UV}$ & ID$_{\rm COMBO 17}$ & ID$_{\rm MUSYC}$ & $z_{\rm p}$ ($z_{\rm spe})$   & SFR ($\my$) & Log ($M_*/\msun$)& 24$\mu$m & $\rm{\chi^2}$ & SpT type\\
\hline
\hline  ... & ... & ... & ... & ... & ... & ... & ... & ...\\
81517   &   63245   &   79110       &   $0.974^{+0.009}_{-0.006}$         & 37.1    &   11.1    &   19.74   &   7.80    &   13   \\
47798   &   20406   &   25420       &   $1.330^{+0.010}_{-0.008}$         & 124.3   &   11.4    &   99    & 3.09    &   13   \\
47839   &   20506   &   25134       &   $0.980^{+0.037}_{-0.035}$         & 18.8    &   10    & 99    & 0.51    &   13   \\
47853   &   20341   &   25125       &   $1.145^{+0.020}_{-0.036}$         & 32.2    &   9.9   & 99    & 0.76    &   13   \\
47897   &   20480   &   --          &   $0.930^{+0.011}_{-0.030}$         & 14.2    &   10.6    &   99    & 8.34    &   13   \\
49702   &   22572   &   27552       &   $0.861^{+0.073}_{-0.049}$         & 19.8    &   10.7    &   20.33   &   2.66    &   13   \\
50258   &   23164   &   28404       &   $0.974^{+0.060}_{-0.023}$         & 17.5    &   10.0    &   99    & 2.24    &   13   \\
50616   &   23587   &   29012       &   (1.046)     &   10.0    &   10.8    &   99    & 1.18    &   11  \\
52269   &   25529   &   31702       &   $0.867^{+0.011}_{-0.012}$         & 34.6    &   10.9    &   20.40   &   2.41    &   13   \\
 ... & ... & ... & ... & ... & ... & ... & ... & ...\\
\hline
\end{tabular}
\textrm{} \label{tab:LBGs}
\end{table*}

\section{{\it HST} images for the LBG sample}
\label{sec:HST}

\subsection{Images}
\label{subsec:images}

The Galaxy Evolution from Morphologies and SEDs  (GEMS) survey
(\citealt{Rix04}) together with the central region from the Great
Observatories Origins Deep Survey (GOODS) project
(\citealt{Dickinson04}) is centered on the CDF-S area.
It is $90\%$ field of the present study for LBGs that is covered by
the GEMS and GOODS surveys. So we can find {\it HST} optical images
for LBGs with very high quality from both the GEMS and the GOODS
surveys through the coordinates of their optical counterparts to
study quantitatively their morphological properties. A brief
introduction of the GEMS and GOODS surveys is as follows.

The GEMS is a large-area of 800 $\rm arcmin^2$, two-color in F606W
($V$ band) and F850LP ($z$ band), imaging survey with the Advanced
Camera for Surveys (ACS) on the {\it HST} (\citealt{Rix04}).
Centered on the CDF-S, it covers an area of $\sim 28'\times 28'$, to
a depth of $m_{AB}\rm (F606W)=28.3$ and $m_{AB} \rm (F850LP)=27.1$
in $5\sigma$ for compact sources. In its central ($\sim 1/4$) field,
there are 15 tiles that the GEMS incorporates ACS imagings from the
GOODS project.
%
The GOODS is a survey of approximately 300 $\rm
arcmin^2$ in two fields of the HDF-N and CDF-S. It provides ACS
images in four bands of F435W(B), F606W(V), F775W(i), and F850LP(z).
To assure data homogeneity, the GEMS group has reduced the first
epoch of GOODS-S data at the center of the overall GEMS area taking
exactly the same processes as the  reduction for the GEMS data,
although GOODS images are a bit shallower than those of the GEMS.
Details of the GEMS\footnote{http://archive.stsci.edu/prepds/gems/}
and GOODS\footnote{http://www.stsci.edu/science/goods/} surveys can
be found from their websites, respectively.

According to the coordinates of optical counterparts of individual
LBGs through the cross-identifications in Sect 2, we find 336 LBGs
in the GEMS and GOODS-S fields.
The number fraction of LBGs within the fields of the GEMS and
GOODS-S surveys is similar to the fraction of the sky coverage.
 It implies that morphological studies of these 336 LBGs
can represent all LBGs of our sample.
Images of F606W and F850LP from the GOODS-S and GEMS surveys can be
downloaded from the GEMS website$^2$.
In the overlapped region between the GOODS-S and the GEMS which is a
small fraction, images with better qualities are chosen.
Note that for our LBG sample at redshift $0.7 \la z \la 1.4$, images
of F606W intense to represent star-forming regions while images of
F850LP  mainly reflect distributions of colder stars.

\begin{figure}
\resizebox{\hsize}{!}
{\includegraphics{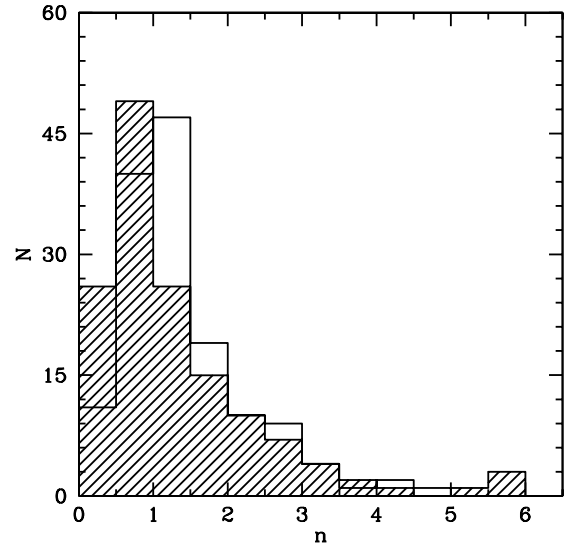}}\\
\caption{The distributions of $\sersic$ indices n for 142 LBGs with
reliable GALFIT results in F606W and F850LP as the hatched and empty
histograms, respectively. }
 \label{fig:sersic}
\end{figure}

\begin{figure}
\resizebox{\hsize}{!} {\includegraphics{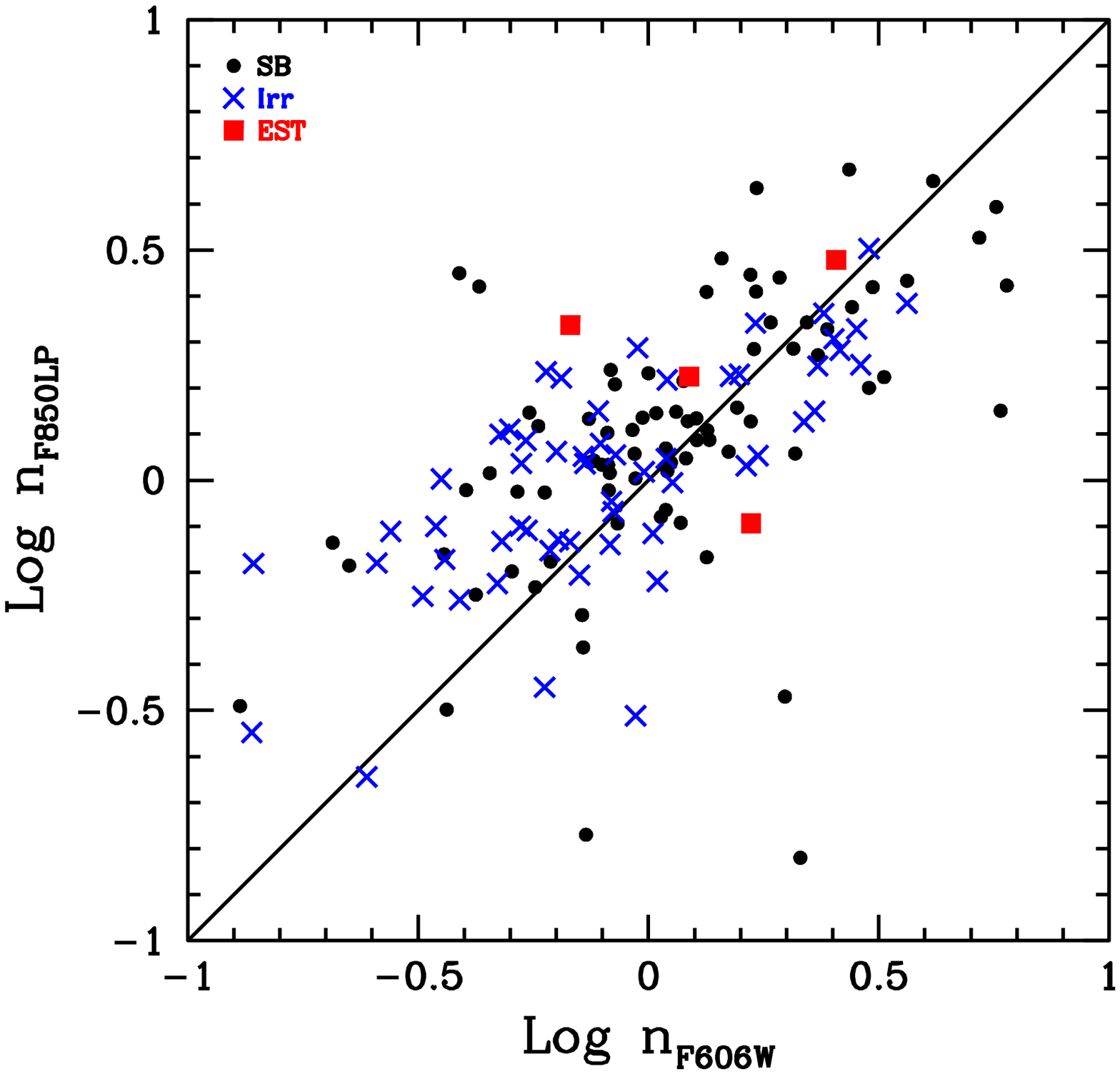}}
\caption{Comparison between $\sersic$ indices for LBGs obtained in
F606W and F850LP bands with
%
dots, crosses and squares denoting LBGs in the SB, Irr and EST
groups respectively, while the solid line denotes where they would
agree.}\label{fig:sv-sz}
\end{figure}

\begin{figure}
\resizebox{\hsize}{!}
{\includegraphics{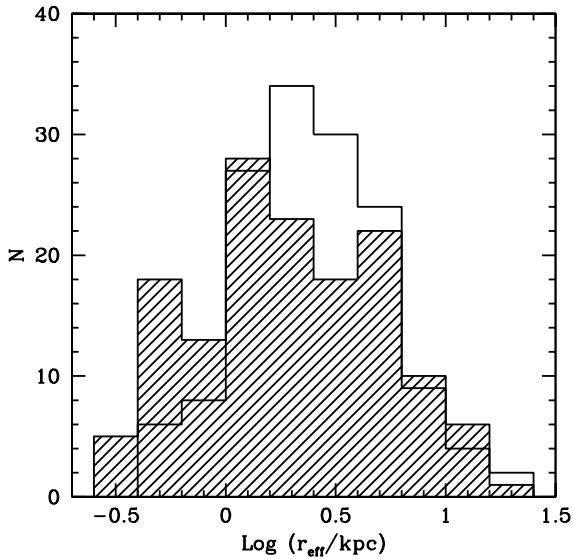}}\\
\caption{The distributions of sizes for 142 LBGs with reliable
GALFIT results in F606W and F850LP as the empty and hatched
histograms, respectively. } \label{fig:r_eff}
\end{figure}

\subsection{Visual classifications of morphologies}
\label{subsect:gallery}

In the present subsection, an image gallery of LBGs in variety of
shapes with classifications by eye is established firstly. We cut
areas of $5''\times 5''$ individually as counter images of LBGs
centered on their COMBO-17 coordinates. Since LBGs in our sample are
at $z\sim 1$, images of $5''\times 5''$
(corresponding to $40\kpc \times 40\kpc$ at $z\sim 1$) are large
enough to study their morphologies. The widely used software
SExtractor\footnote{http://www.astromatic.net/software/sextractor}
\citep{Bertin96} is performed to detect LBGs in images
 with the detection threshold being chosen as $3\sigma$
for F606W and F850LP respectively (see also in next section).
As pointed out by \citet{Caldwell08}, the GEMS can reliably detect
most of COMBO 17 galaxies with magnitudes $R<24$ and its detection
ability decreases significantly for objects with $R>24$. Since there
are 59 LBGs in the GEMS and GOOD-S fields with their $R$ band
magnitudes fainter than 24, images of some LBGs locating within the
GEMS and GOODS surveys are missed.

Totally, there are 309 and 279 LBGs detected by SExtractor
successfully in their F606W and F850LP counter images, respectively,
with 277 LBGs being commonly detected in both bands.
%
 That more LBGs detected in F606W is consistent with
that LBGs are bright in F606W, since they are star forming galaxies
at $z\sim 1$ and the depth of F606W is a bit deeper. Spatial
distributions of detected/undetected LBGs in the GOOD-S and the GEMS
fields are roughly the same. It implies that the slightly different
observational depths between GOODS and the GEMS do not influence the
following morphological studies for our LBG sample.

For easy comparison, LBGs detected in images of both bands are taken
for morphological studies in the present paper. For 59 LBGs
undetected or only detected in one band,
they display low SFRs/stellar masses as
expected. Most of them are classified as starburst type LBGs, in
consistence with Fig. \ref{fig:mass_sfr}.
For visual classifications of LBGs morphologies, we follow the
prescriptions of \citet{Elmegreen09}.
Note that two LBGs with their UV IDs of 74154 and 81517 in B07 are
found to have many bad pixels in their images in both bands although
they are detected by SExtractor. These two LBGs are ruled out from
the following morphological studies.

It is found by eye that 10 LBGs show anomalous morphologies, which
display very chaotic shapes, chain-like structures and contain
multiple bright knots. These LBGs are classified visually as
``chain" LBGs in the present paper. As examples, images of three
``chain" LBGs in F606W and F850LP are displayed in Fig.
\ref{fig:figgalaxy1} respectively. Several bright knots of star
forming regions in images for ``chain" LBGs suggest that they are
ongoing merging processes in a very young and active dynamical stage
\citep{Abraham99}. Their visual morphologies are consistent with
their spectral types of starbursts or irregulars with relatively
high star formation rates and massive stellar masses.

7 LBGs can be clearly seen by eye in their counter images of both
bands as large spiral galaxies individually with two prominent
components of disks and bright central bulges.  These galaxies are
classified as ``spiral" LBGs with examples shown in
Fig.\ref{fig:figgalaxy2}. It is found that all of them have high
SFRs and massive stellar masses being classified in the EST group as
``Sbc/Scd/E" type galaxies during their SED fittings in Sect. 2.

It is also easily found that 23 LBGs clearly show very close
companions in their images of either F606W or F850LP band. We call
these galaxies  ``tadpole" LBGs with  three examples shown in Figs.
\ref{fig:figgalaxy3}. They are classified as starburst/Arp
220/irregular types, consistent with their morphological
classifications by eye.

There are 132 LBGs which display very clumpy star forming regions
spreading wide areas  in their images. We call them ``clump" LBGs
with three examples shown in Fig. \ref{fig:figgalaxy4}. In fact,
they are marginally detected by SExtractor. All of them are
classified as starburst type LBGs with relatively low SFRs and low
stellar masses which are very similar to those LBGs without
detections or with detections available in only one band.

Except the above classified LBGs, the rest 103 LBGs show very
regular shapes in both of their counter images which are called as
``bulge" LBGs. Their examples are shown in Fig.
\ref{fig:figgalaxy5}. In their SED fittings, most of them are
classified as starburst galaxies.

Among 277 common detected LBGs in both bands, 224 (81\%) and 53
(19\%) LBGs are classified as Blue- and Red-LBGs respectively.
Although the fractions are similar to the whole LBG sample, Blue-
and Red-LBGs display very different fractions in the visual
classifications. For instance, 4 and 3 Red-LBGs are ``spiral" and
``chain" LBGs. 5 among 23 ``tadpole" and 28 among 132 ``clump" LBGs
are Red-LBGs. To be concluded, Red-LBGs are always more structured
visually than Blue-LBGs.

We compile the image gallery of 277 LBGs detected in both F606W and
F850LP  in Table 6 with their visual classifications. Quantitative
studies of the morphological properties are in the next section.

\begin{figure}
\resizebox{\hsize}{!} {\includegraphics{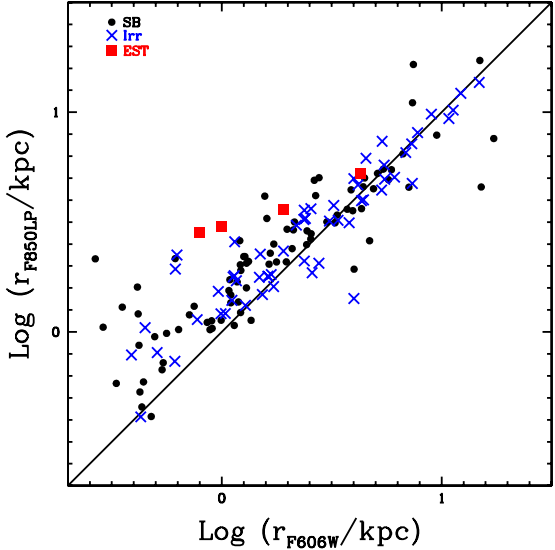}}
\caption{Comparison between sizes of LBGs obtained in F606W and
F850LP with dots, crosses and squares denoting LBGs in the SB, Irr
and EST groups, while the solid line denotes where they would
agree.} \label{fig:rv-rz}
\end{figure}

\begin{figure}
\resizebox{\hsize}{!} {\includegraphics{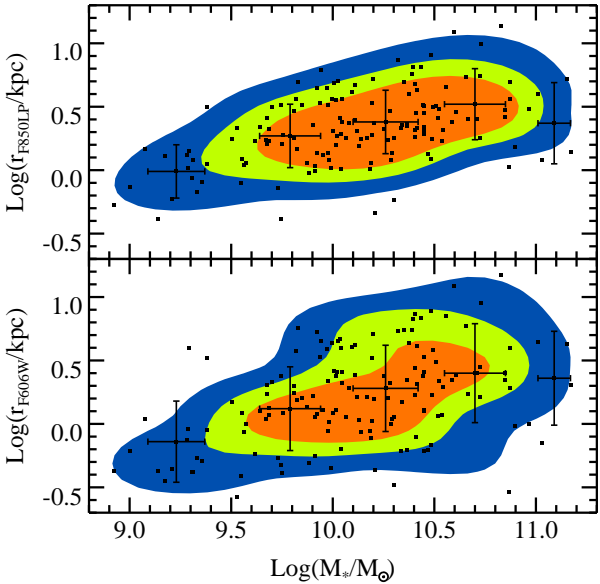}}
\caption{The distributions of LBG and the corresponding density
contours together with the mean values and the standard deviations
as error bars for different mass bins in the size-stellar mass plans
with the results of the F606W and F850LP bands in the upper and
lower panels. Contours from outside to inside in the figure
correspond to the mean number densities of LBGs and the over-dense
contrasts of 2 and 3 times of the mean densities, respectively (see
text for details).}
%
 \label{fig:r_mass}
\end{figure}

\section{Morphological properties of LBGs}
\label{sec:mor}

\subsection{Morphological analysis}
\label{subsec:galfit}

The powerful software GALFIT with the one-$\sersic$-component model,
is adopted to analyze morphological properties of LBGs. GALFIT is a
galaxy/point source fitting algorithm developed by \citet{Peng02,
Peng10}\footnote{
http://users.obs.carnegiescience.edu/peng/work/galfit/galfit.html},
that fits 2D parameterized axially-symmetric directly to images. We
refer \citet{Peng02, Peng10} for details.

In the present paper, we focus on 277 LBGs with images detected in
both F606W and F850LP (see the previous section) bands.  The
$\sersic$ index and the half-light radius, the most important
structure parameters of a galaxy, are obtained through performing
GALFIT to its image.
%
Simple prescriptions, instead of the details for the processes and
other parameters chosen, are as follows. SExtractor \citep{Bertin96}
is applied firstly to individual counter images of LBGs to get their
initial parameters and mask files required by GALFIT, with the
detection threshold being chosen as $3\sigma$ for both F606W and
F850LP bands, respectively. The point spread function (PSF) adopted
here is generated by the TinyTim software \citep{Krist93} which is
designed to reproduce the {\it HST} modeled PSF.

Although 277 LBGs with image detected successfully by SExtractor in
both F606W and F850LP can get their $\sersic$ indices and half-light
radii through GALFIT with the one-$\sersic$-component model, results
of half LBGs are not physically reliable. For instance, the 7
``spiral" LBGs classified in the previous section cannot be reliably
fitted, since they display at least two components in their images.
Obviously, the 10 ``chain" LBGs cannot be fitted using GALFIT, since
they show multiple components with different shapes.
%
There are 87 among 132 ``clump" LBGs with their residual images
still having several bright knots after GALFIT performance.
It implies that they display complicate structures.
For another 5 among 132 ``clump" LBGs, they are very faint with low
surface brightness and detected marginally by SExtractor. So it is
very difficult for GALFIT to get their true physical parameters
subtracted from the sky background. These LBGs are also excluded in
morphological studies below although their goodness of GALFIT seems
acceptable.
Moreover, two LBGs with their UV IDs of 74154 and 81517 in B07 must
be ruled out since their images show many bad pixels in both bands.
Totally, 142 LBGs, consisting of 127 Blue- and 15 Red-LBGs, have
reliable results by GALFIT with their images detected in both F606W
and F850LP bands.

It is important to understand if these 142 LBGs can represent
morphological properties of all LBGs at $z\sim 1$ without
significant bias. The distributions of redshifts, spectral types,
SFRs and stellar masses $M_*$ for these 142 LBGs with reliable
GALFIT results are also plotted in Figs. \ref{fig:zphot_144_all},
\ref{fig:spt}, \ref{fig:sfr} and \ref{fig:star} as dashed
histograms, respectively, to compare with those for all LBGs. It can
be seen that the distributions of
these 142 LBGs are very similar to those of all LBGs. So, we take
them reasonably as the morphological sample, which can represent the
whole LBGs sample, to study morphological properties of LBGs below.

Details of morphological properties of 277 LBGs with images detected
in both F606W and F850LP bands are summarized individually in Table
6. The table consists in two rows for each LBG. Column 1 in the
first row denotes UV IDs of LBGs in B07, which is the same as the
first column in Table \ref{tab:LBGs}. Columns 2 and 3 in the first
row list $\sersic$ indices in F606W and F850LP with their
corresponding half-light radii (sizes) obtained by GALFIT in Columns
4 and 5, respectively. Columns 6 and 7 in the first row list
indicator flags of GALFIT results and visual classifications of LBG
images. Note: flag (0/1) denotes a reliable/unreliable result of the
GALFIT fitting; type (1-5) denotes 1: tadpole; 2: chain; 3: clump;
4: bulge and 5: spiral LBGs with 6 denoting bad pixel images,
respectively. In the second row, LBG images detected by SExtractor,
modeled by GALFIT, and residual images with the upper and lower
panels for F606W and F850LP bands are shown respectively. Figures of
corresponding SED fittings are in the right-hand side of the second
row. Full table can be downloaded at http://202.121.53.133/ZhuChen/.

\subsection{$\sersic$ indices and sizes}
\label{subsec:sersic}

The $\sersic$ profile is one of the most popular prescriptions to
study morphologies of galaxies. The $\sersic$ index n is often used
to describe the concentration of the light profile for a galaxy. It
displays a steeper inner profile and a more extended outer wing when
n is larger. The effective radius $r_{\rm eff}$ describes the size
of a galaxy which contains its half of the light within $r_{\rm
eff}$. Usually, galaxies with $\rm n \la 2$ and $\rm n > 2$ are
classified as disky types and elliptical types respectively (see
\citealt{Peng02, Peng10}).

The distributions of $\sersic$ indices for 142 LBGs, i.e., the
morphology sample, obtained by GALFIT in the images of F606W and
F850LP bands
are shown in Fig. \ref{fig:sersic} as the hatched and empty
histograms, respectively. It is found that about 80\% of the LBGs
have their $\sersic$ indices smaller than 2 with a median values of
around 1 for both bands. We conclude that most LBGs are disklike
late type galaxies. It is consistent with \citet{Burgarella06,
Burgarella08} who suggested that $\sim 75\%$ LBGs at $z\sim 1$ are
disklike after partly analyzing their counterparts from GOODS data.
%
\citet{Haberzettl12} found similar result that the
majority of LBG sample at $z\sim 2$ shows disk-like structure.

\citet{Giavalisco96} found that star-forming galaxies (LBGs at
$z\sim 3$) in their sample are compact with sizes and scale lengths
similar to the present-day bulges or intermediate-luminosity
spheroids, while they are often surrounded by lower surface
brightness nebulosities. However, multi-color morphologies of some
of their LBGs clearly show similarly irregular. Here deeper images
with higher resolution and S/N provided by the GOODS and GEMS
surveys make us possible to get more reliable conclusions,
especially for outskirts of LBGs which could not be detected clearly
at that moment and play important roles in morphology analysis.

For a disk galaxy, its central concentration of stars in the space
distribution always increases with age and that of star formation
activity evolves in an inverse way. For interacting disk galaxies,
the induced star formation activities usually  occur more outskirts
than their stars (\citealt{Rawat09}). Note that $\sersic$ indices in
F606W and F850LP reflect somehow central concentrations of star
formation activities and stars in space distributions respectively
for LBGs at $z\sim 1$. It is interesting to compare the resulted
$\sersic$ indices of 142 LBGs in F606W images with those in F850LP
images, which is in Fig. \ref{fig:sv-sz}. Same as in Fig.
\ref{fig:mass_sfr}, LBGs in the SB, Irr and EST groups are labeled
as different symbols for illustration.

As can be seen from the figure that LBGs in individual groups have
similar distributions in the diagram. Most of the disk-like LBGs
with $\rm n_{850LP} \la 2$ have their $\sersic$ indices in F850LP
larger than those in F606W. It implies that their stellar
distributions are more central concentrated than their star
formation activities, similar to the local ``old" disk galaxies or
interacting disk galaxies. Furthermore, LBGs with $\rm n_{850LP} \la
\rm n_{F606W} \la 2$ might be very young disk galaxies according to
the analysis above.

Another important physical quantity obtained by GALFIT is the
effective radius (size) of an LBG,  within which half of its
luminosity is included. As for $\sersic$ indices, we plot
respectively the distributions of sizes in F606W  and F850LP for the
LBG morphology sample as the hatched and empty histograms in Fig.
\ref{fig:r_eff}. It can be seen from the figure that the median
values of sizes for LBGs are $2.34\kpc$  and $2.68\kpc$ in F606W and
F850LP, respectively, larger than the median size of LBGs in UV at
$z\sim 3$ (\citealt{Shu01}) and similar to that of local disk
galaxies (\citealt{Shen03}) if the cosmic evolution is simply taken
into account as $(1+z)^{-1}$.
%
%
Assuming that sizes of LBGs are resulted from angular momentums of
their host halos, the Gaussian-like distributions of the logarithmic
sizes for two bands  can be easily explained by the Gaussian
distribution for the spin parameters in logarithm of halos
(\citealt{Cole96}).

Same as Fig. \ref{fig:sv-sz}, comparison between the sizes of LBGs
obtained in F606W and F850LP is shown in Fig. \ref{fig:rv-rz} As can
be seen from the figure that the distributions of LBGs in individual
groups are similar. For LBGs with $r_{\rm F606W} \ga 3\kpc$, $r_{\rm
F606W}$ agrees with $r_{\rm F850LP}$ well. For LBGs with $r_{\rm
F606W} \la 2\kpc$, $r_{\rm F606W}$ is often smaller than $r_{\rm
F850LP}$. It implies that smaller LBGs must display more compact in
F606W, i.e., more compact star formation regions, so that they can
have SFRs higher enough to be selected as LBGs.

\section{Correlations between photometric and morphological
properties} \label{sec:cor}

As mentioned in the previous section, 142 LBGs with reliable
morphological parameters
in the fields of the GOODS and GEMS can represent all LBGs at $z\sim
1$ for their morphological studies. Among these 142 LBGs, 79, 59, 4
are in the SB, Irr and EST group as classified according to their
spectral types during their SED fittings (see Sect. 3.1). We
investigate correlations between their photometric and morphological
properties in the present section. To limit the length of the pager,
only figures with strong correlations are shown.

\subsection{Spectral types vs $\sersic$ indices and sizes}

Since the number of LBGs in the EST group is too small to have
statistical significance, we only focus the discussions on LBGs in
the SB and Irr groups. The mean $\sersic$ indices of LBGs in the SB
and Irr groups are 1.30 and 1.43, 1.07 and 1.17 for the F606W and
F850LP bands respectively.
%
Moreover, the mean sizes for LBGs in the SB and Irr
groups are 1.82$\kpc$ and 2.15$\kpc$, 3.09$\kpc$ and 3.35$\kpc$ for
the F606W and F850LP bands respectively.
It implies that LBGs in the SB group on average show more compact
distributions of both star formation activities and stars than LBGs
in the Irr group. This is consistent with the implications obtained
from the ``downsizing" effect in Sect. \ref{sub:down} and the recent
study of \citet{Elbaz11}.
%
In fact,
same conclusion can be drawn
from the visual classifications in Sec. \ref{sec:HST}.

\subsection{SFRs vs $\sersic$ indices and sizes}
\label{subsec:SFRvsMor}

Since the number of LBGs in the morphological sample is only 142, we
discuss correlations between SFRs and morphological parameters,
i.e., $\sersic$ indices and sizes, as a whole, instead of grouping
LBGs into different spectral types.
%
It is found that both $\sersic$ indices $\rm
n_{F606W}$ and sizes $r_{\rm F606W}$ of LBGs in the F606W band
display weak correlations with scatters of increasing with SFRs.
Since images of the F606W band represent distributions of star
formation activities for LBGs at $z\sim 1$, this implies that more
compact LBGs in F606W on average have higher SFRs.
It is also in consistence with \citet{Elbaz11} and the implication
from the ``downsizing" effect that LBGs with higher SFRs are on
average more compact (see Sect. \ref{sub:down}).

On the other hand neither $\sersic$ indices $\rm n_{F850LP}$ nor
sizes $r_{\rm F850LP}$ of LBGs in the F850LP band show any
significant correlations with SFRs with large scatters. This is
consistent with that images of the F850LP band represent stellar
distributions of LBGs but their star formation activities.


\subsection{Stellar masses vs $\sersic$ indices and sizes}
\label{subsec:MvsMor}

Because of large scatters, no significant correlations between
$\sersic$ indices in the F606W/F850LP bands and stellar masses $M_*$
for LBGs can be found. For the F606W band, this can be simply
explained as its irrelevance to stellar distributions for LBGs at
$z\sim 1$. Since images of the F850LP band reflect stellar
distributions of LBGs, the insignificant correlation between $n_{\rm
F850LP}$ and $M_*$ suggests that their stellar distributions
%
may have similar profiles. LBGs with more massive stellar masses
mainly because they have larger sizes (see below). This implies that
mean stellar surface densities in central regions of LBGs are
similar to have a small dynamic range as their SFR surface
densities, which is consistent with implications from the
``downsizing" effect in Sect. \ref{sub:down}.

Finally, strong correlations between sizes $r_{\rm F606W}$/$ r_{\rm
F850LP}$ in the F606W/F850LP bands and stellar masses $M_*$, i.e.,
the size-stellar mass relations, for LBGs are found and shown in
Fig. \ref{fig:r_mass}, respectively. In the figure we plot the
distributions of LBGs and the corresponding density contours
together with the mean values and the standard deviations as error
bars for different mass bins. Same as Fig. \ref{fig:color_mass},
contours from outside to inside correspond to the mean number
densities of LBGs and the over-dense contrasts of 2 and 3 times of
the mean densities, respectively.

It can be found that sizes in the both F606W and F850LP bands
increase with stellar masses for LBGs significantly. Since there is
no significant correlation between $\sersic$ indices and stellar
masses, LBGs with more massive stellar masses certainly have larger
sizes  in the F850LP band. The correlation between sizes in the F606
band and their stellar masses for LBGs can be easily obtained from
their size  - stellar mass relation in the F850LP band and the
``downsizing' effect in Sect. \ref{sub:down}. Ignoring the
contribution of the most massive bin due to few LBGs in Fig.
\ref{fig:r_mass}, we can easily obtain the slopes of $0.37\pm 0.04$
and $0.36 \pm 0.04$ with the same corresponding coefficients of 0.99
for the size-stellar mass relations in the F606W and F850LP bands,
respectively. It is in consistence with \citet{Mosleh11} and
\citet{Ichikawa12}. Together with the least square fitting for the
``downsizing" effect, it is again implies that the mean rest-frame
UV surface densities, i.e., the mean SFR surface densities, of LBGs
in their central regions display a small dynamic range.


\section{Conclusions}
\label{sec:con}

\citet{Burgarella07} defined a preliminary sample of 420 LBG
candidates, with the completeness $\sim 80\%$ down to $NUV \sim
24.8$, at $z\sim 1$ by $FUV-NUV
>2$ from the deep {\it GALEX} observation in the field of the CDF-S
according to their photometric redshifts from COMBO-17. Four
confirmed AGNs are included. Although the global accuracies of
photometric redshifts in the both COMBO 17 and MUSYC catalogs are
pretty high, differences of photometric redshifts for LBG candidates
between these two catalogs are significant. Together with the MUSYC
{\it BVR} selected catalog, we re-estimate their photometric
redshifts using broader multi-wavelength data from UV, optical to
NIR by the code Hyperz. Comparisons between re-estimated photometric
redshifts and spectroscopic redshifts if available show that our
photometric redshift determinations are fairly good. After carefully
cross-identifying LBG candidates one-to-one with their optical
counterparts and considering their re-estimated photometric
redshifts and SEDs, we refine a new updated sample of 383 LBGs at
$0.7\la z \la 1.4$, with two AGNs in this redshift range being
excluded. This means that the AGN contamination is about 0.5\% which
is smaller than that of $\sim3\%$ for LBGs at $z\sim 3$
(\citealt{Steidel03, Lehmer08}).

According to SED fittings, 260 and 111 LBGs are classified as
starburst galaxies and irregulars with 11 being classified as
Sbc/Scd types and only one as an elliptical.
%
To be easily compared by future studies, the averaged
SEDs of LBGs are presented for SB/Irr types and Blue-/Red-LBGs,
respectively.
The ages of LBGs spread from several Myr to 1.5Gyr with a median
value of $\sim50$Myr, comparable to \citet{Haberzettl12} for LBGs at
$z\sim 2$ and \citet{Papovich01} for LBGs at $z\sim 3$. The median
dust attenuations in the rest-frame FUV for Blue- and Red-LBGs are
1.7 and 2.5 magnitudes, respectively, in good agreement with the
previous study by B07.
Based on the $1/V_{\rm max}$ method and considering uncertainties of
individual luminosity bins due to uncertainties of photometric
redshifts,
%
we obtain the rest-frame FUV luminosity function (LF) of LBGs  with
the completion down to $\rm M_{1800\AA} = -19$. By adopting the
Levenberg-Marquardt method with the consideration of the two
dimensional error bars, a Schechter parametrization is applied to
the LF and we get the best-fit parameters of $\alpha = -1.61\pm 0.40
$, $\rm M^* = -20.40 \pm 0.22$ and $\phi^* = (0.89\pm 0.30)\times
10^{-3}\mpc^{-3}{\rm dex^{-1}}$, respectively. According to
\citet{Arnouts05}, it implies that more than half of the NUV
selected galaxies at $z \sim 1$ are selected as LBGs. Simple
comparisons with previous studies on FUV LFs at different redshifts
and brief discussions are presented.


Dust-corrected star formation rates (SFRs) of LBGs are from $4\my$
to $220\my$, with a median value of $\sim 25\my$ which is similar to
that for LBGs at $z\sim 3$ (\citealt{Shapley01, Steidel03}) and a
bit smaller than those for LBGs at $z\sim 2$
(\citealt{Haberzettl12}) and $z\sim 5$ (\citealt{Verma07}).
Stellar masses of individual LBGs are obtained according to their
rest-frame $K$-band luminosities and the calculated mass-to-light
ratios in $K$ band through their SED fittings. Results are similar
to those if simply taking a constant mass-to-light ratio in $K$ band
of $M_*/L_K = 0.5M_\odot/L_{\odot,K}$.
Stellar masses $M_*$ of LBGs distribute from $2.3\times 10^8 \msun$
to $4 \times 10^{11} \msun$, with a median value of $\sim 10^{10}
\msun$ which is comparable to those of LBGs at $z\sim 2$
(\citealt{Haberzettl12}) and $z\sim 3$ (\citealt{Shapley01,
Papovich01}) and about 10 times bigger than that of LBGs at $z\sim
5$ (\citealt{Verma07}). As expected, the mean SFR of LBGs in the SB
group is larger than that of LBGs in the Irr group. The mean stellar
mass of Red-LBGs is larger than that of Blue-LBGs.


It is found that LBGs in the Irr and EST groups
distribute in the SFR-$M_*$ diagram along the ``main sequence" of
star forming galaxies suggested by \citet{Elbaz11} with the
considerations of uncertainties and the redshift range.
%
%
Most of LBGs in the SB group locate in the starburst region, away
from the ``main sequence".
Distances to the ``main sequence", i.e., the inverse way of star
forming timescales, increase from earlier to later spectral types
for LBGs on average.
Since median $M_*$ of LBGs increases very rapid from $z\sim 5$ to
$z\sim 3$ and then varies slowly to $z\sim 1$ together with that
median SFRs are similar for LBGs at different redshifts, it suggests
that the star formation mode for LBGs at $z>3$ is mainly starburst
with the significance increasing with $z$, and it evolves to be more
significant to the quenching mode after $z\sim 3$. This is in
consistence with that the epoch of major mergers ends at $z\sim 2$
(\citealt{Conselice03,Conselice11}). We predict that fraction of
starburst galaxies increases with redshift for LBGs at $z>3$.

Taking the observational limit of $NUV \la 26^{\rm m}.2$ into
account, the ``downsizing" effect is clearly found for LBGs at
$z\sim 1$ and the significance of the effect decreases for LBGs from
the SB, Irr to EST groups. Detailed discussions on the implications
of the  ``downsizing" effect for galaxy formation and evolution,
especially for the structure parameters such as compactness and
surface densities of galaxies, are presented. Taking the observed
size-stellar mass relation into account, we suggest that LBGs with
larger SFRs are on average more compact. Mean rest-frame UV surface
densities in central regions of LBGs are suggested to be insensitive
to their SFRs, in consistence with \citet{Giavalisco96},
\citet{Barden05} and \citet{Akiyama08}. The size-stellar mass
relation can be directly resulted from the ``downsizing" effect.
Comparisons with previous studies on the ``downsizing" effect at
different redshifts are presented. It is found that the
``downsizing" effect is more significant with the increasing of $z$
till $z\sim 1$.

In the rest-frame color $(U-B)$-$M_*$ diagram, LBGs at $z\sim 1$
distribute in the blue cloud. It is interesting that most of LBGs
locate in the lower part of the blue cloud. It suggests that LBGs
display lower dust attenuation on average than star forming galaxies
selected by the other methods. This is consistent with
\citet{Buat05, Buat12}, \citet{Reddy09} and the references therein.
Considering LBGs of starburst types with low stellar masses, we
suggest that LBGs may evolve along the blue cloud and finally arrive
in the red sequence.

Since the GEMS and GOODS-S surveys by HST covers $\sim 90\%$ of the
field of our LBG sample, high quality images in F606W (V band) and
F850LP (z band) are downloaded from The Multimission Archive at
STScI for morphological studies of LBGs. The number fraction of LBGs
within the field of the GEMS and GOODS-S surveys is similar to the
fraction of the sky coverage.
The counter images of LBGs are analyzed using SExtractor. There are
309 and 279 LBGs successfully detected in the images of F606W and
F850LP respectively, with 277 being commonly detected in both bands.
We establish an image gallery of these 277 LBGs by visually
classifying their images individually as types of ``chain",
``spiral", ``tadpole", ``bulge" and ``clump", respectively. It is
concluded that Red-LBGs are more structured than Blue-LBGs.

The powerful software GALFIT of the one-component model is applied
to LBG images availably detected in both F606W and F850LP bands. A
morphological sample of 142 LBGs with reliable GALFIT results is
established to represent all LBGs. We find more than $80\%$ LBGs in
the morphological sample with their $\sersic$ indices smaller than 2
in both bands. It is concluded that LBGs at $z\sim 1 $ are dominated
by disklike galaxies, consistent with \citet{Burgarella06,
Burgarella08}. Similar result is obtained by \citet{Haberzettl12}
for LBGs at $z\sim 2$. The distributions of half-light radii (sizes)
in F606W and F850LP are similar, with their median values of
$2.34\kpc$ and $2.68\kpc$ respectively. If the cosmic evolution
factor is simply taken as $(1+z)^{-1}$, the median sizes of LBGs are
comparable to the local disk galaxies.

Correlations between photometric and morphological properties of
LBGs are investigated. Note that LBGs display strong correlations
between their half-light radii and $M_*$, i.e., the size-stellar
mass relation, in both bands. Similar slopes of $0.036\pm 0.04$ in
two bands, in consistence with \citet{Mosleh11} and
\citet{Ichikawa12}, are obtained with their linear fitting
coefficients all to be 0.99.
For $\sersic$ indices, a weak correlations between $n_{\rm F606W}$
and SFRs suggests that more compact LBGs have higher SFRs.
Insignificant correlation between $n_{\rm F850LP}$ and $M_*$ implies
that stellar distributions of LBGs may have similar profiles.
%
It should be pointed out that correlations between photometric and
morphological properties for LBGs can be explained through the
implications of the ``downsizing" effect.


\section*{Acknowledgments}
We would like to thank thorough reviewing and valuable comments from
the anonymous referee. ZC is very grateful to the financial support
by DB and the LAM for the visit to LAM and is also grateful to DB
and VB for hospitalities during the stay in Marseille. This work is
partly supported by the Chinese National Nature Science foundation
Nos. 10878003, 10833005 \& 10778725, Shanghai Science Foundations
and Leading Academic Discipline Project of Shanghai Normal
University (DZL805). Some/all of the data presented in this paper
were obtained from the Multimission Archive at the Space Telescope
Science Institute (MAST). STScI is operated by the Association of
Universities for Research in Astronomy, Inc., under NASA contract
NAS5-26555. Support for MAST for non-HST data is provided by the
NASA Office of Space Science via grant NNX09AF08G and by other
grants and contracts.


\newpage
\begin{table*}\label{tab:tt}
\caption{Details of morphological properties of 277 LBGs with images
detected in both F606W and F850LP bands.
Column 1 in the first row denotes UV IDs of LBGs in B07, which is
the same as the first column in Table \ref{tab:LBGs}. Columns 2 and
3 in the first row list $\sersic$ indexes in F606W and F850LP with
their corresponding half-light radii (sizes) obtained by GALFIT in
Columns 4 and 5, respectively. Columns 6 and 7 in the first row list
indicator flags of GALFIT results and visual classifications of LBG
images. Note: flag (0/1) denotes a reliable/unreliable result of the
GLAFIT fitting; type (1-5) denotes 1: tadpole; 2: chain; 3: clump;
4: bulge and 5: spiral LBGs with 6 denoting bad pixel images,
respectively. In the second row, LBG images detected by SExtractor,
modeled by GALFIT, and residual images with the upper and lower
panels for F606W and F850LP bands are shown respectively. Figures of
corresponding SED fittings are in the right-hand side of the second
row with arrows denoting upper limits of the flux detections. Full
table can be downloaded at http://202.121.53.133/ZhuChen/.}

\begin{tabular}{|C{1.2cm}|C{1.2cm}|C{1.2cm}|C{1.2cm}|C{1.2cm}|C{1.2cm}|C{1.2cm}C{1.2cm}C{1.2cm}C{1.2cm}|}
\multicolumn{10}{c}{continue $...$}\\
\hline $ID_{UV}$ & $n_{\rm F606W}$ & $n_{\rm F850LP}$ & $r_{\rm
F606W} {\rm (kpc)}$ &
$r_{\rm F850LP} {\rm (kpc)}$ & flag & type & & & \\

\hline 47839 & 99 & 99 & 99 & 99 & 1 & 3 & & &\\
\hline
\multicolumn{10}{|c|}{}\\
\multicolumn{2}{|c}{\includegraphics[width=3cm,height=3cm]{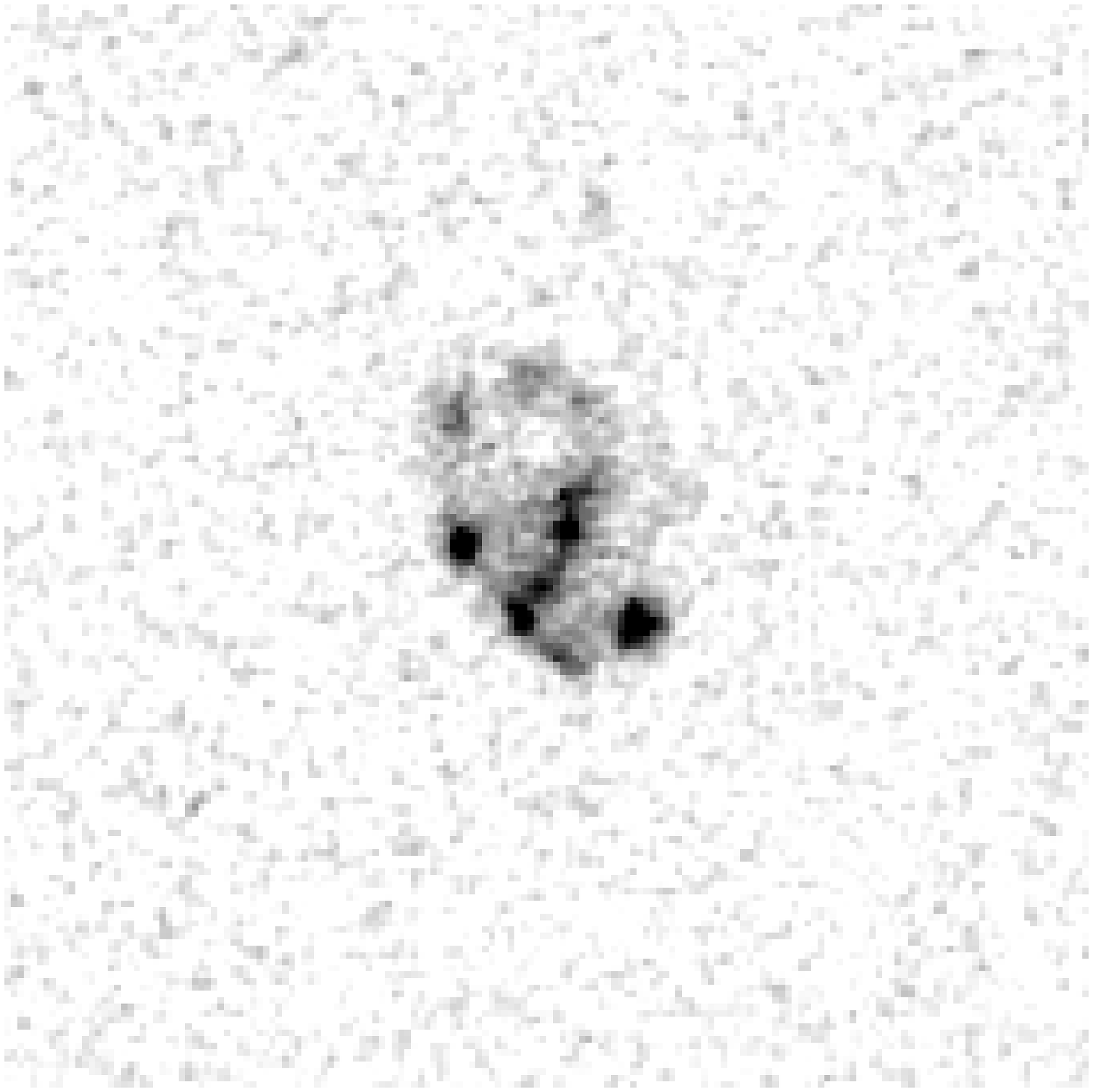}}&
\multicolumn{2}{c}{\includegraphics[width=3cm,height=3cm]{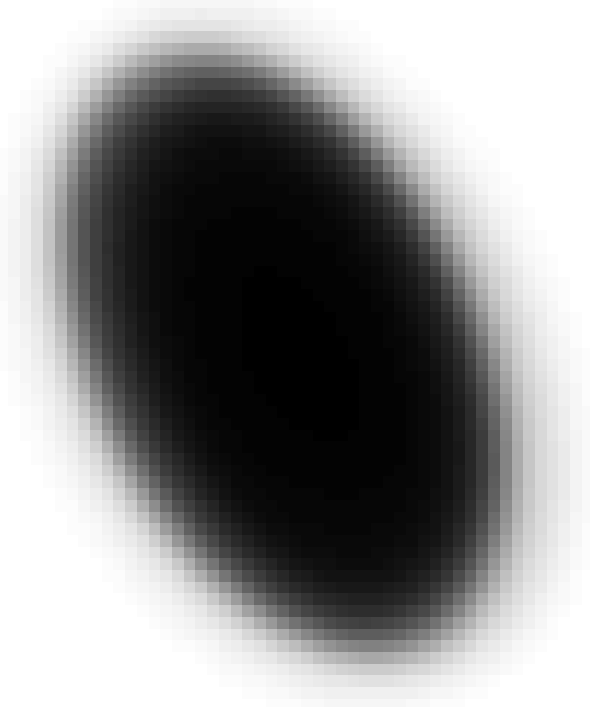}}&
\multicolumn{2}{c}{\includegraphics[width=3cm,height=3cm]{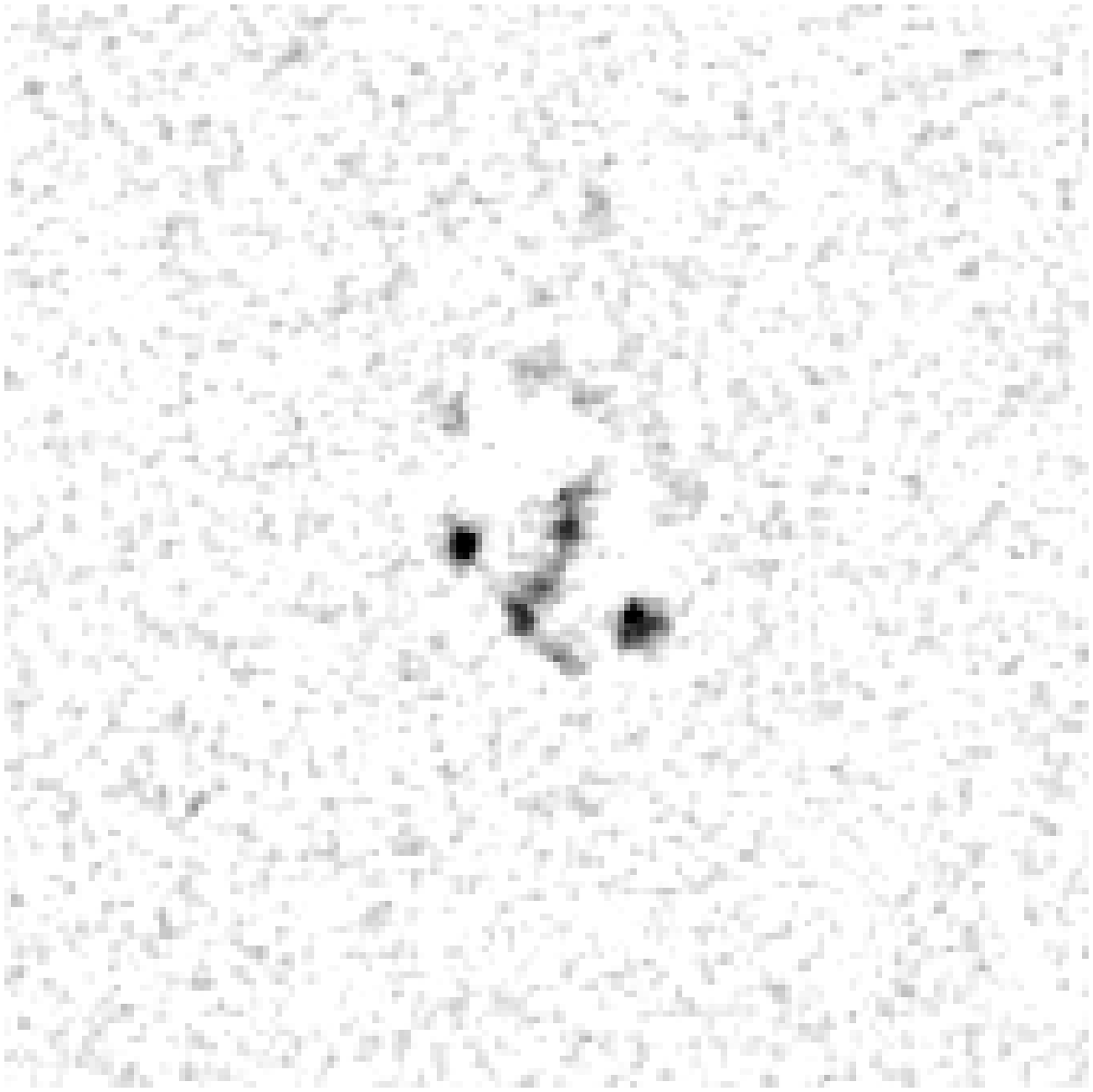}}&
\multicolumn{4}{c|}{\multirow{2}{*}[1.5cm]{\includegraphics[width=6cm,height=4cm]{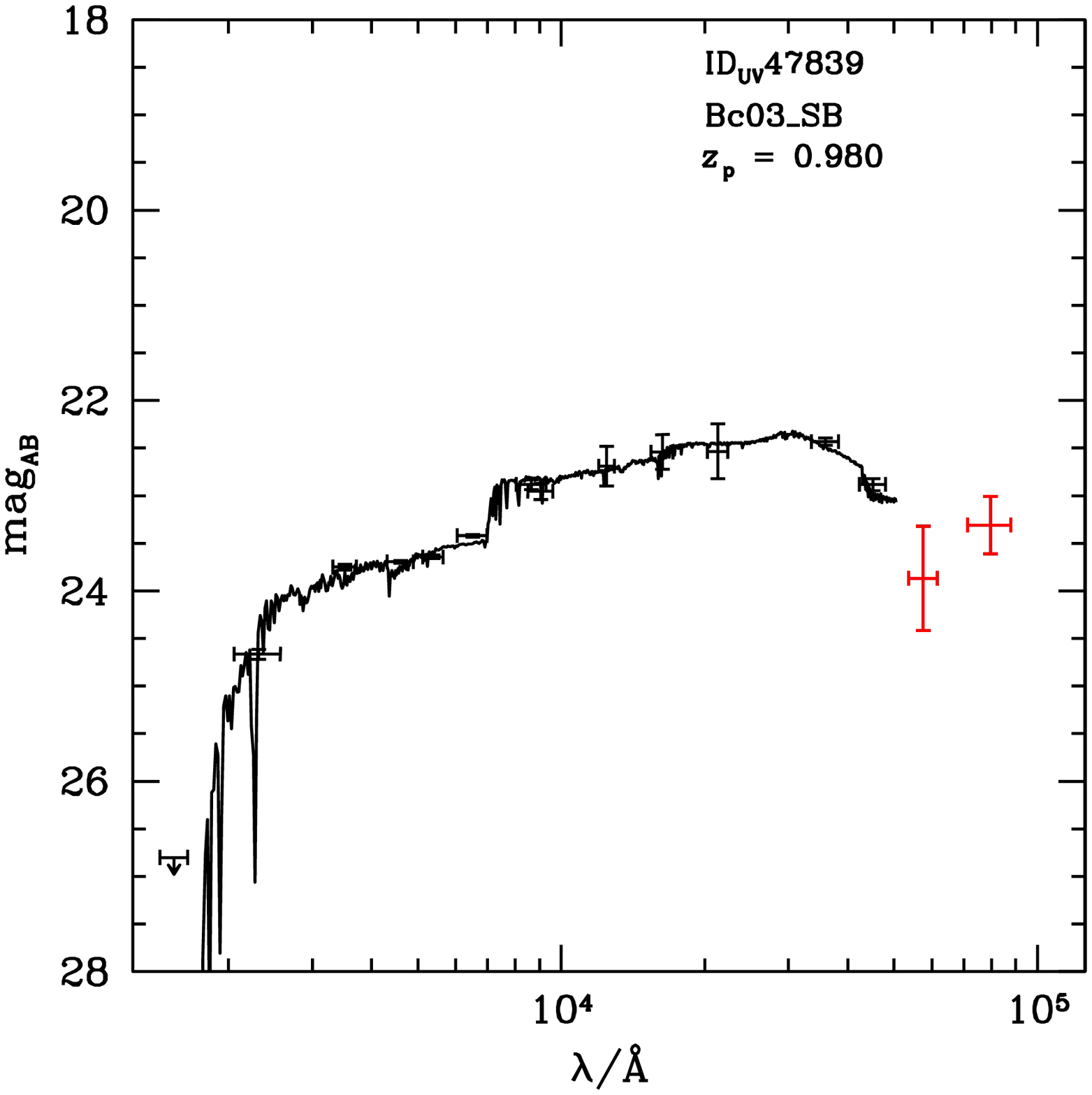}}}\\
\multicolumn{2}{|c}{\includegraphics[width=3cm,height=3cm]{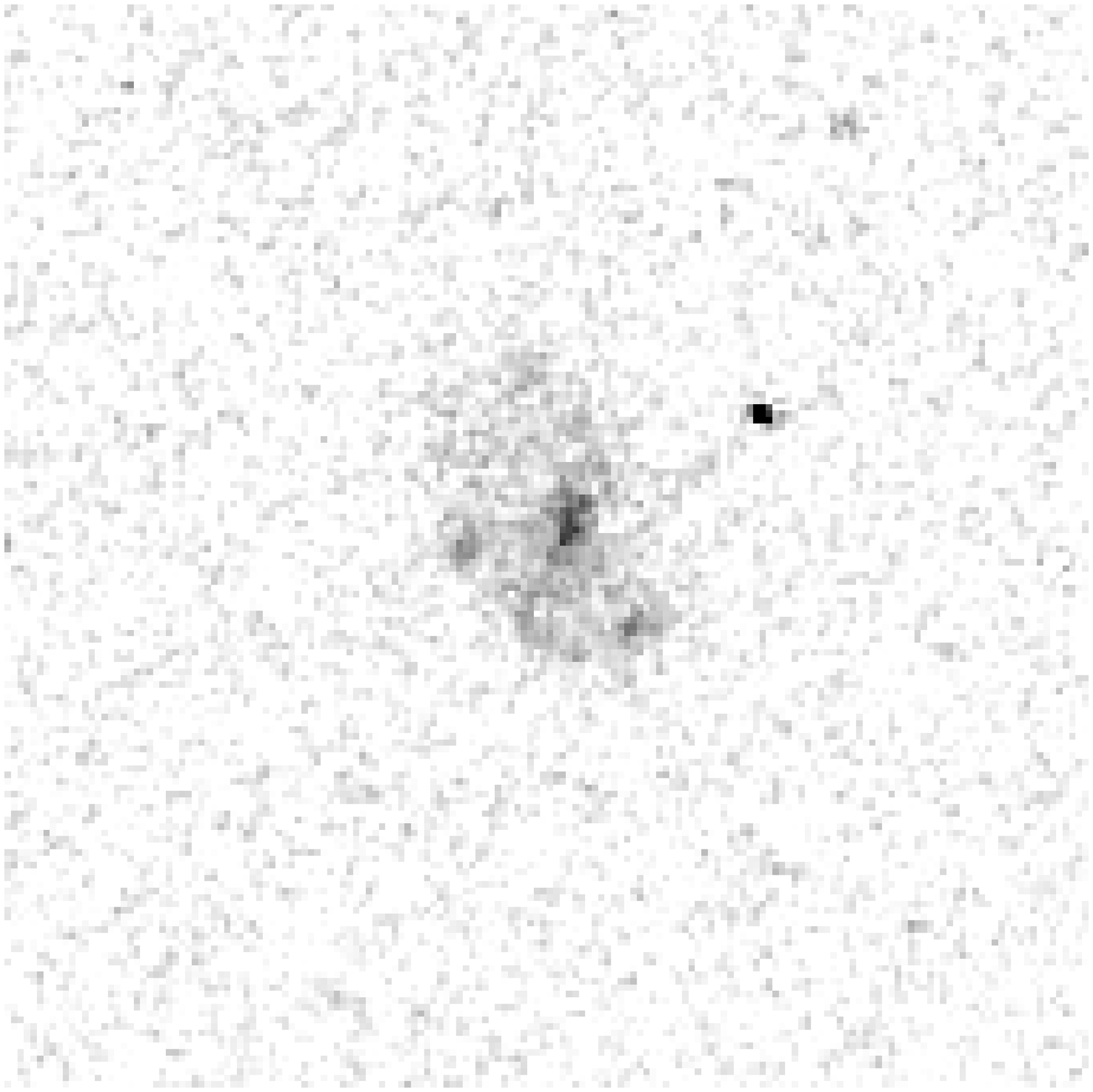}}&
\multicolumn{2}{c}{\includegraphics[width=3cm,height=3cm]{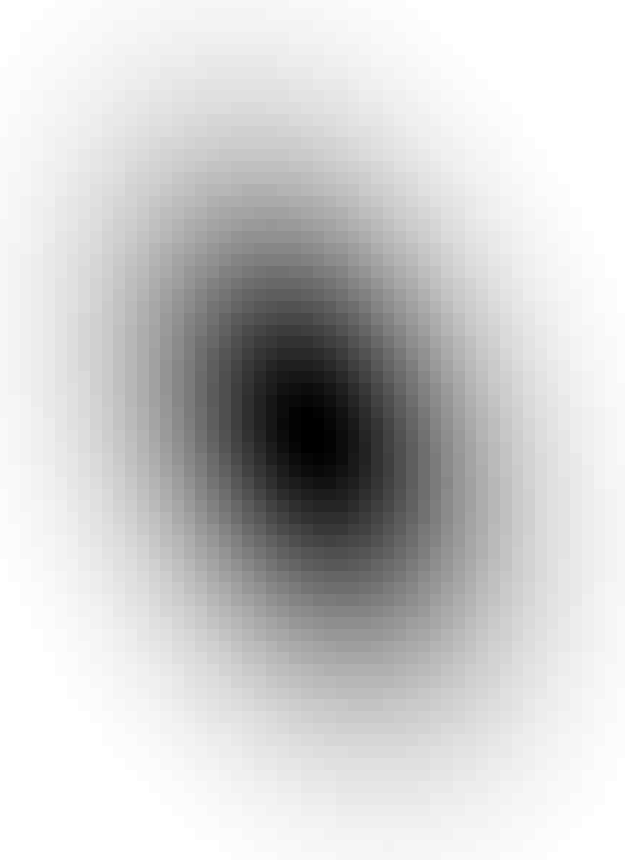}}&
\multicolumn{2}{c}{\includegraphics[width=3cm,height=3cm]{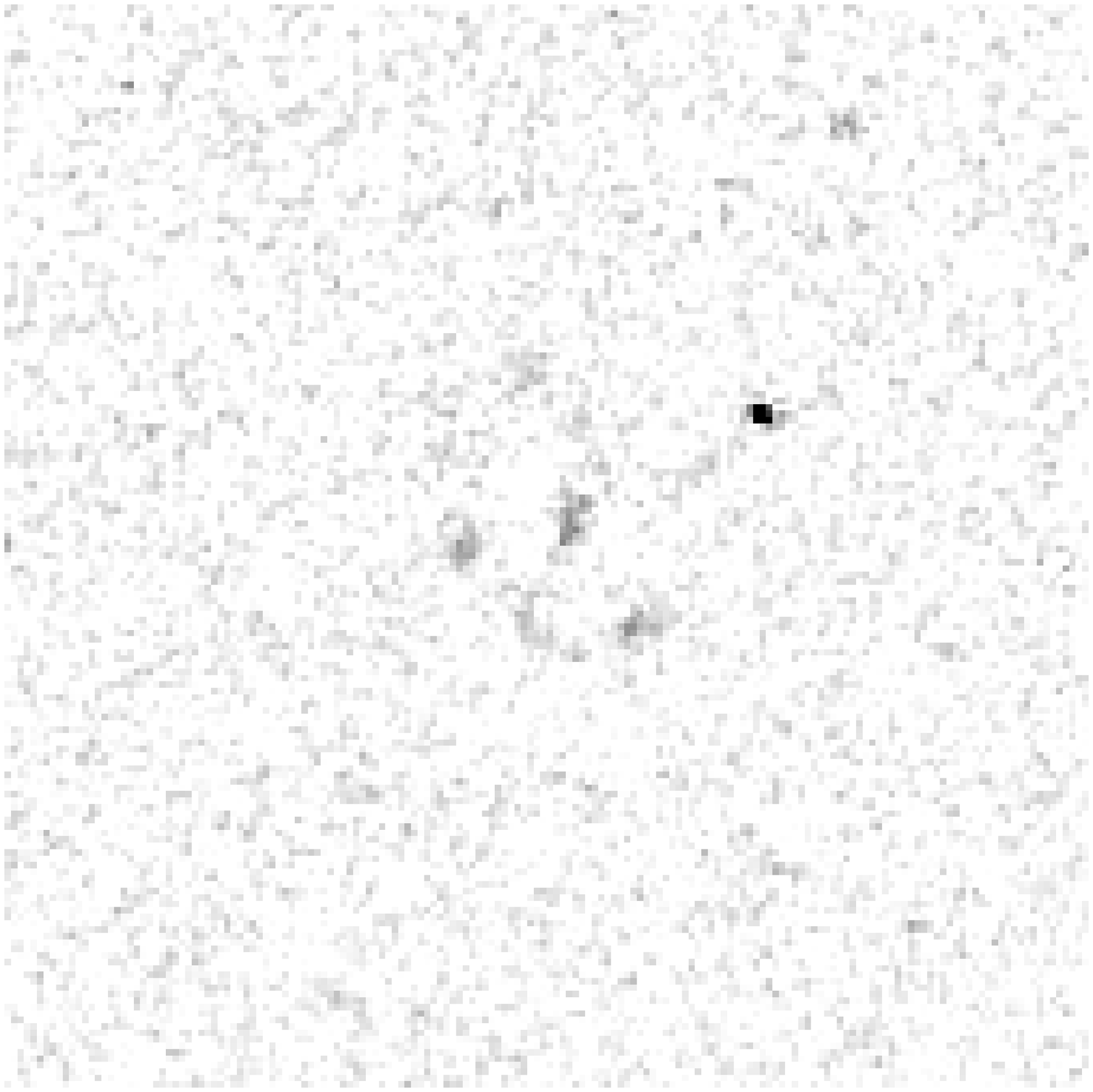}}&
\multicolumn{4}{c|}{}\\

\hline $ID_{UV}$ & $n_{\rm F606W}$ & $n_{\rm F850LP}$ & $r_{\rm
F606W} {\rm (kpc)}$ &
$r_{\rm F850LP} {\rm (kpc)}$ & flag & type & & & \\

\hline 47897 & 5.69 & 3.92 & 7.10 & 4.56 & 0 & 4 & & &\\
\hline
\multicolumn{10}{|c|}{}\\
\multicolumn{2}{|c}{\includegraphics[width=3cm,height=3cm]{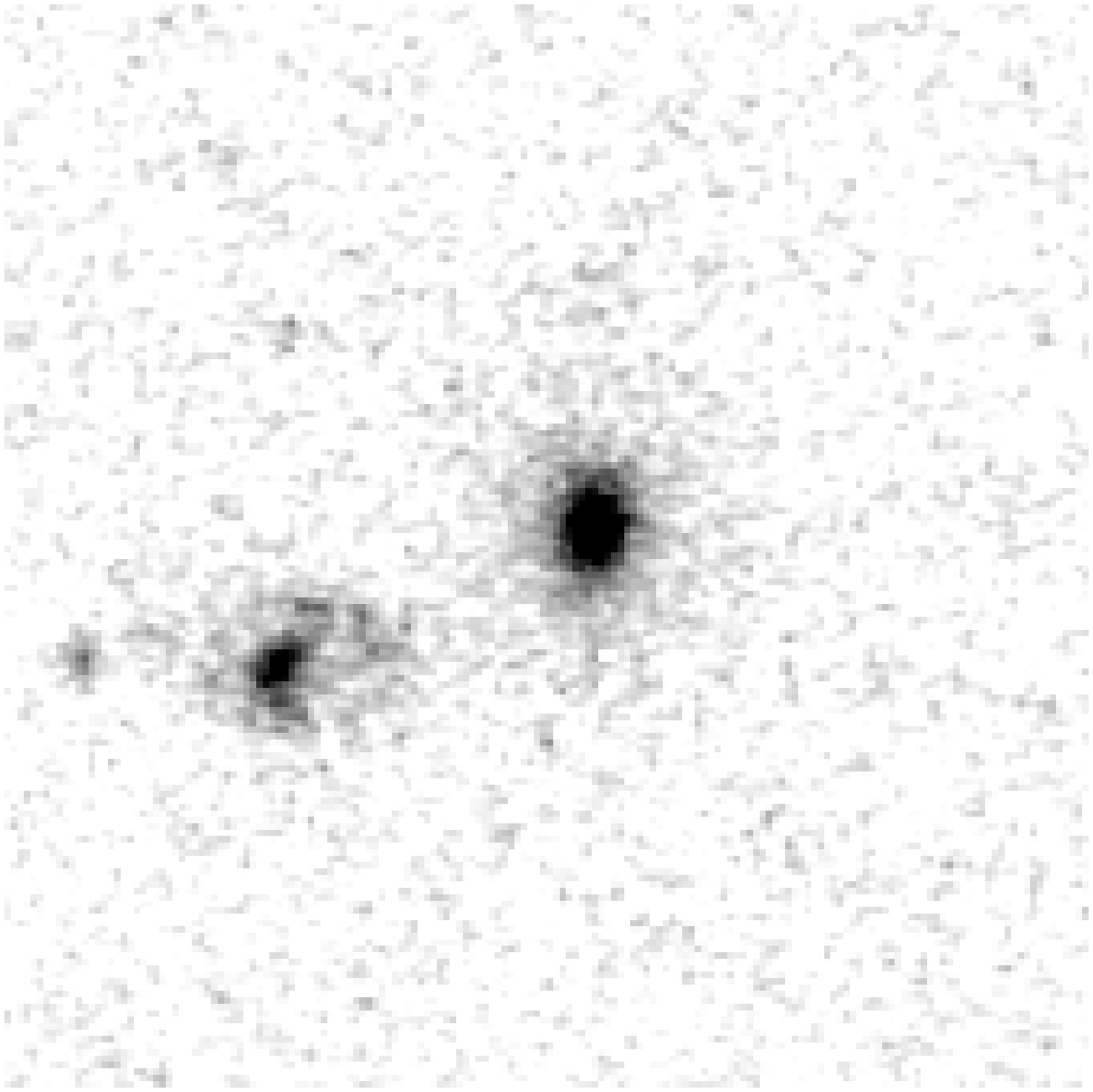}}&
\multicolumn{2}{c}{\includegraphics[width=3cm,height=3cm]{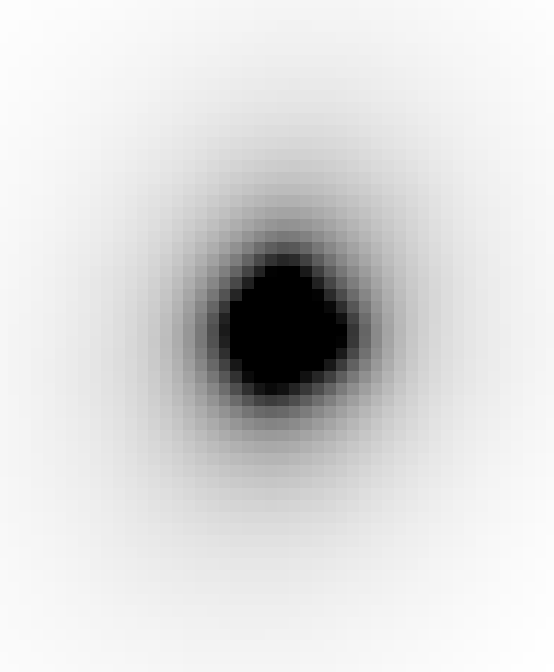}}&
\multicolumn{2}{c}{\includegraphics[width=3cm,height=3cm]{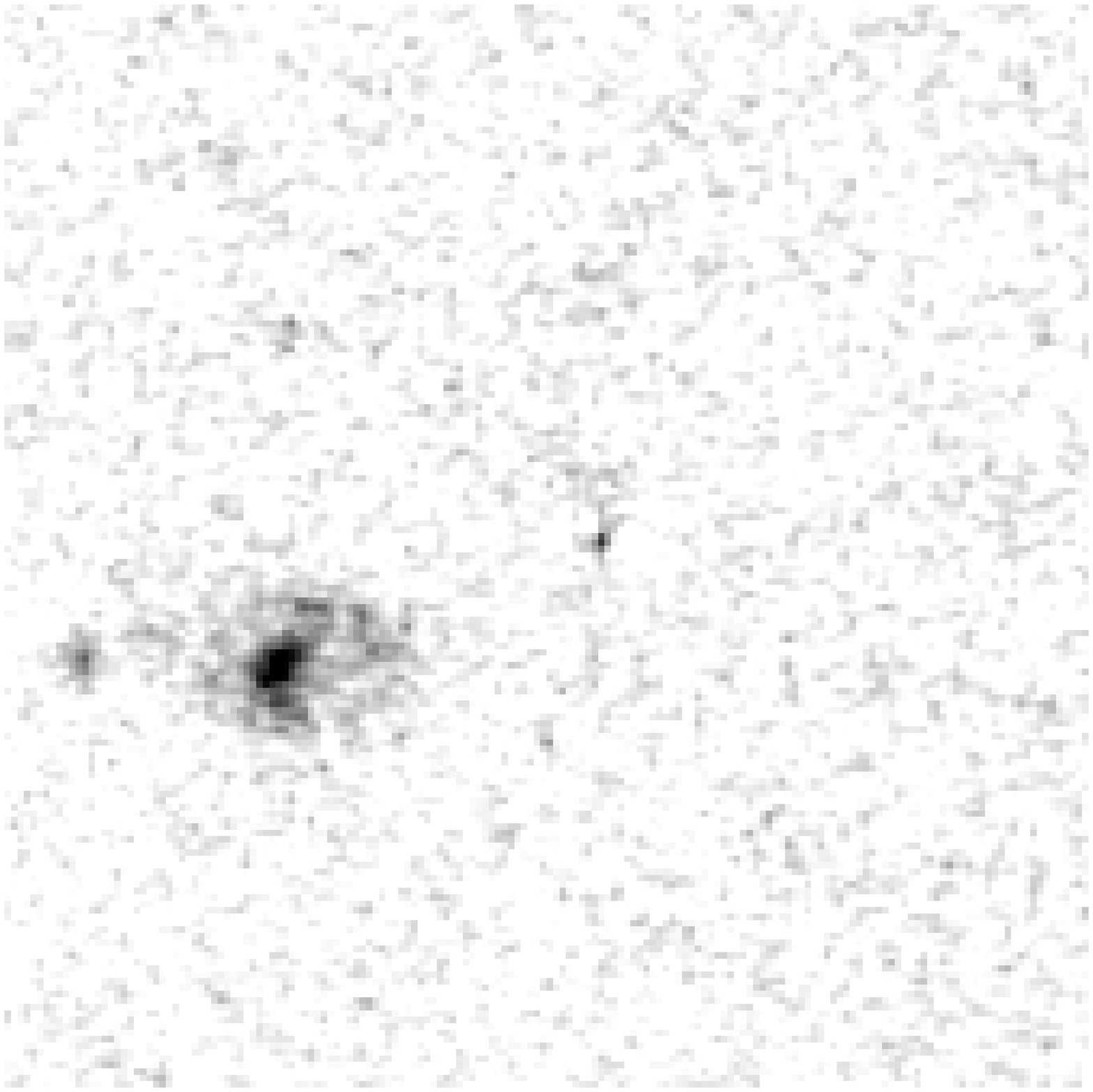}}&
\multicolumn{4}{c|}{\multirow{2}{*}[1.5cm]{\includegraphics[width=6cm,height=4cm]{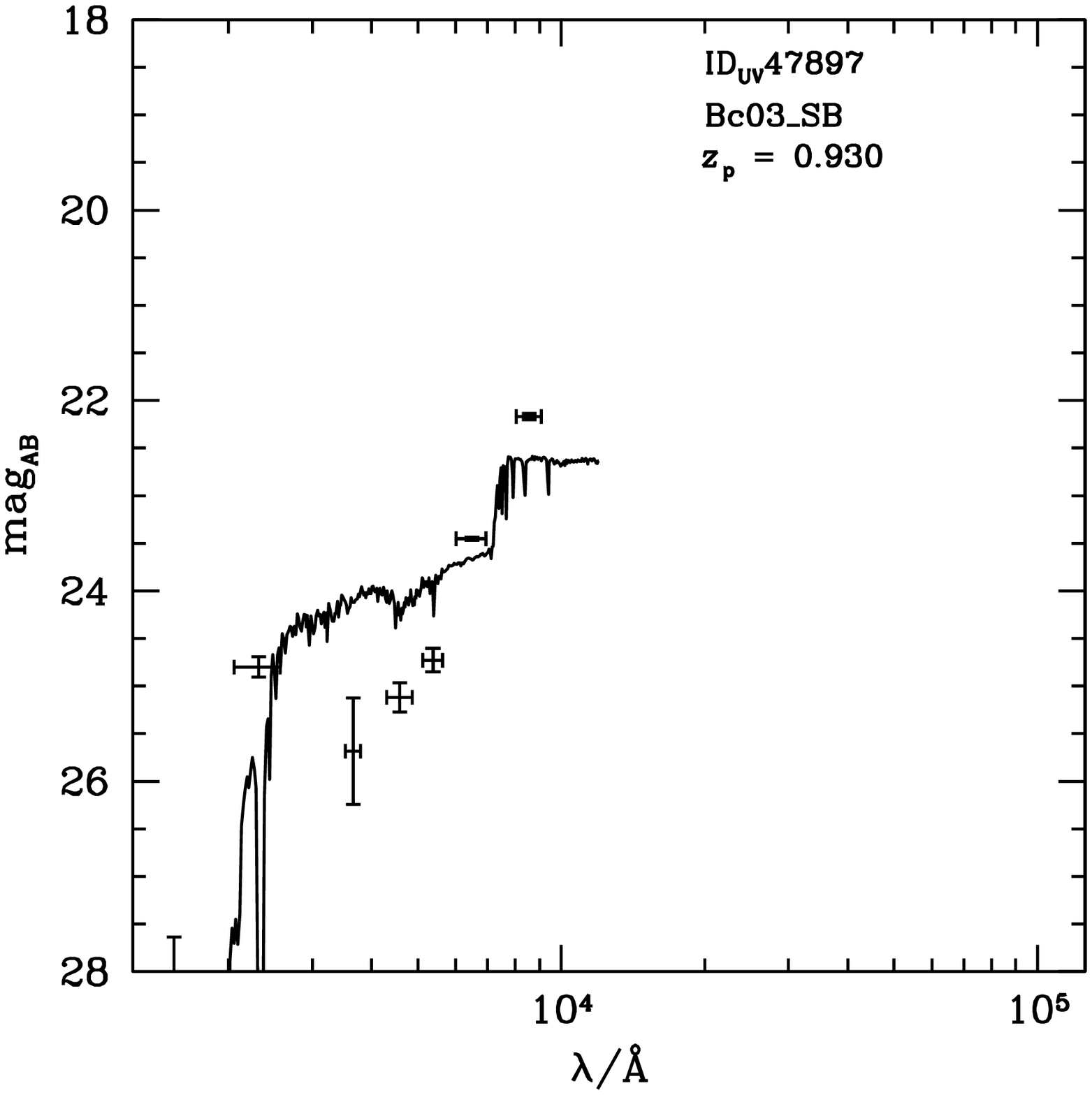}}}\\
\multicolumn{2}{|c}{\includegraphics[width=3cm,height=3cm]{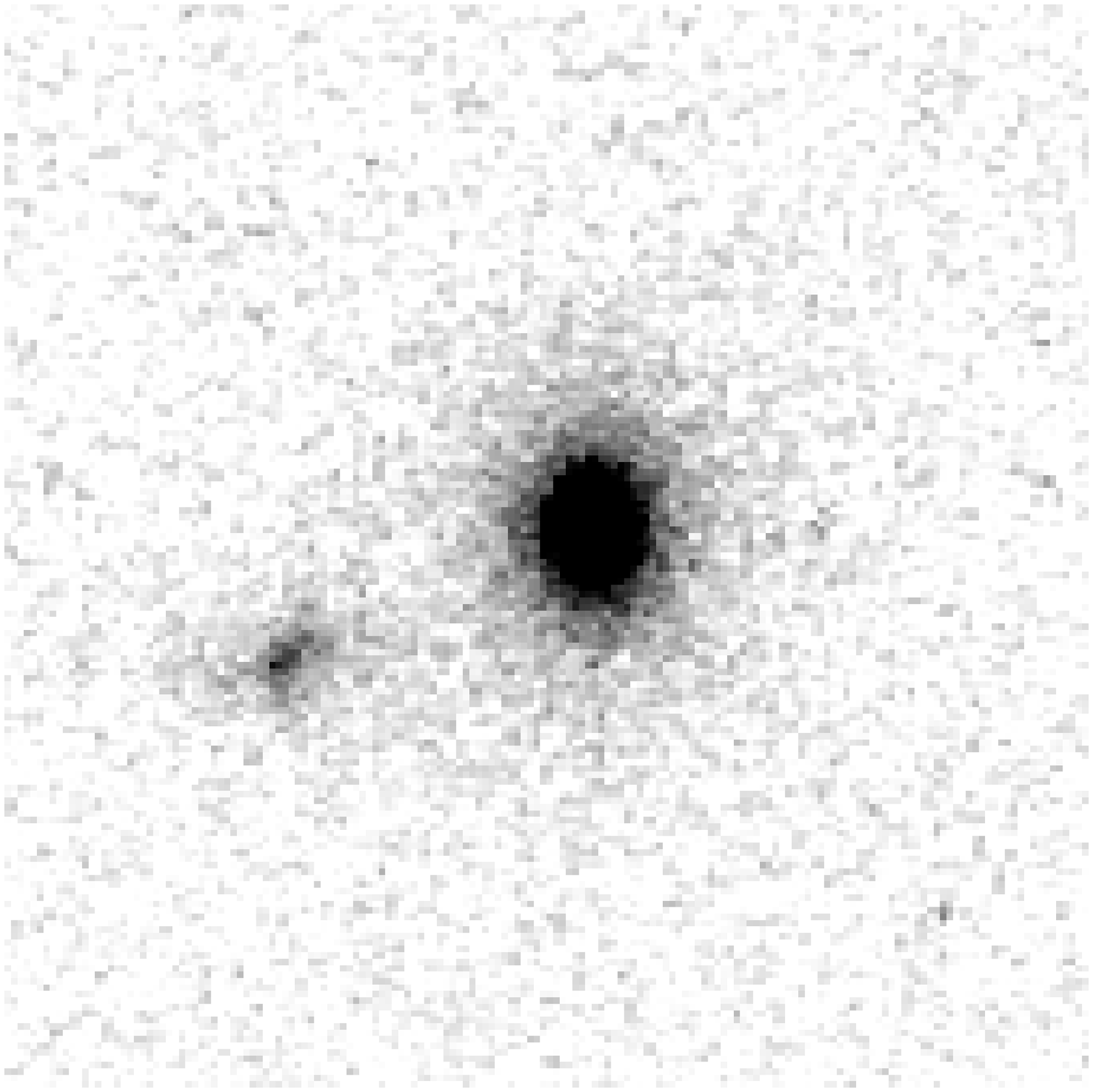}}&
\multicolumn{2}{c}{\includegraphics[width=3cm,height=3cm]{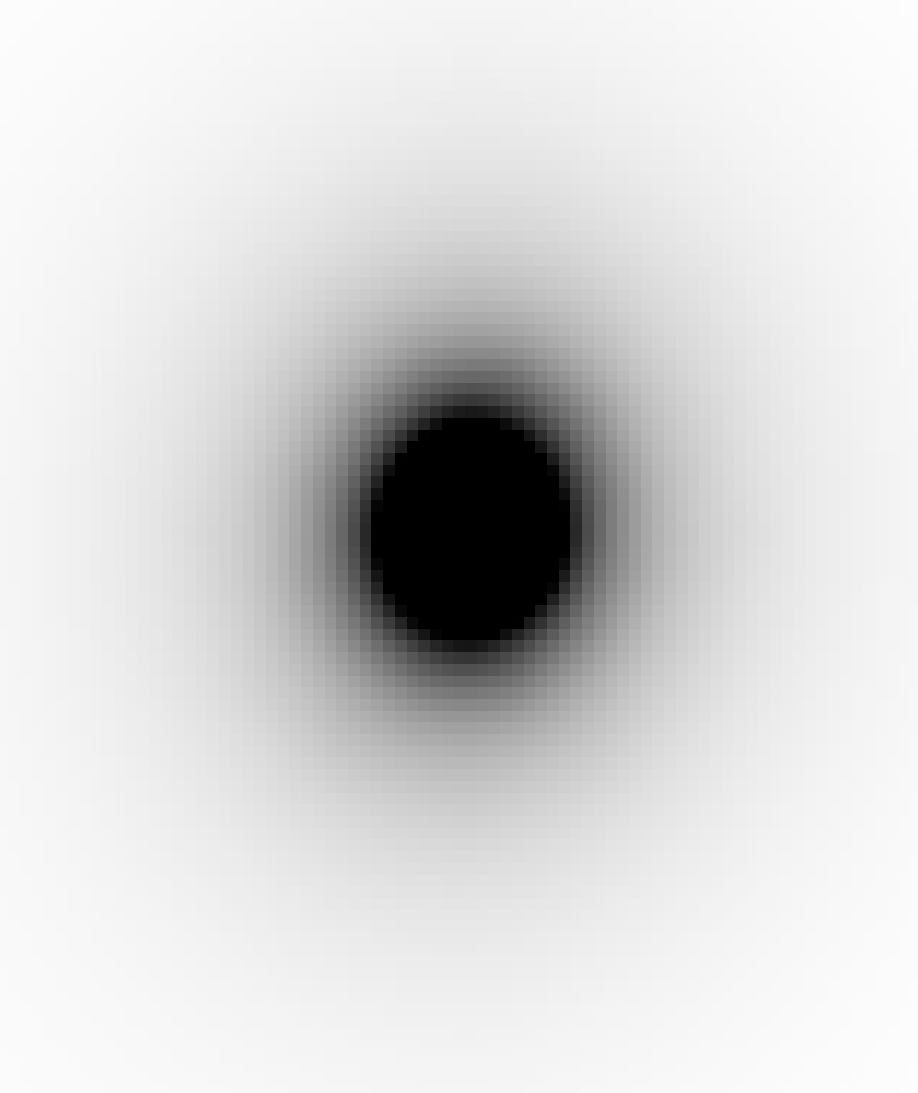}}&
\multicolumn{2}{c}{\includegraphics[width=3cm,height=3cm]{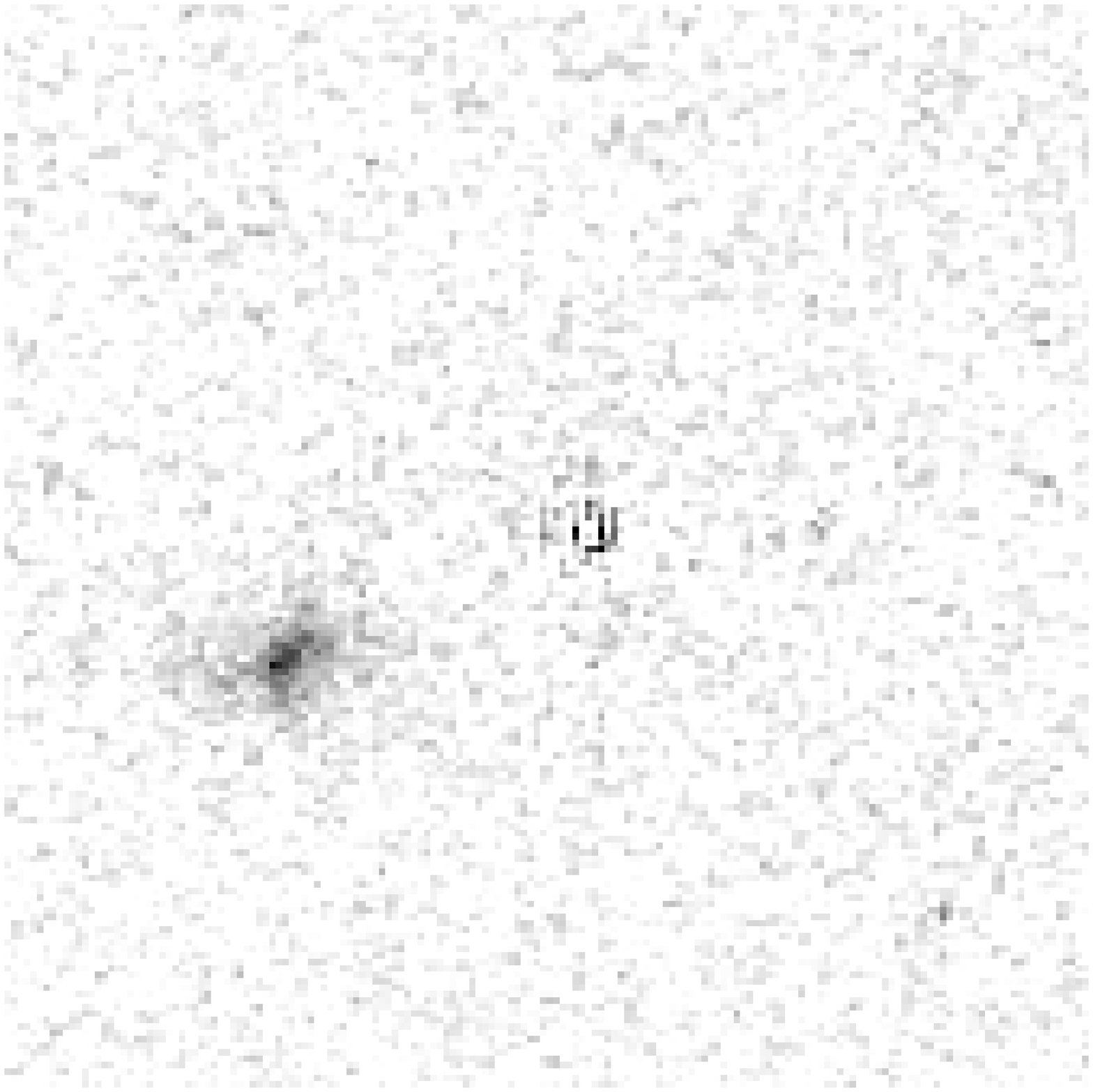}}&
\multicolumn{4}{c|}{}\\

\hline $ID_{UV}$ & $n_{\rm F606W}$ & $n_{\rm F850LP}$ & $r_{\rm
F606W} {\rm (kpc)}$ &
$r_{\rm F850LP} {\rm (kpc)}$ & flag & type & & & \\

\hline 49702 & 99 & 99 & 99 & 99 & 1 & 3 & & &\\
\hline
\multicolumn{10}{|c|}{}\\
\multicolumn{2}{|c}{\includegraphics[width=3cm,height=3cm]{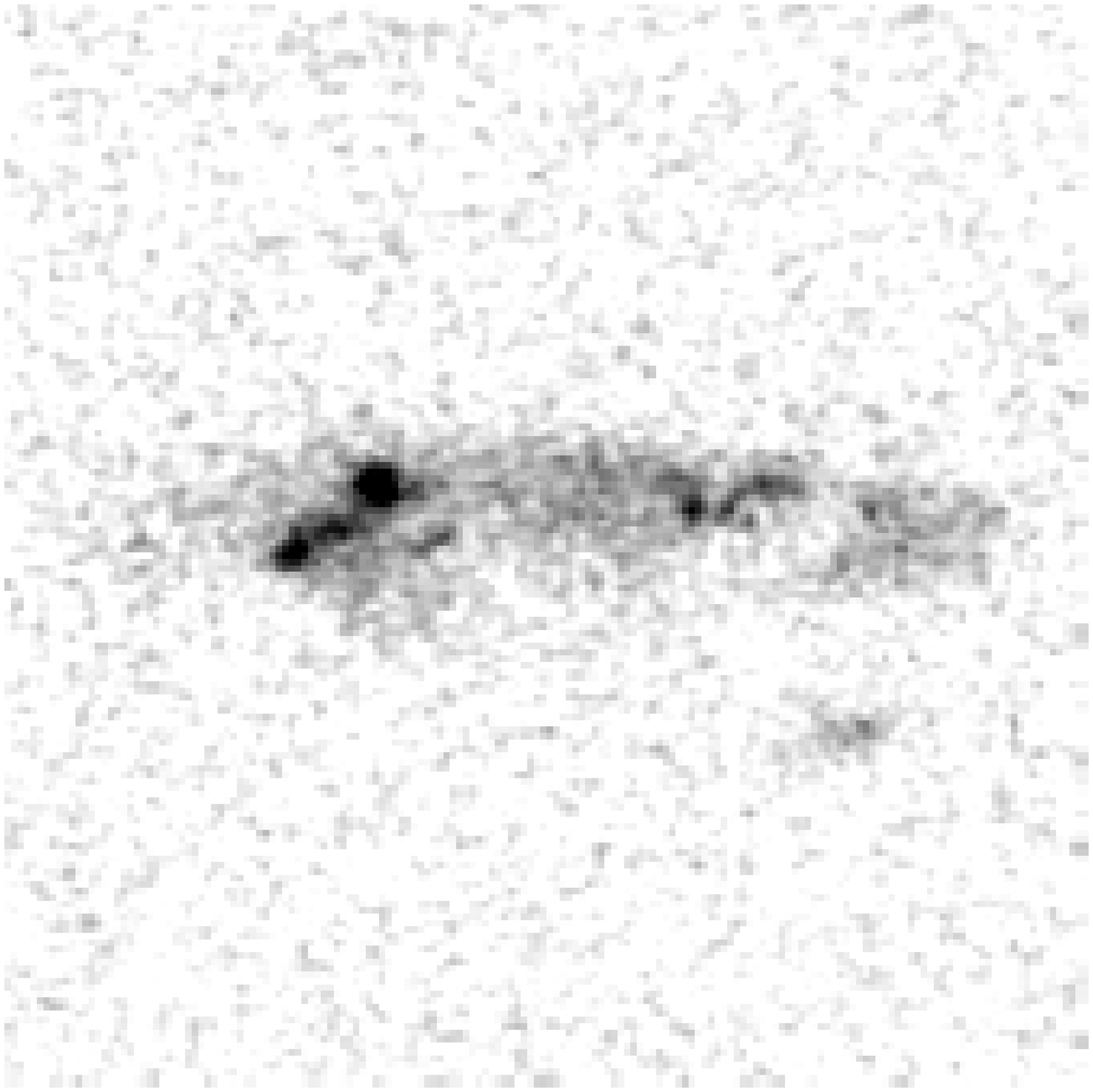}}&
\multicolumn{2}{c}{\includegraphics[width=3cm,height=3cm]{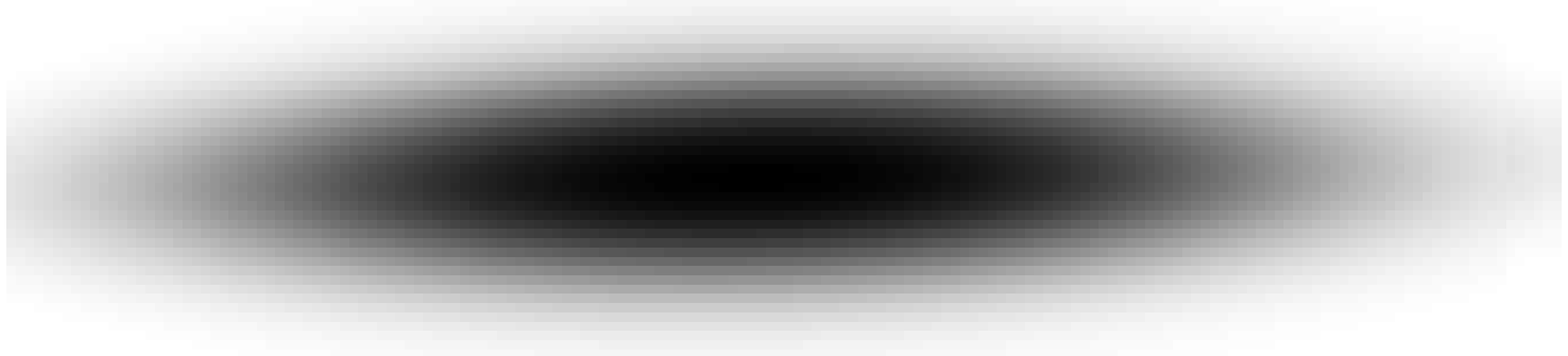}}&
\multicolumn{2}{c}{\includegraphics[width=3cm,height=3cm]{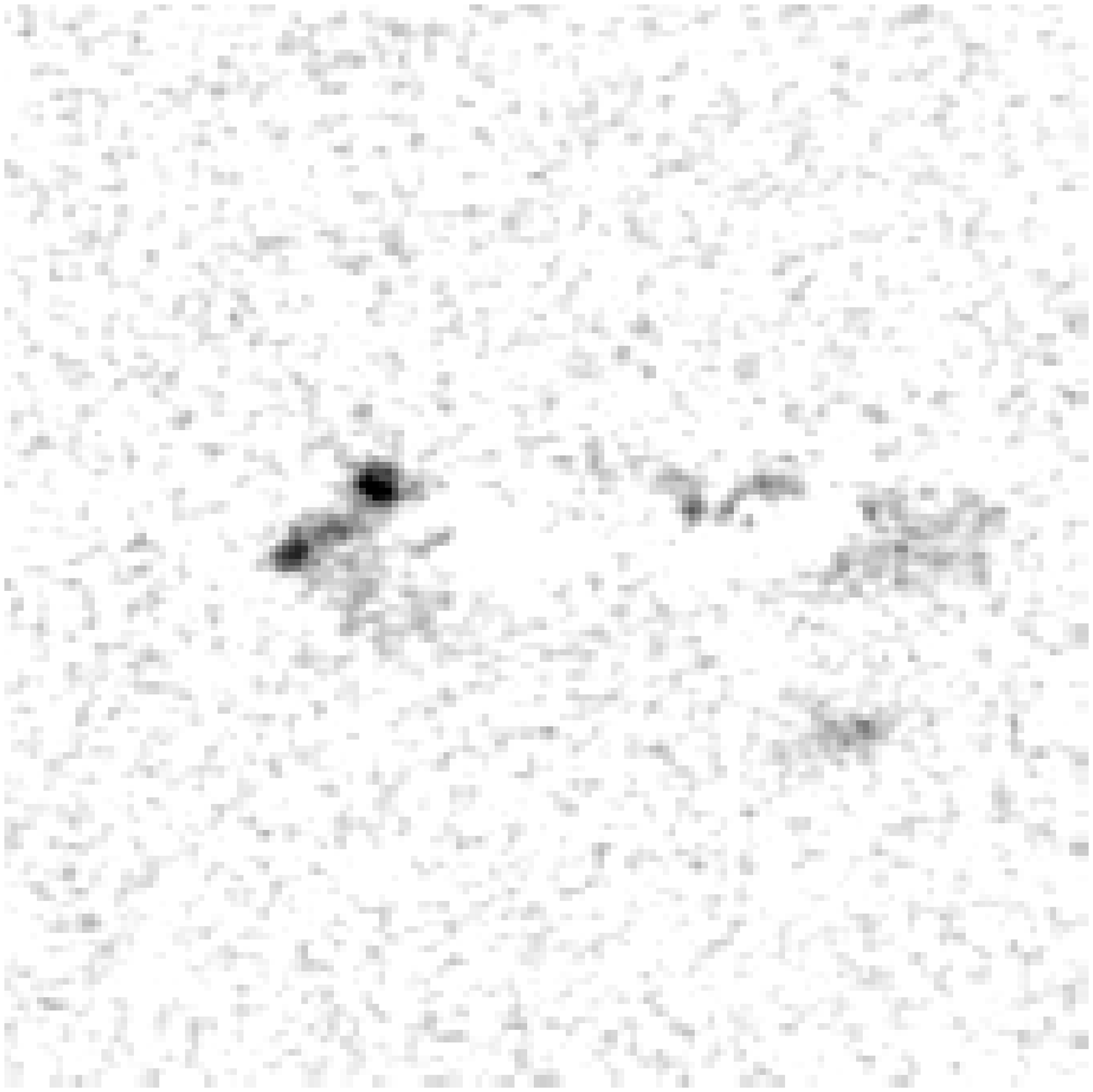}}&
\multicolumn{4}{c|}{\multirow{2}{*}[1.5cm]{\includegraphics[width=6cm,height=4cm]{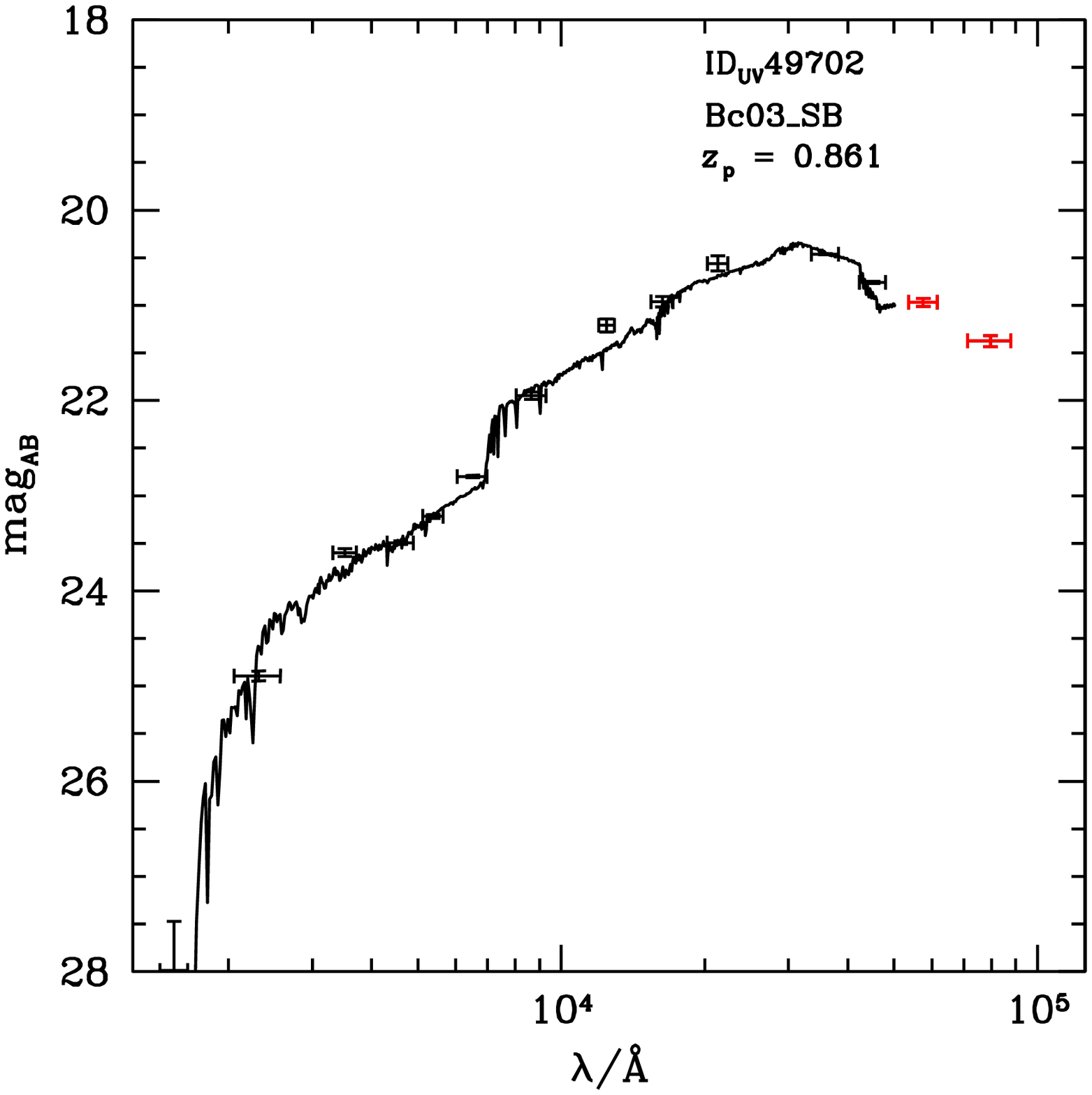}}}\\
\multicolumn{2}{|c}{\includegraphics[width=3cm,height=3cm]{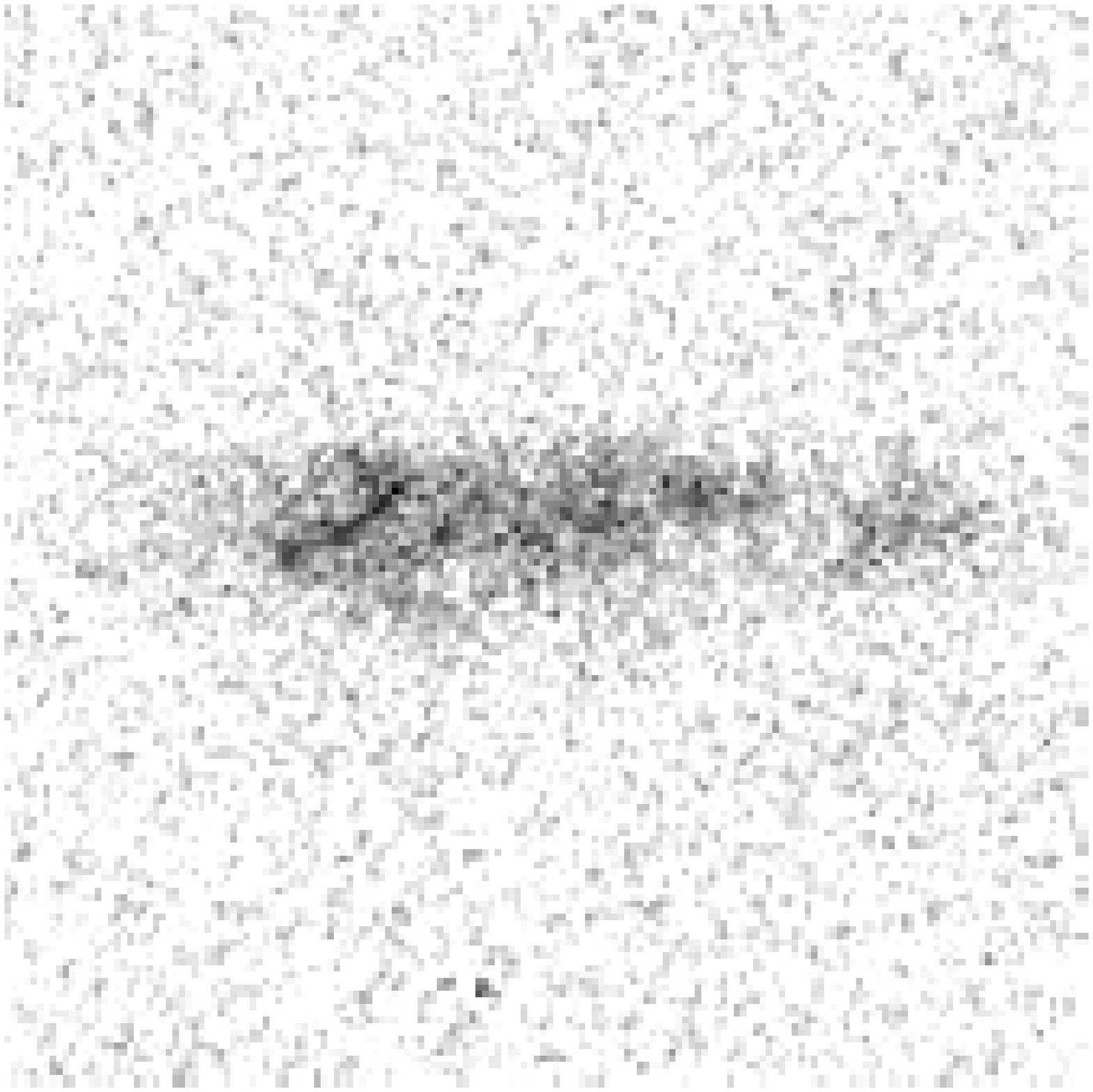}}&
\multicolumn{2}{c}{\includegraphics[width=3cm,height=3cm]{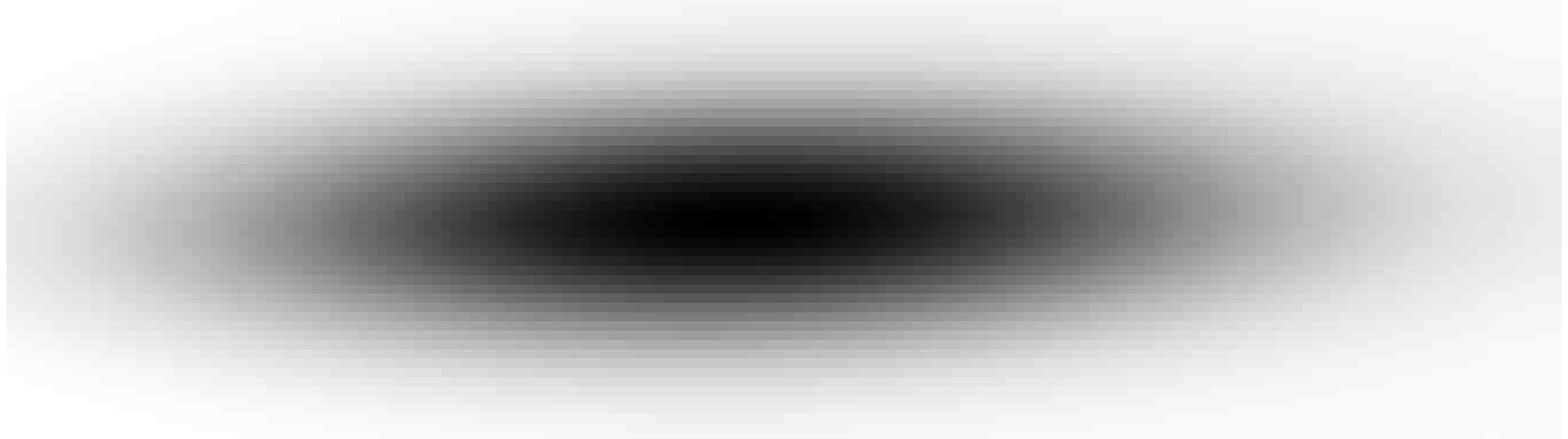}}&
\multicolumn{2}{c}{\includegraphics[width=3cm,height=3cm]{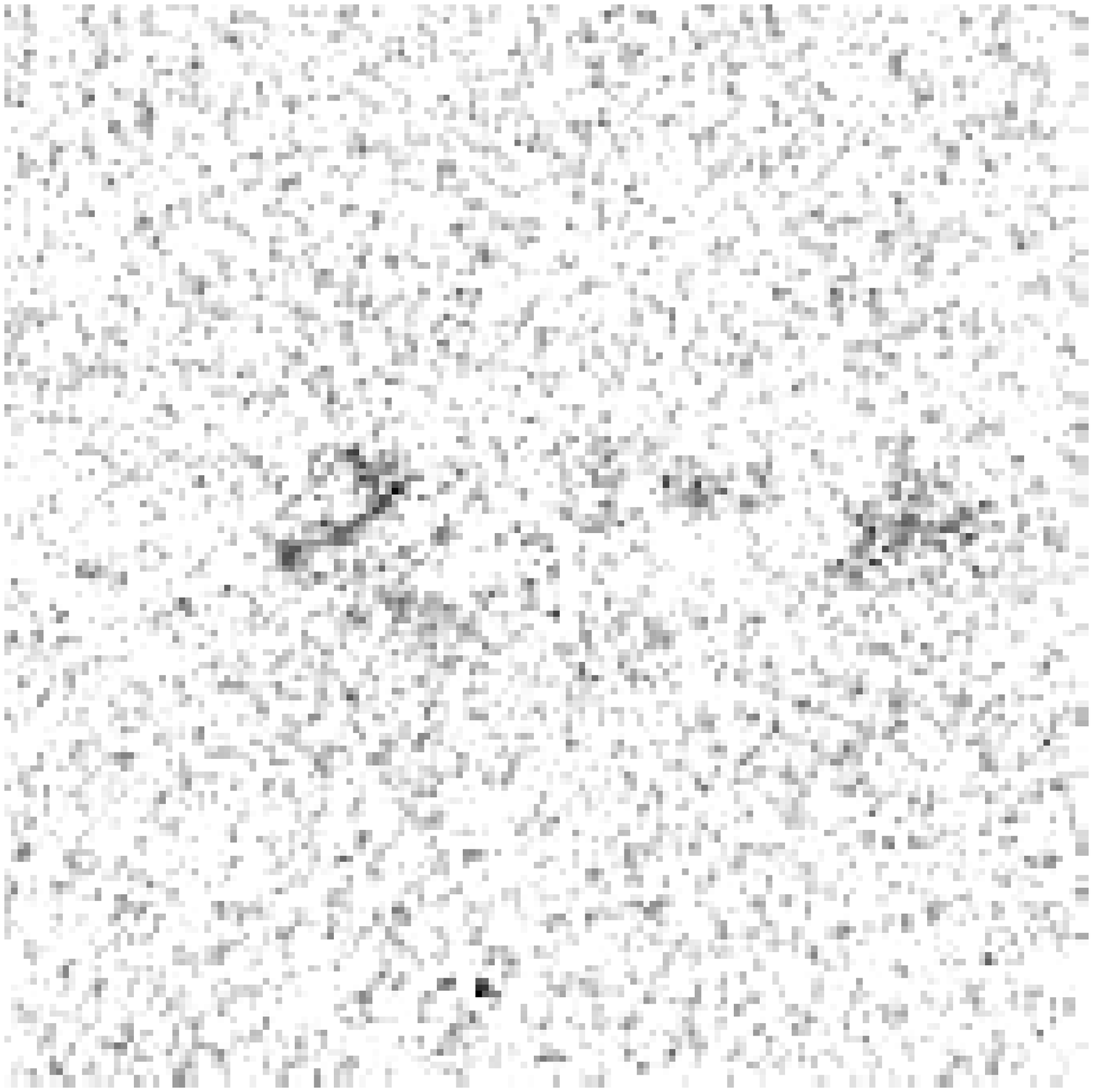}}&
\multicolumn{4}{c|}{}\\
\hline
\multicolumn{10}{c}{continued on next page $...$}\\
\end{tabular}
\end{table*}

\newpage
\begin{table*}
\begin{tabular}{|C{1.2cm}|C{1.2cm}|C{1.2cm}|C{1.2cm}|C{1.2cm}|C{1.2cm}|C{1.2cm}C{1.2cm}C{1.2cm}C{1.2cm}|}
\multicolumn{10}{c}{continue $...$}\\
\hline $ID_{UV}$ & $n_{\rm F606W}$ & $n_{\rm F850LP}$ & $r_{\rm
F606W} {\rm (kpc)}$ &
$r_{\rm F850LP} {\rm (kpc)}$ & flag & type & & & \\

\hline 50258 & 0.55 & 1.40 & 4.70 & 2.60 & 0 & 3 & & &\\
\hline
\multicolumn{10}{|c|}{}\\
\multicolumn{2}{|c}{\includegraphics[width=3cm,height=3cm]{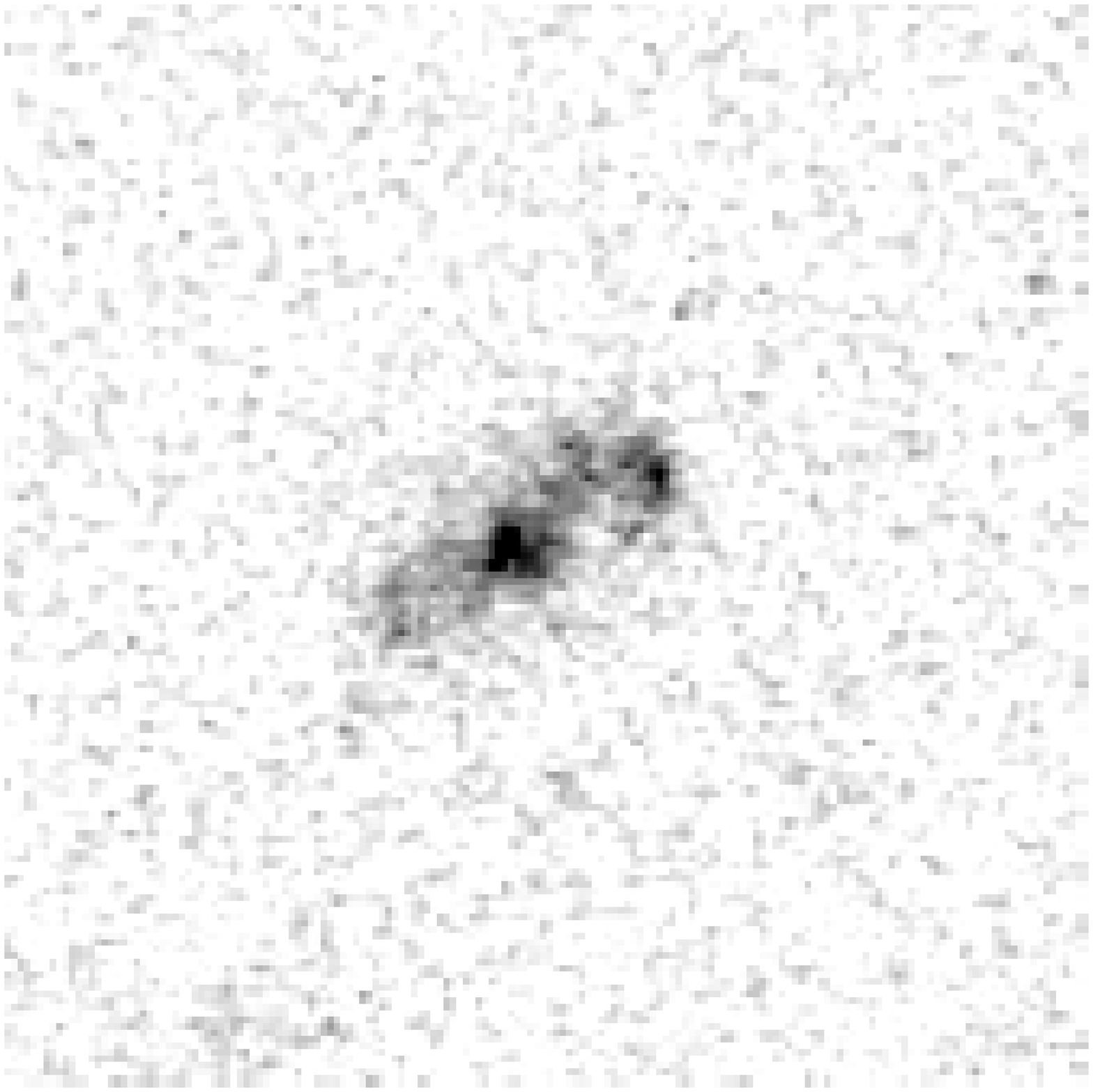}}&
\multicolumn{2}{c}{\includegraphics[width=3cm,height=3cm]{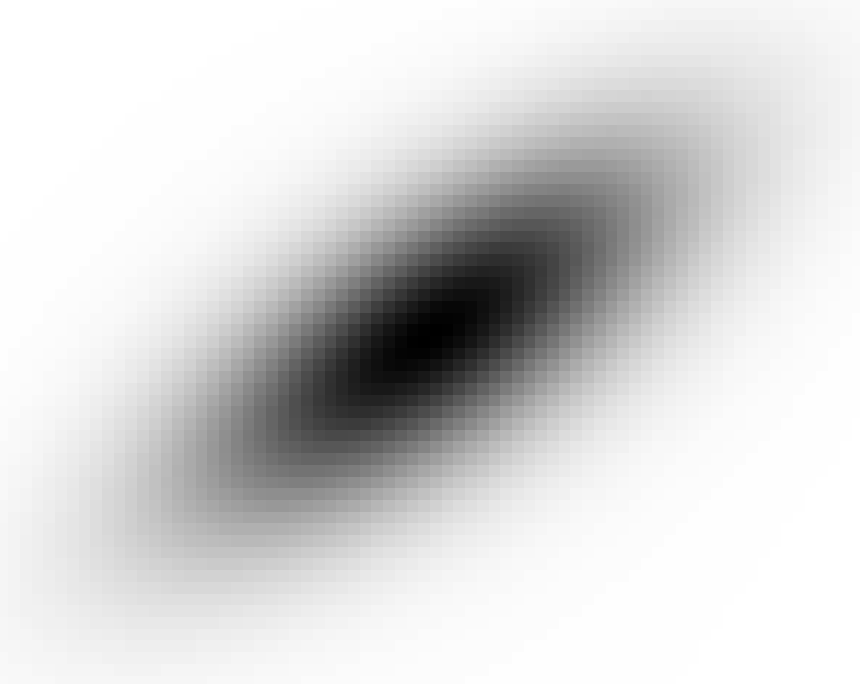}}&
\multicolumn{2}{c}{\includegraphics[width=3cm,height=3cm]{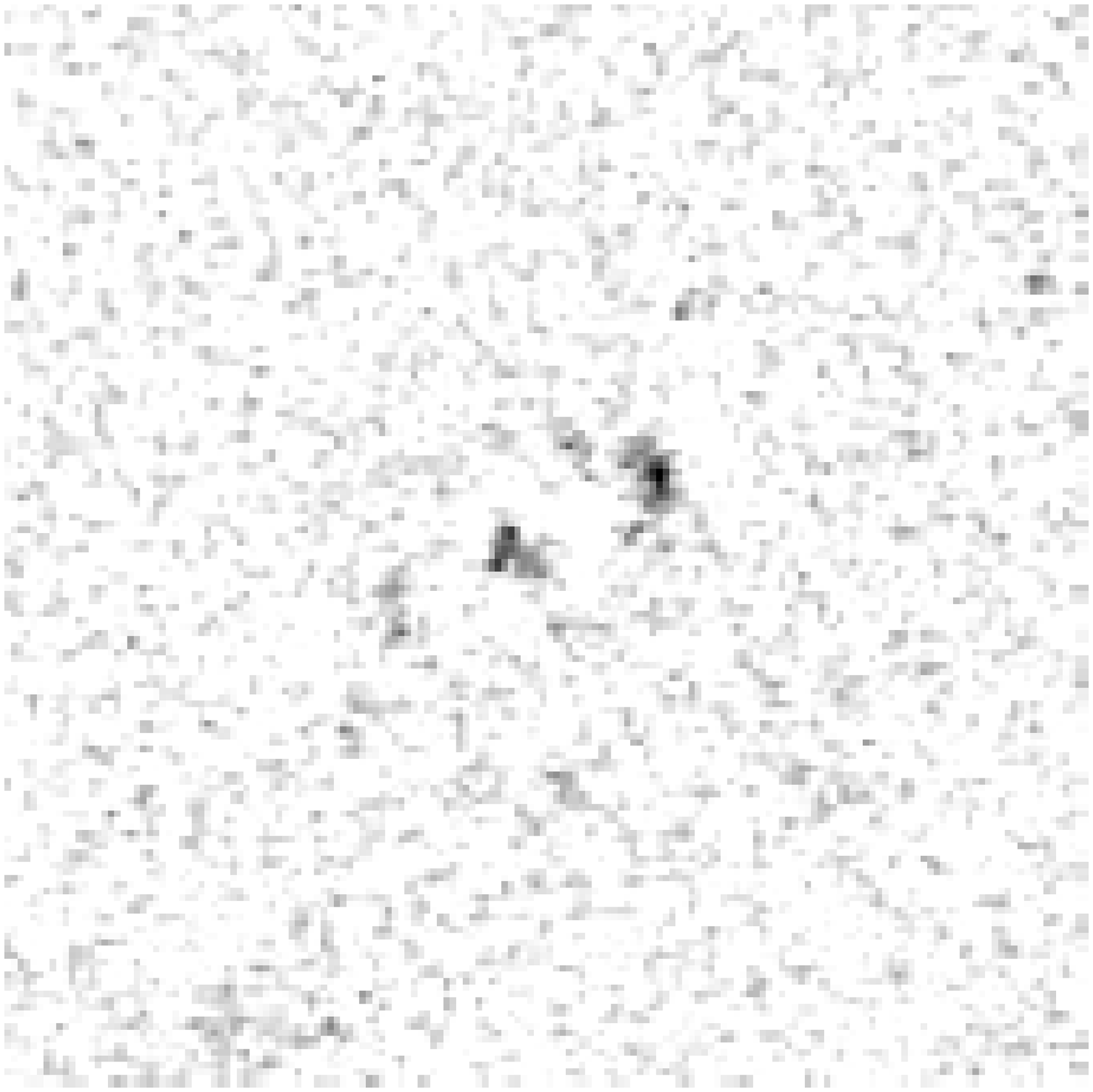}}&
\multicolumn{4}{c|}{\multirow{2}{*}[1.5cm]{\includegraphics[width=6cm,height=4cm]{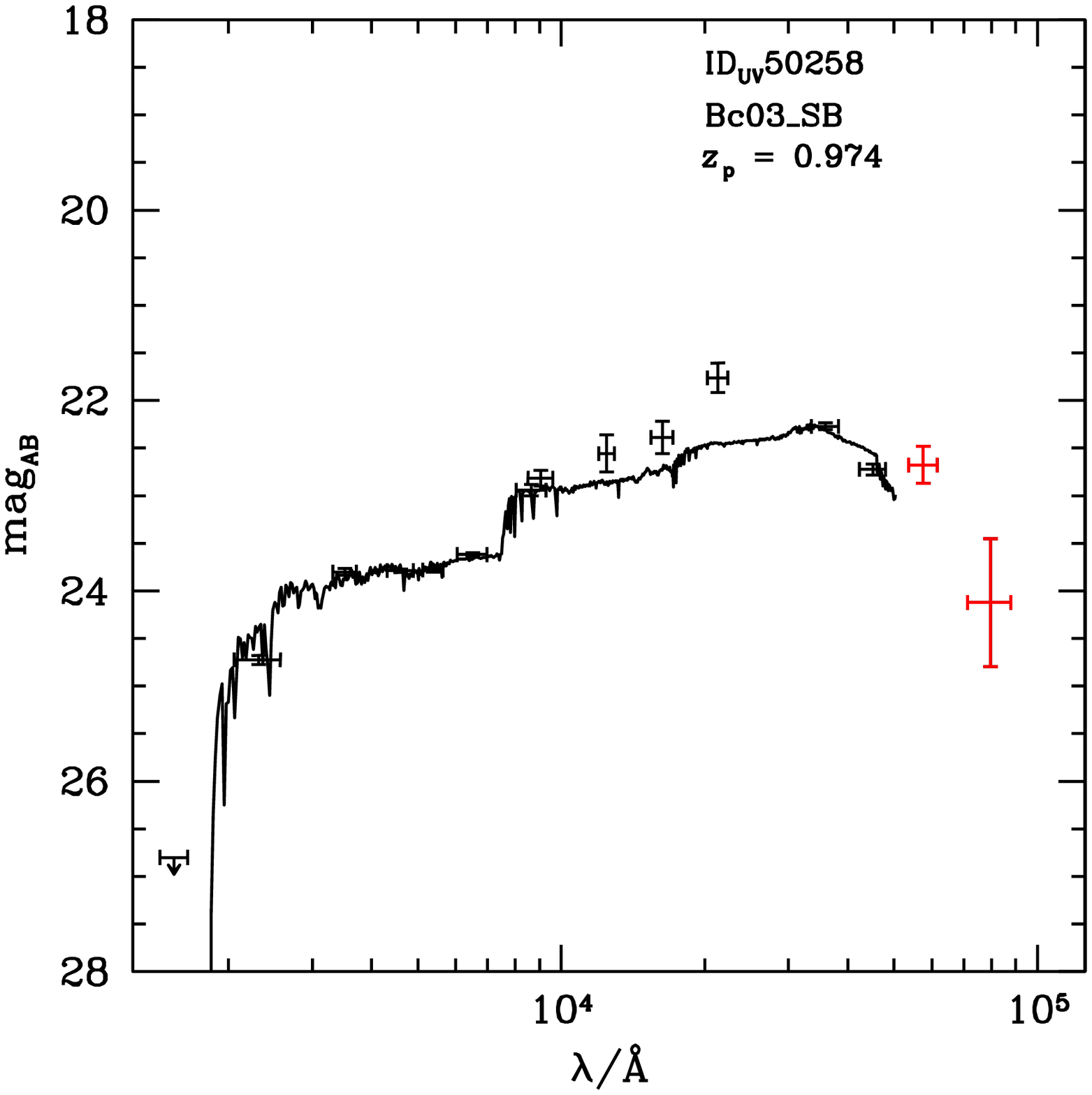}}}\\
\multicolumn{2}{|c}{\includegraphics[width=3cm,height=3cm]{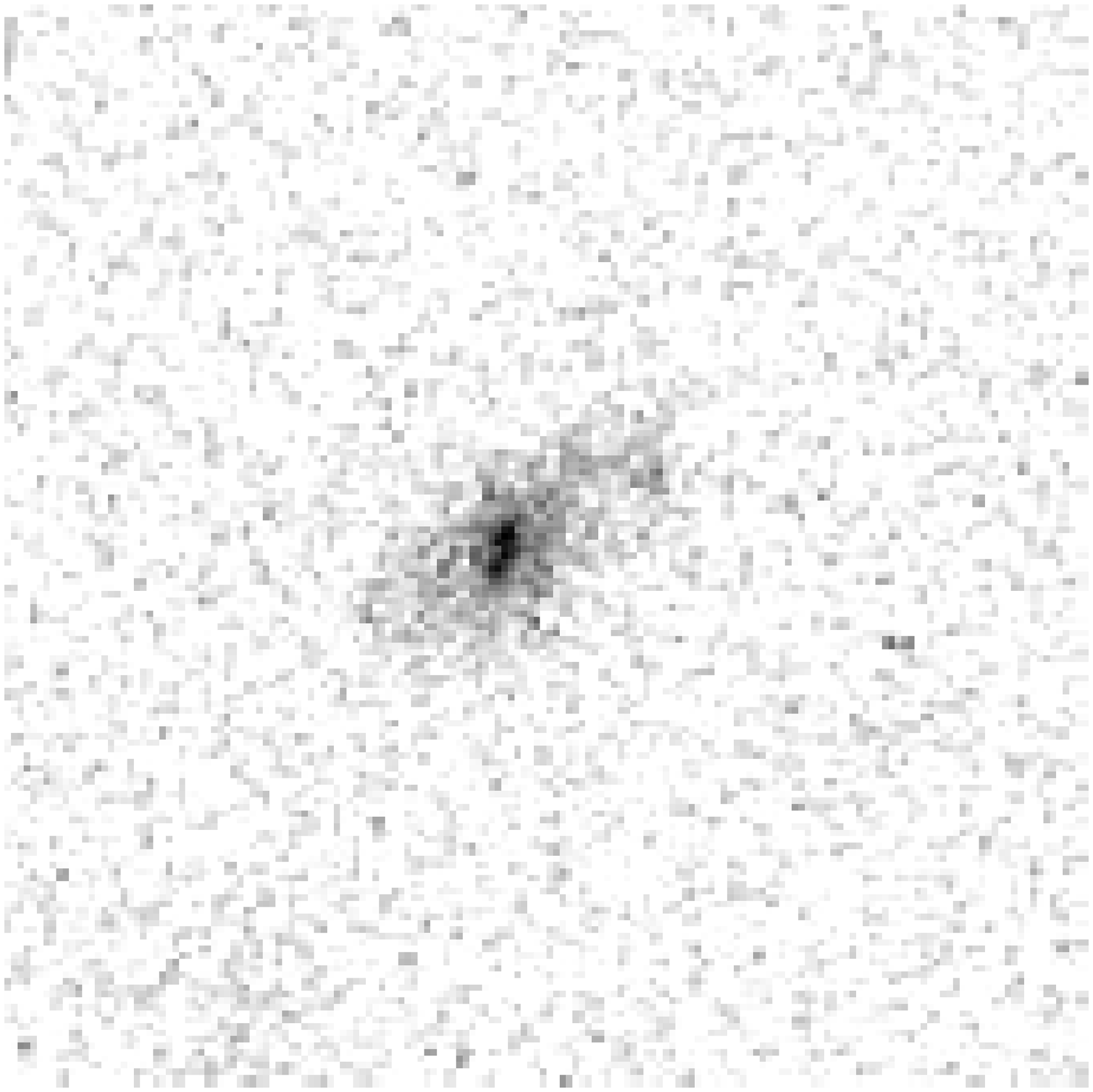}}&
\multicolumn{2}{c}{\includegraphics[width=3cm,height=3cm]{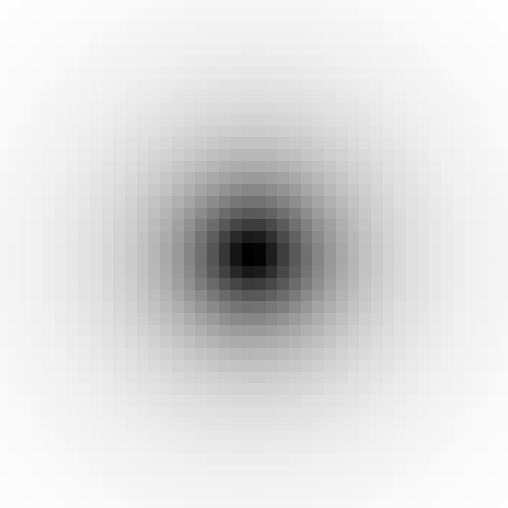}}&
\multicolumn{2}{c}{\includegraphics[width=3cm,height=3cm]{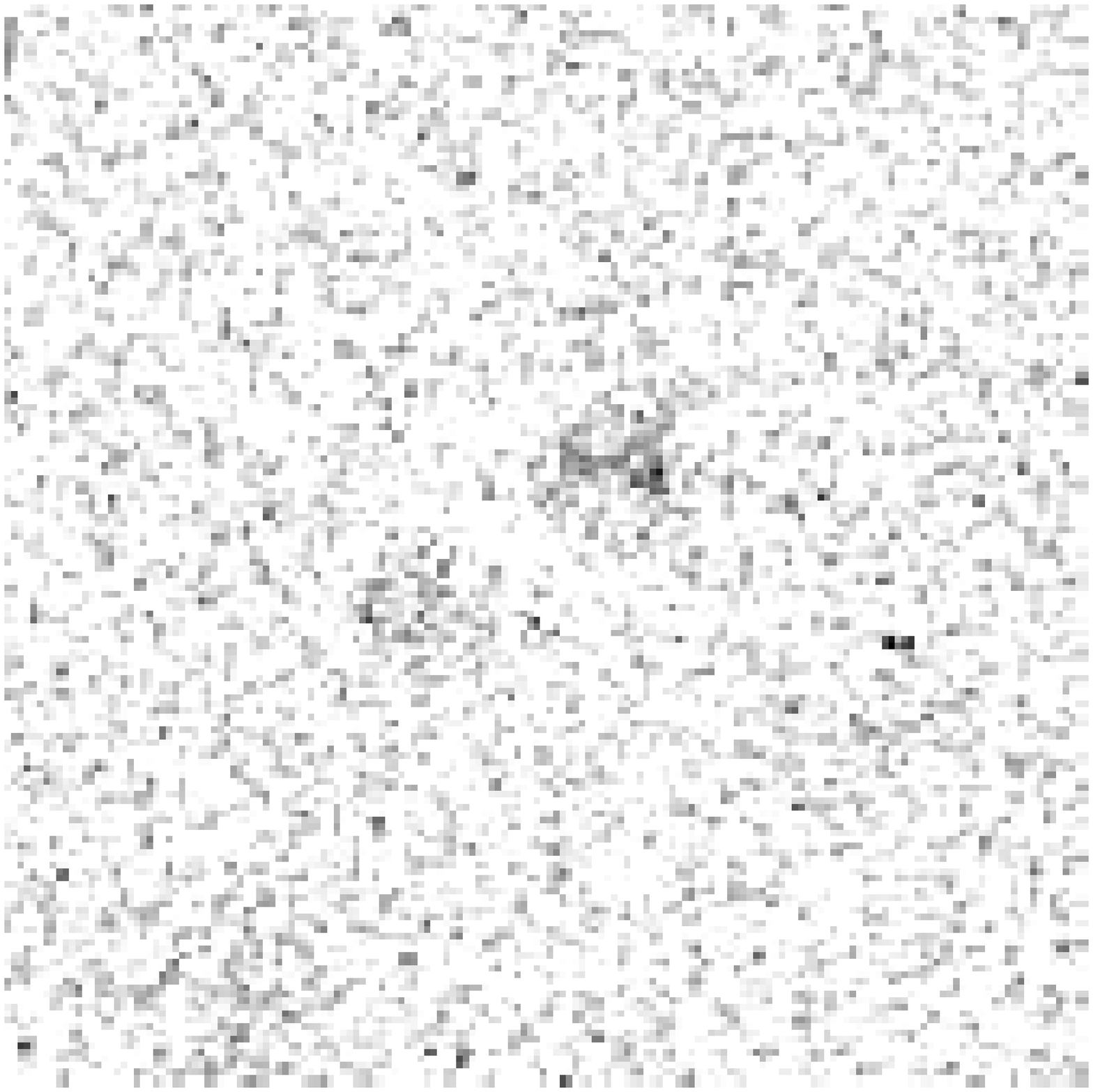}}&
\multicolumn{4}{c|}{}\\

\hline $ID_{UV}$ & $n_{\rm F606W}$ & $n_{\rm F850LP}$ & $r_{\rm
F606W} {\rm (kpc)}$ &
$r_{\rm F850LP} {\rm (kpc)}$ & flag & type & & & \\

\hline 50616 & 99 & 99 & 99 & 99 & 1 & 3 & & &\\
\hline
\multicolumn{10}{|c|}{}\\
\multicolumn{2}{|c}{\includegraphics[width=3cm,height=3cm]{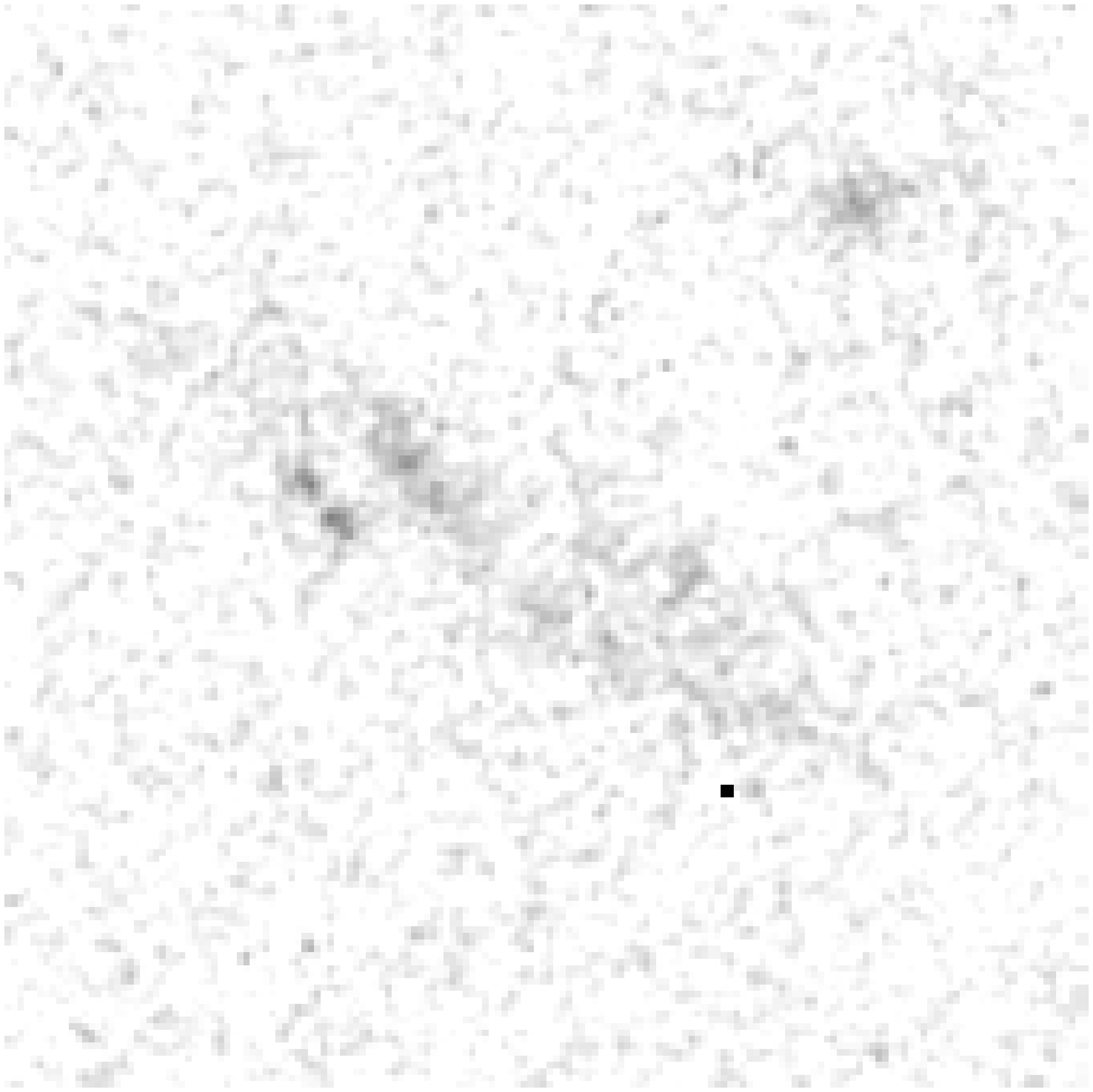}}&
\multicolumn{2}{c}{\includegraphics[width=3cm,height=3cm]{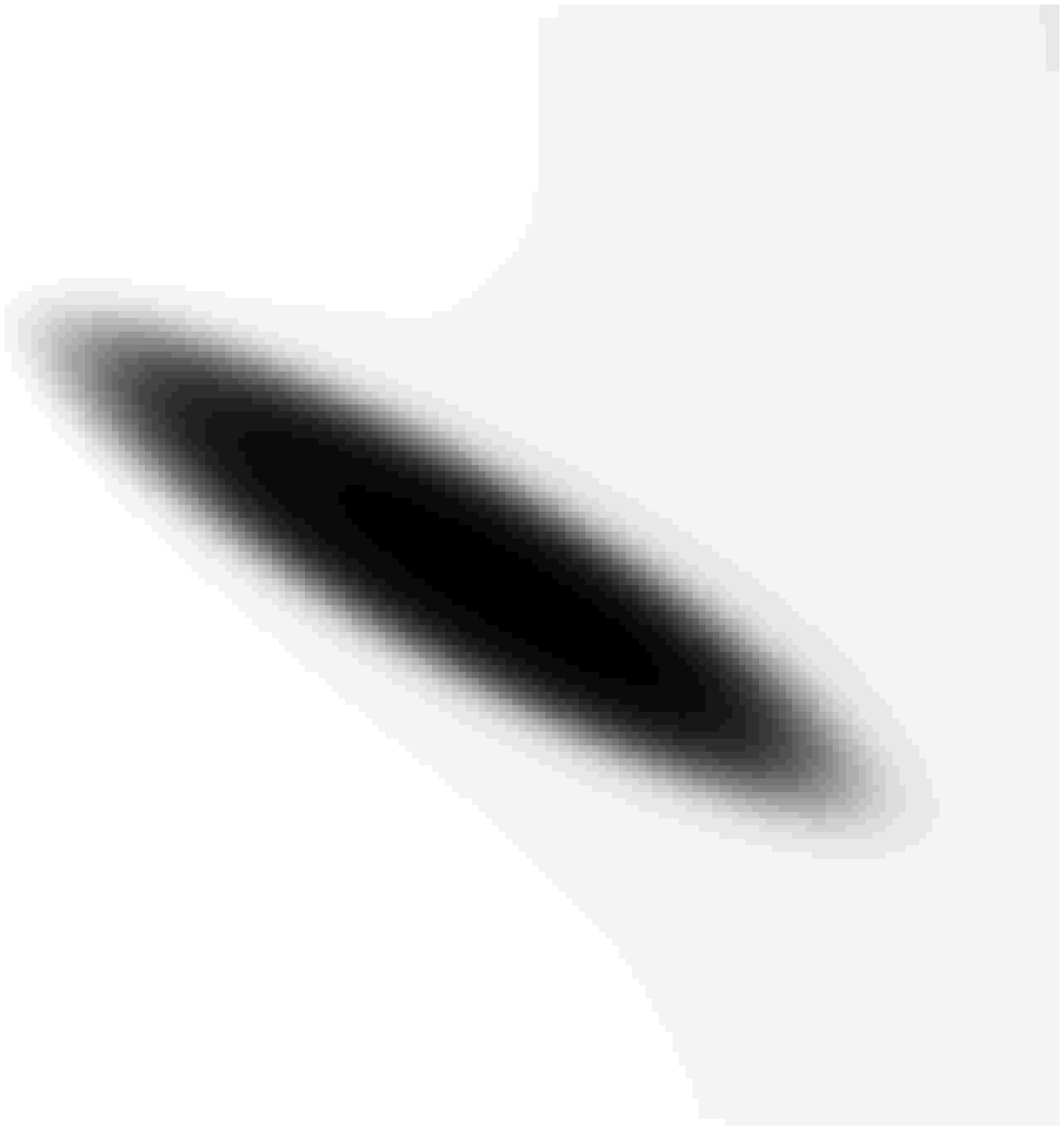}}&
\multicolumn{2}{c}{\includegraphics[width=3cm,height=3cm]{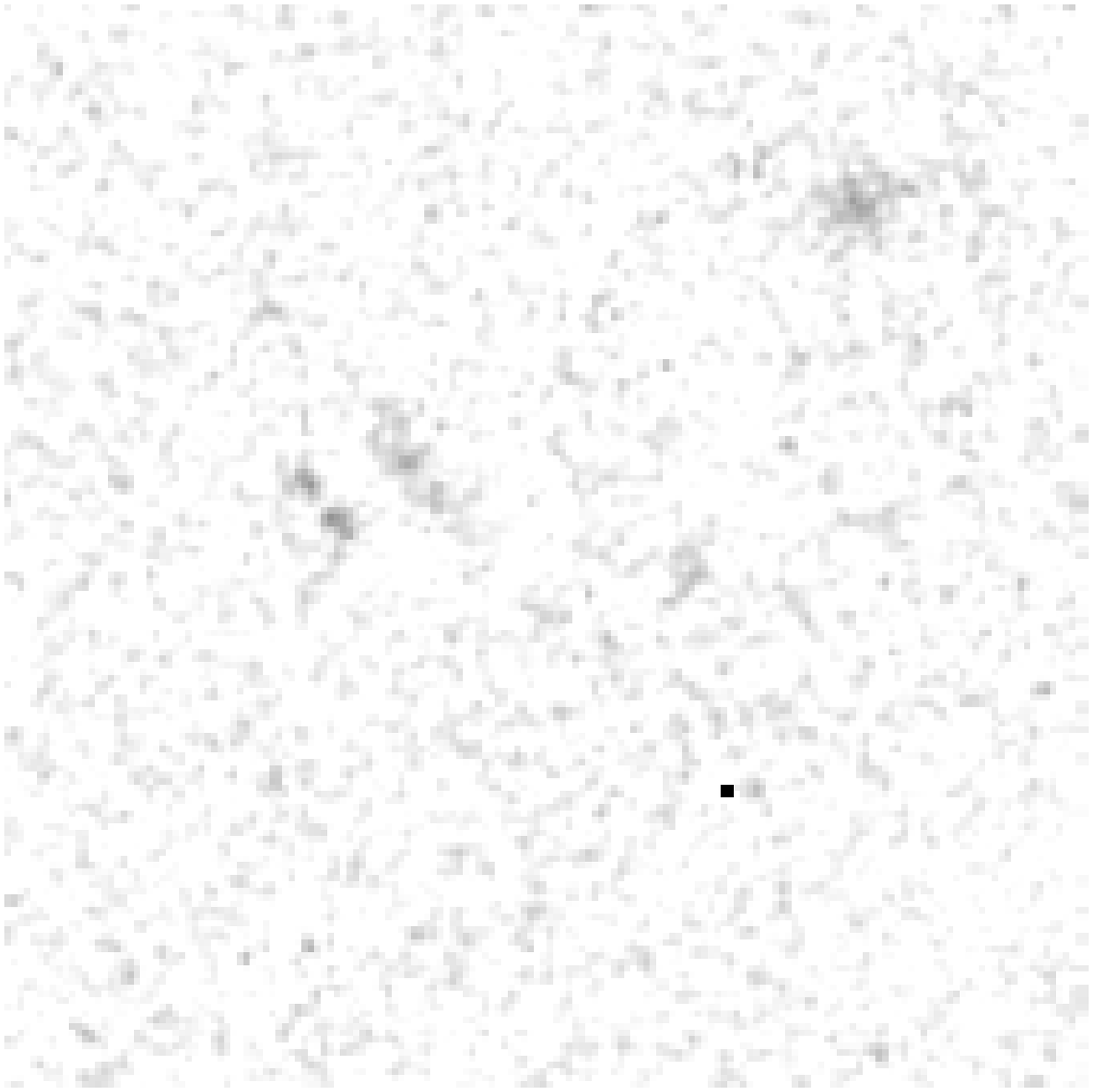}}&
\multicolumn{4}{c|}{\multirow{2}{*}[1.5cm]{\includegraphics[width=6cm,height=4cm]{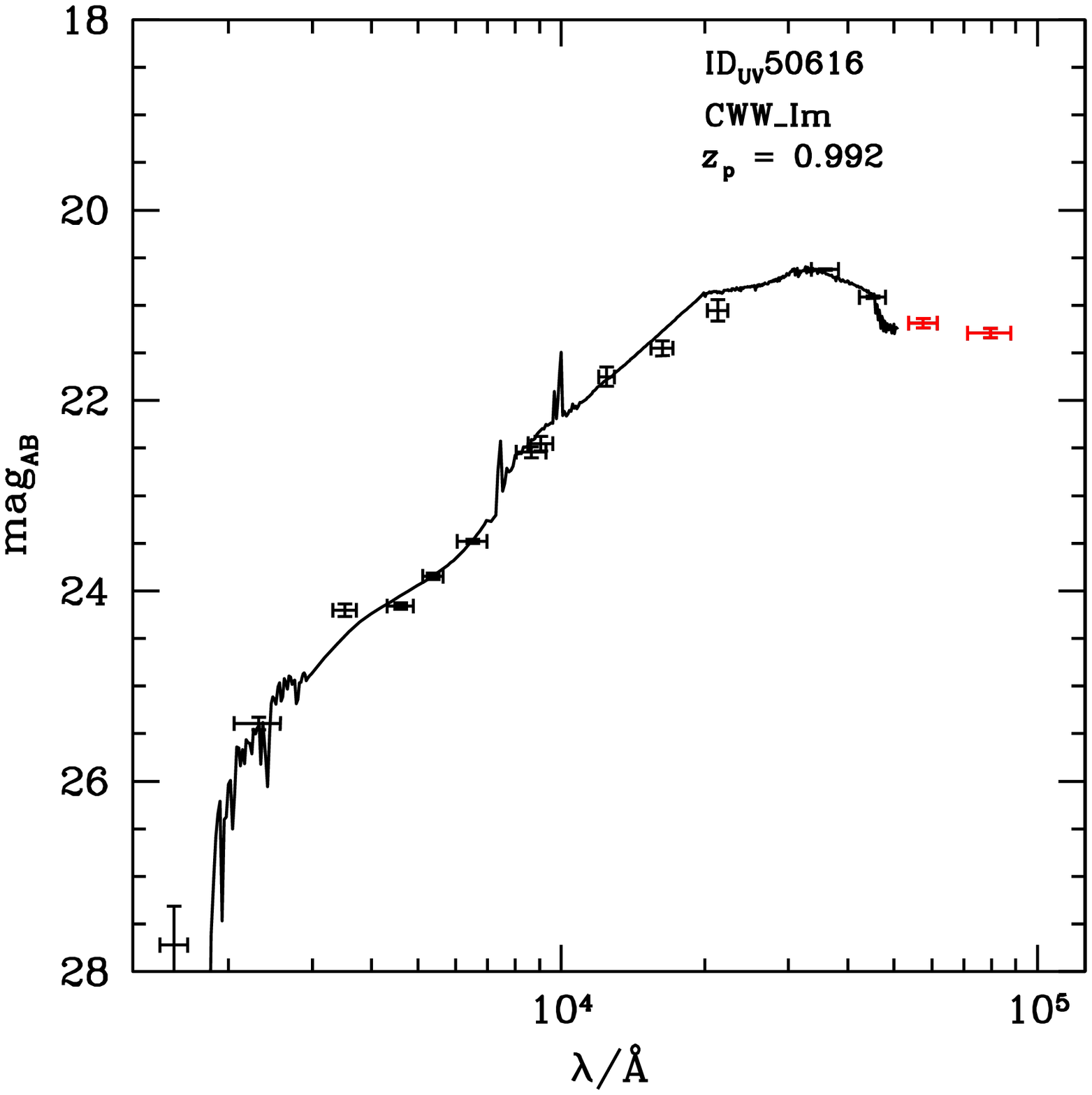}}}\\
\multicolumn{2}{|c}{\includegraphics[width=3cm,height=3cm]{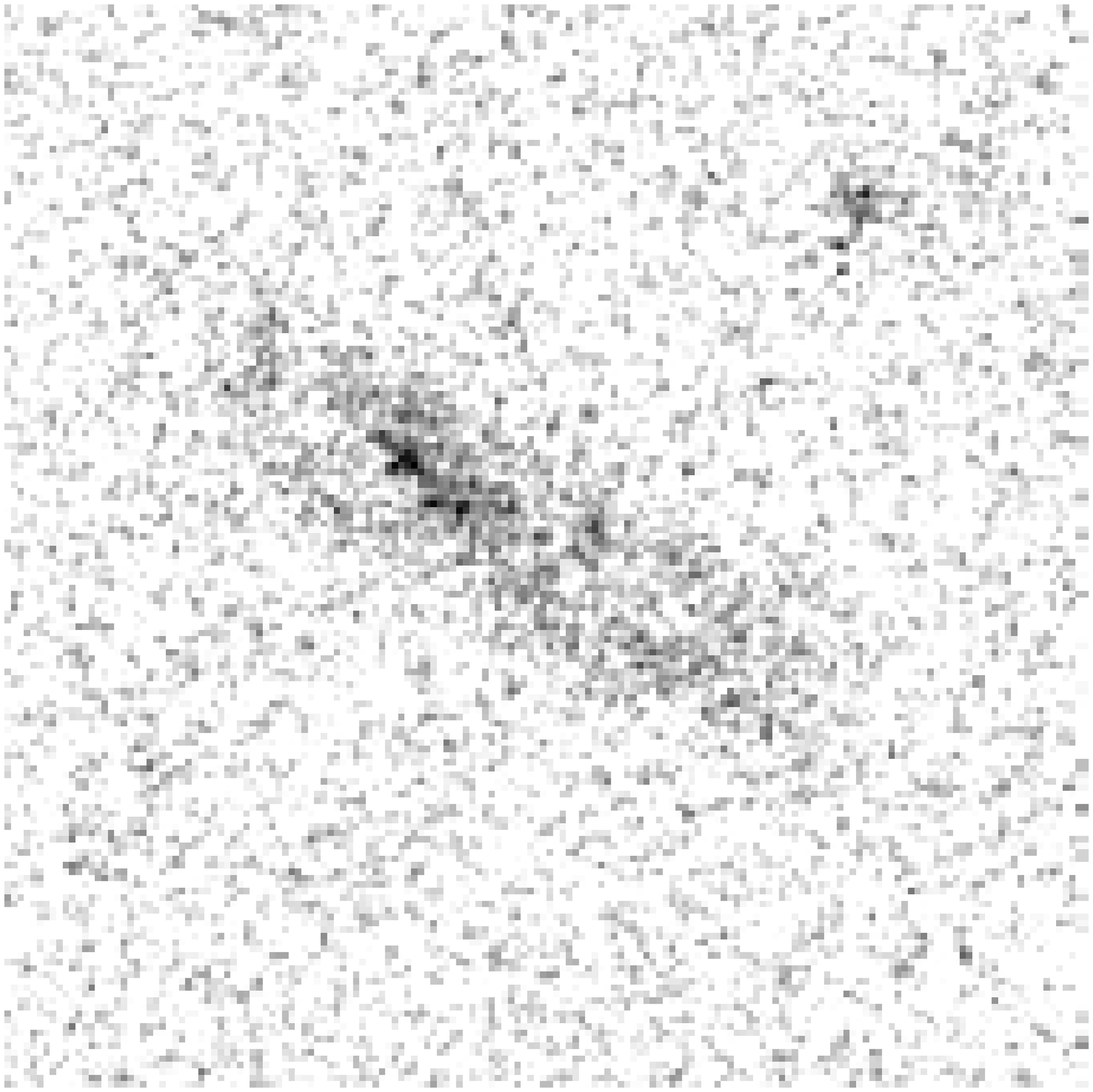}}&
\multicolumn{2}{c}{\includegraphics[width=3cm,height=3cm]{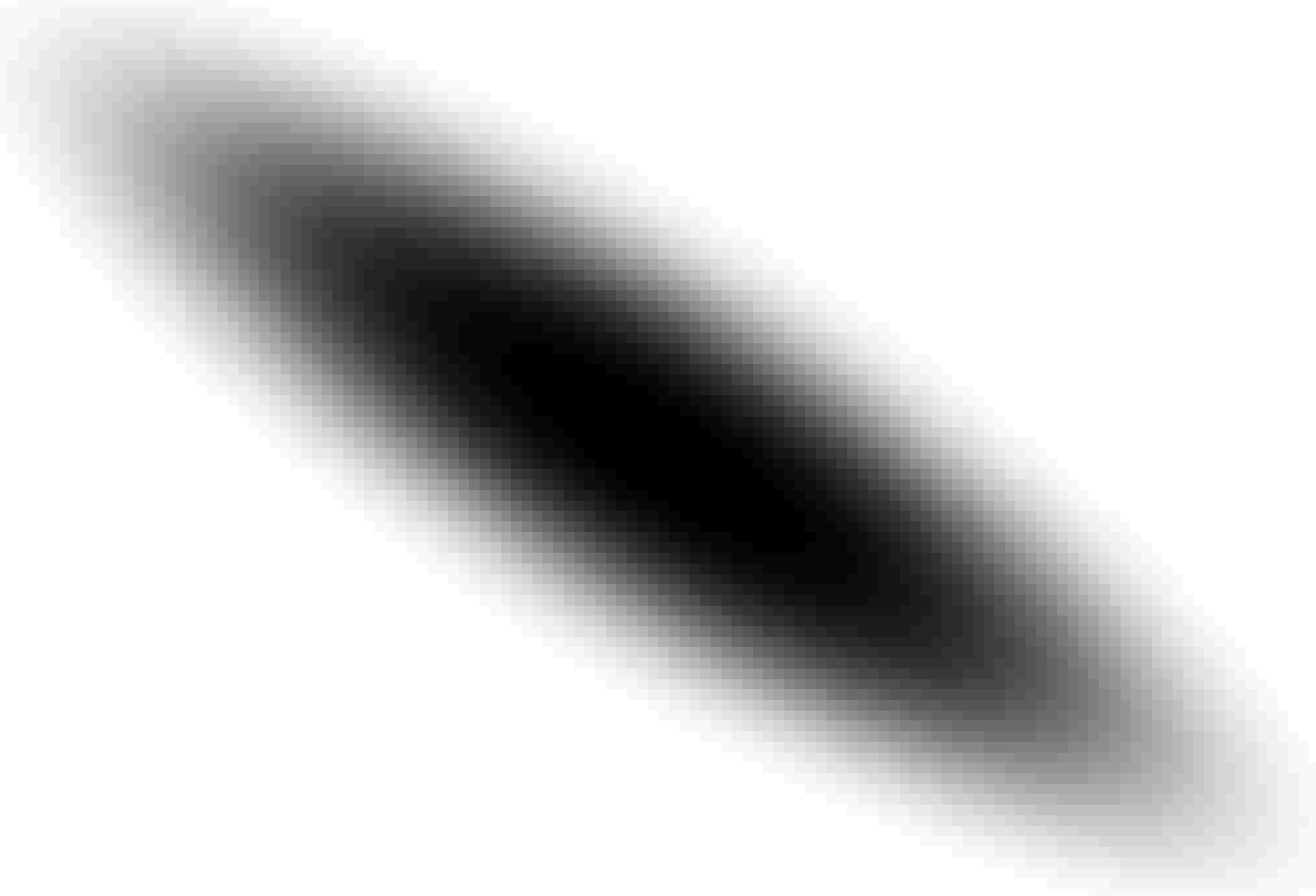}}&
\multicolumn{2}{c}{\includegraphics[width=3cm,height=3cm]{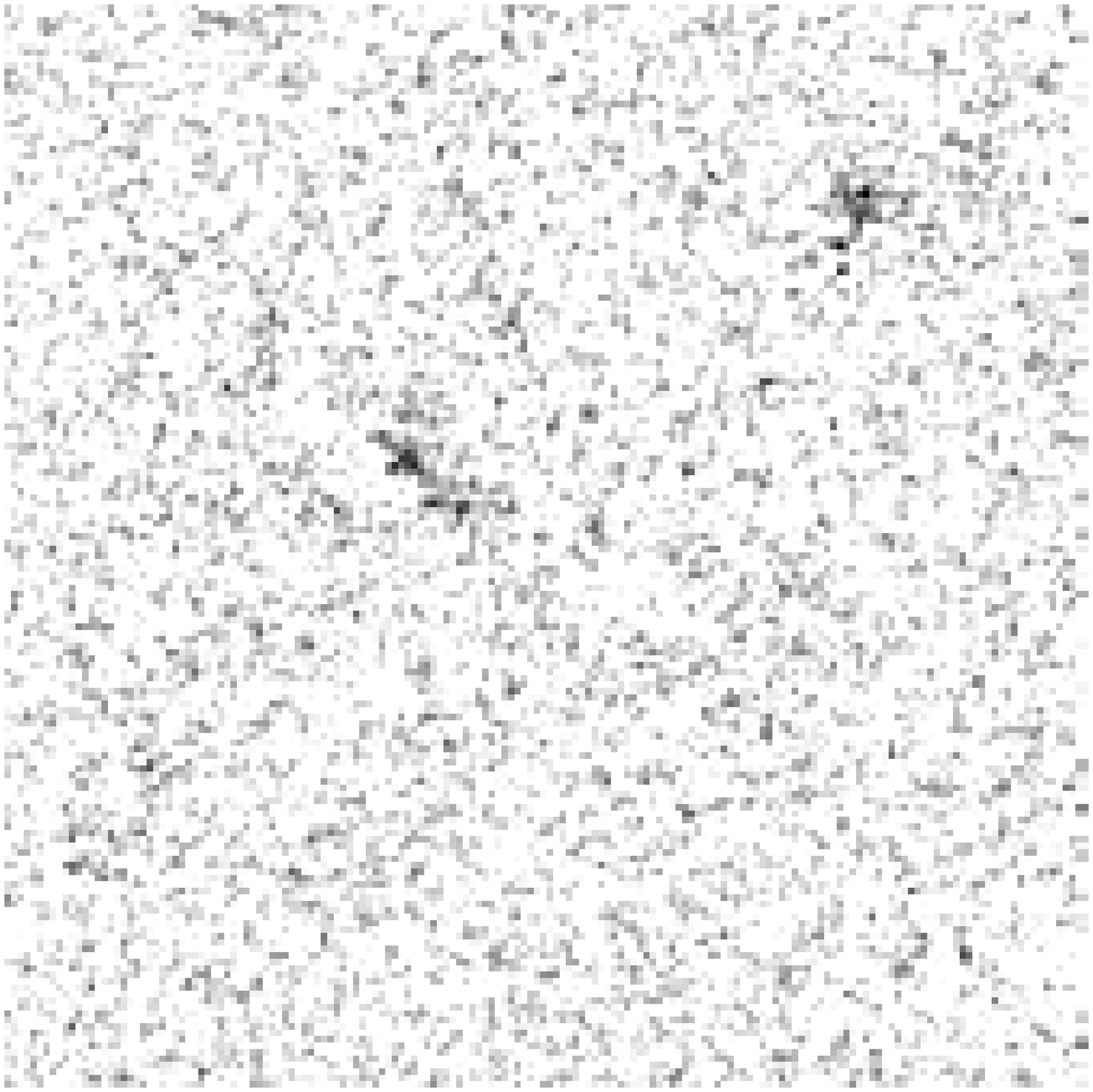}}&
\multicolumn{4}{c|}{}\\

\hline $ID_{UV}$ & $n_{\rm F606W}$ & $n_{\rm F850LP}$ & $r_{\rm
F606W} {\rm (kpc)}$ &
$r_{\rm F850LP} {\rm (kpc)}$ & flag & type & & & \\

\hline 52269 & 99 & 99 & 99 & 99 & 1 & 3 & & &\\
\hline
\multicolumn{10}{|c|}{}\\
\multicolumn{2}{|c}{\includegraphics[width=3cm,height=3cm]{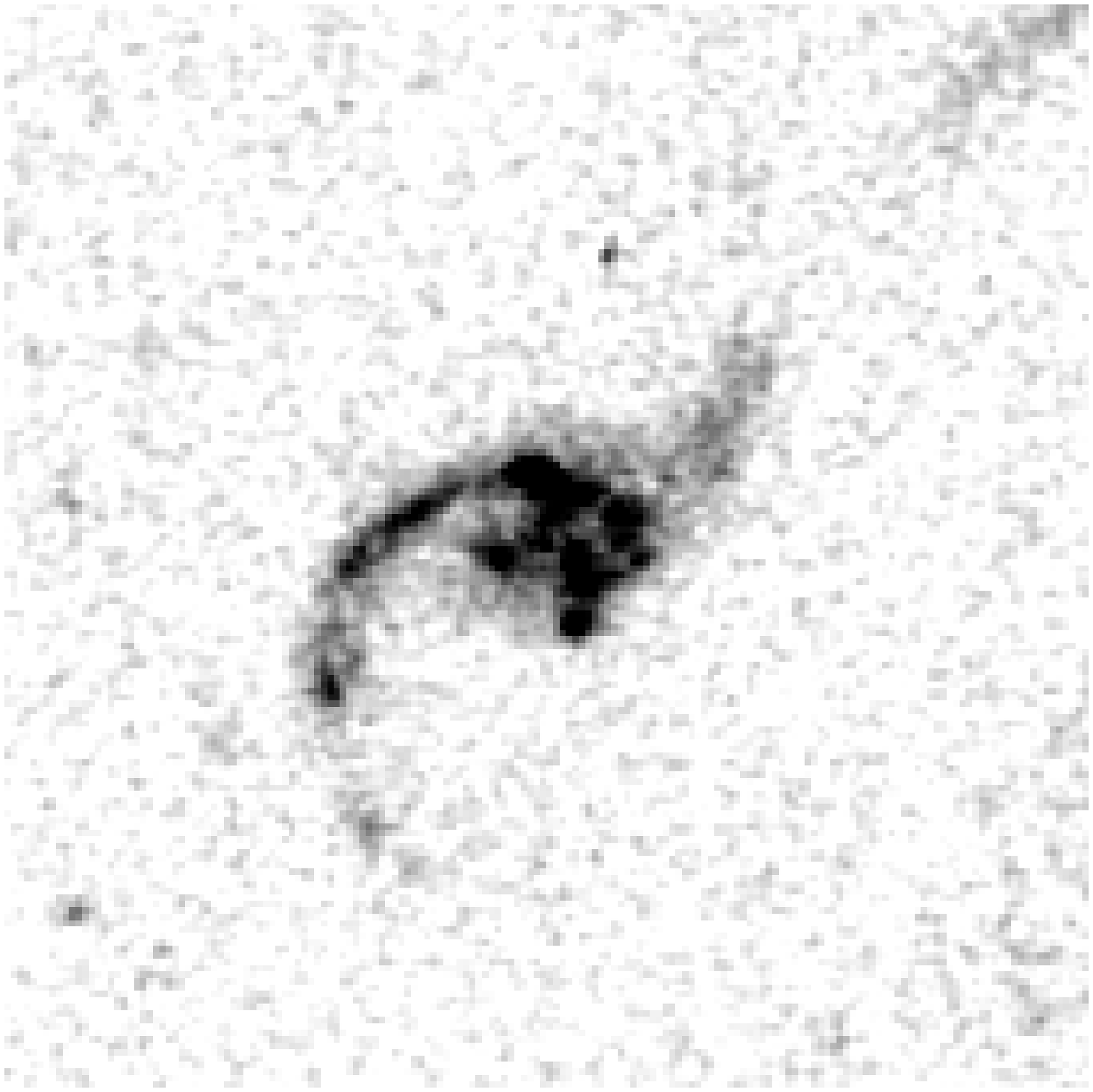}}&
\multicolumn{2}{c}{\includegraphics[width=3cm,height=3cm]{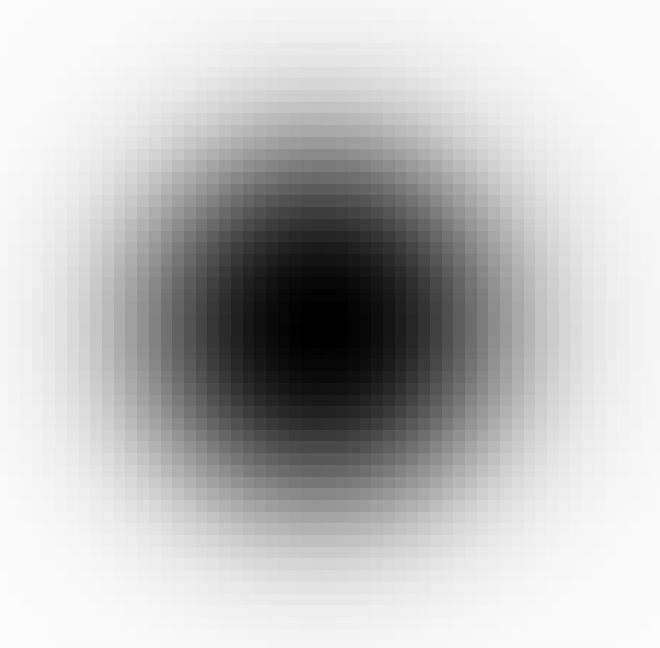}}&
\multicolumn{2}{c}{\includegraphics[width=3cm,height=3cm]{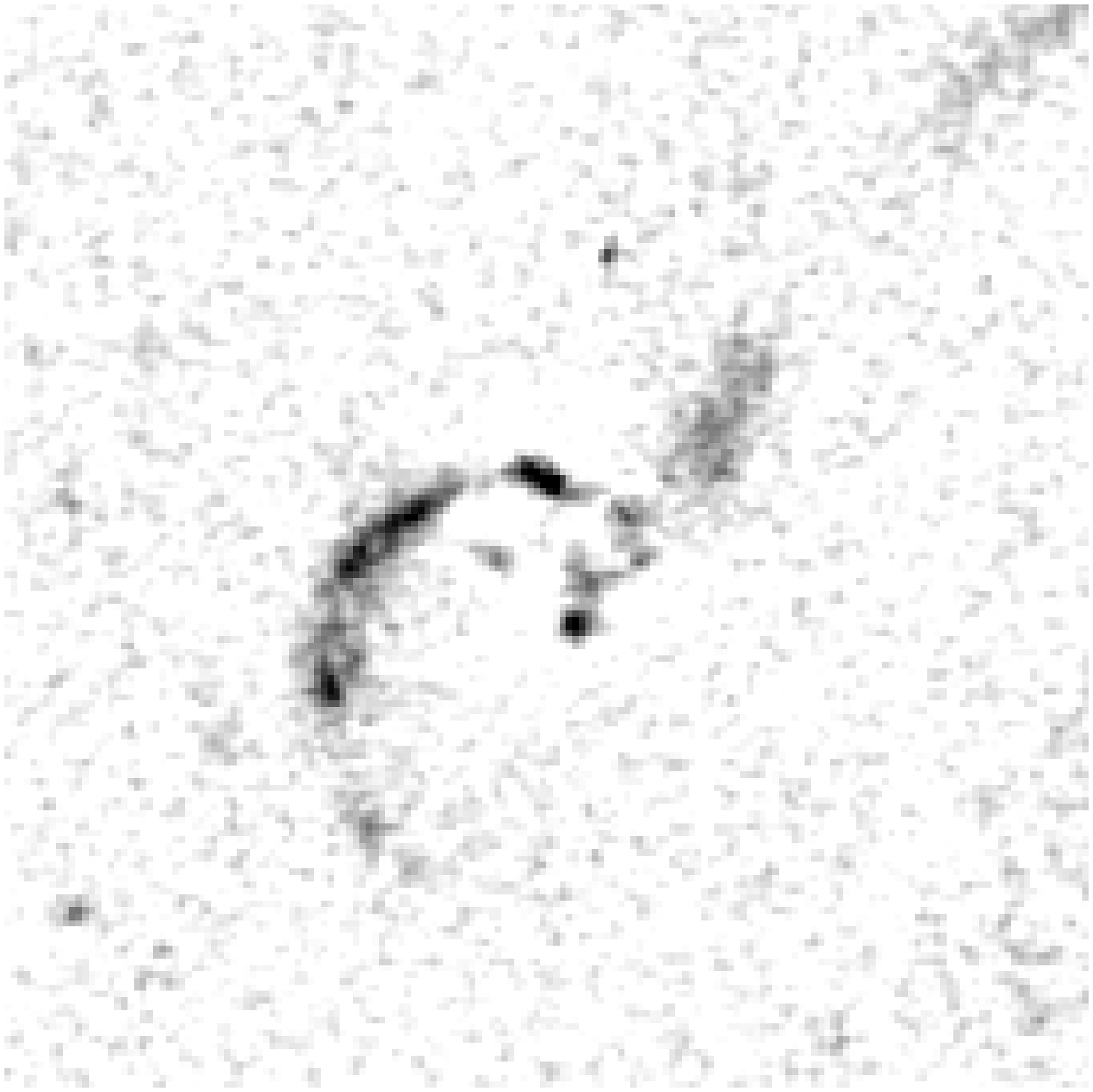}}&
\multicolumn{4}{c|}{\multirow{2}{*}[1.5cm]{\includegraphics[width=6cm,height=4cm]{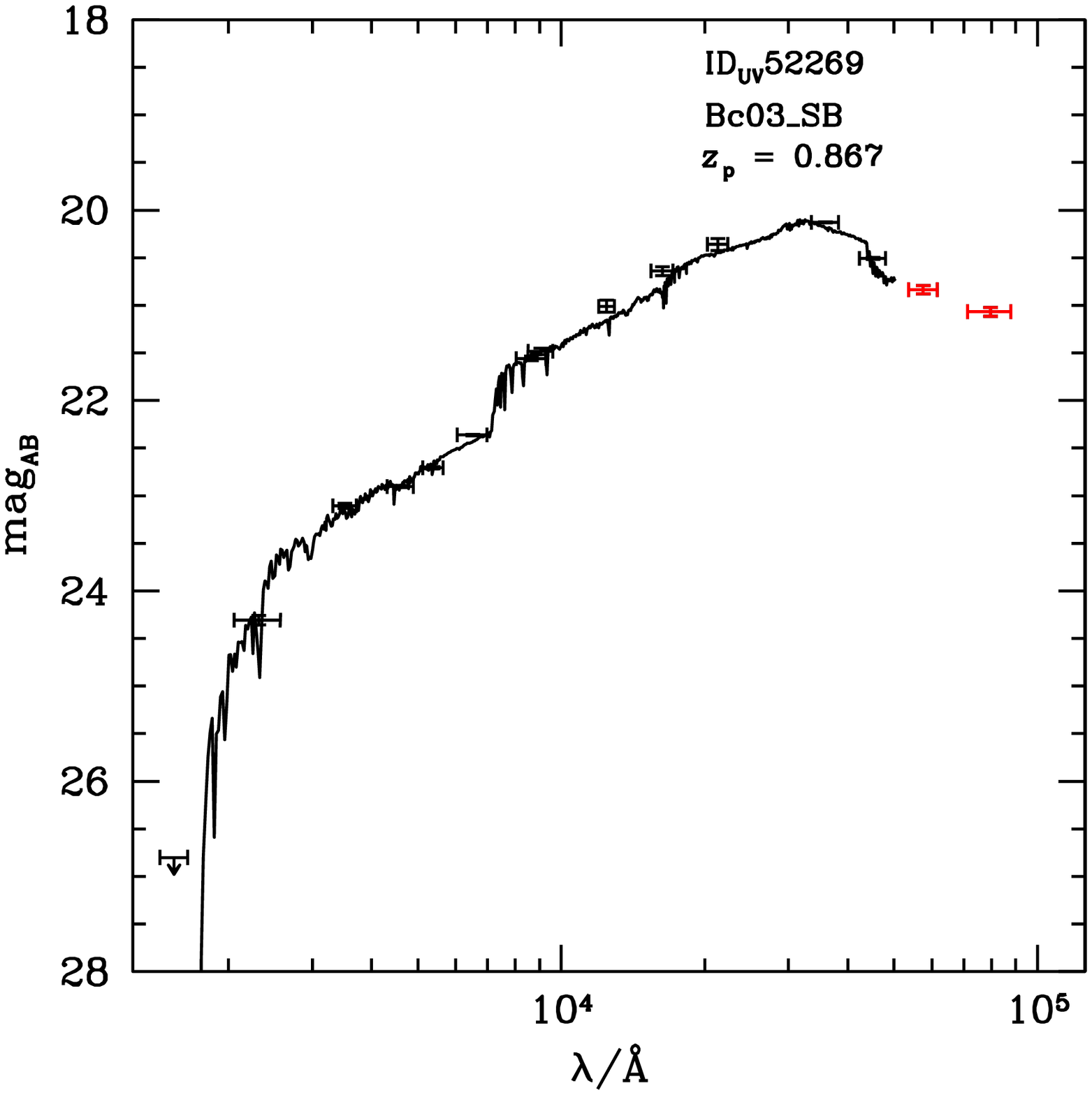}}}\\
\multicolumn{2}{|c}{\includegraphics[width=3cm,height=3cm]{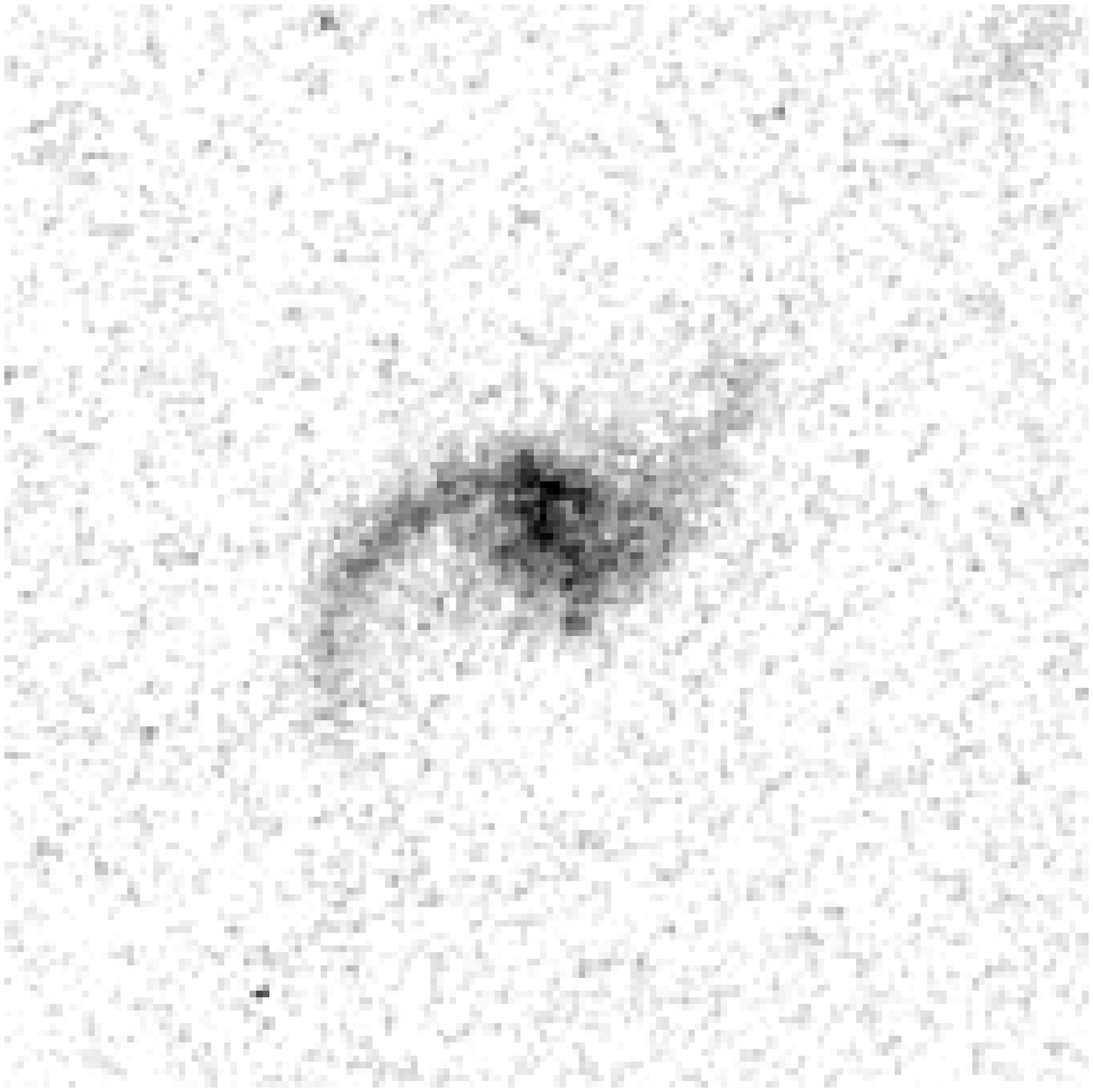}}&
\multicolumn{2}{c}{\includegraphics[width=3cm,height=3cm]{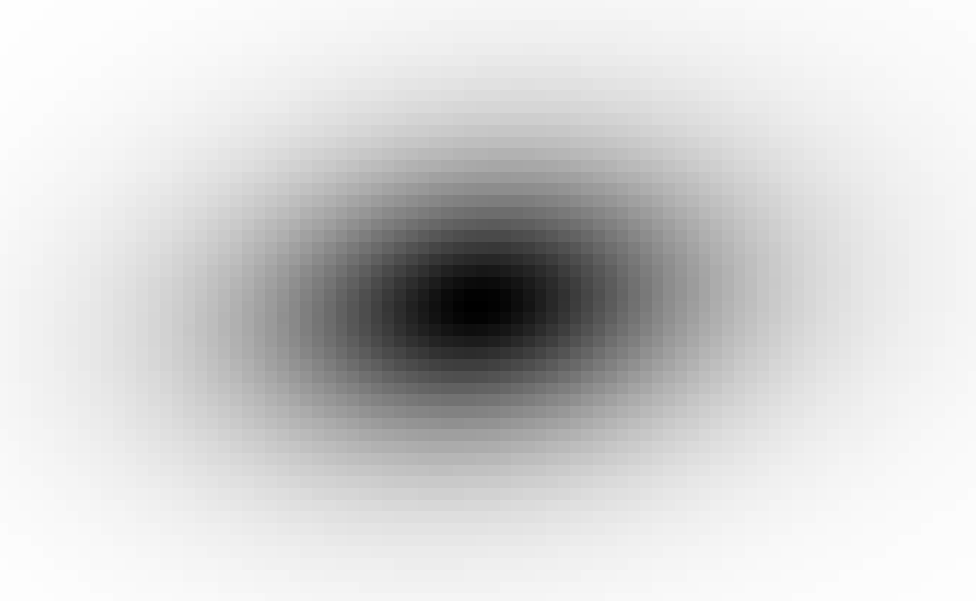}}&
\multicolumn{2}{c}{\includegraphics[width=3cm,height=3cm]{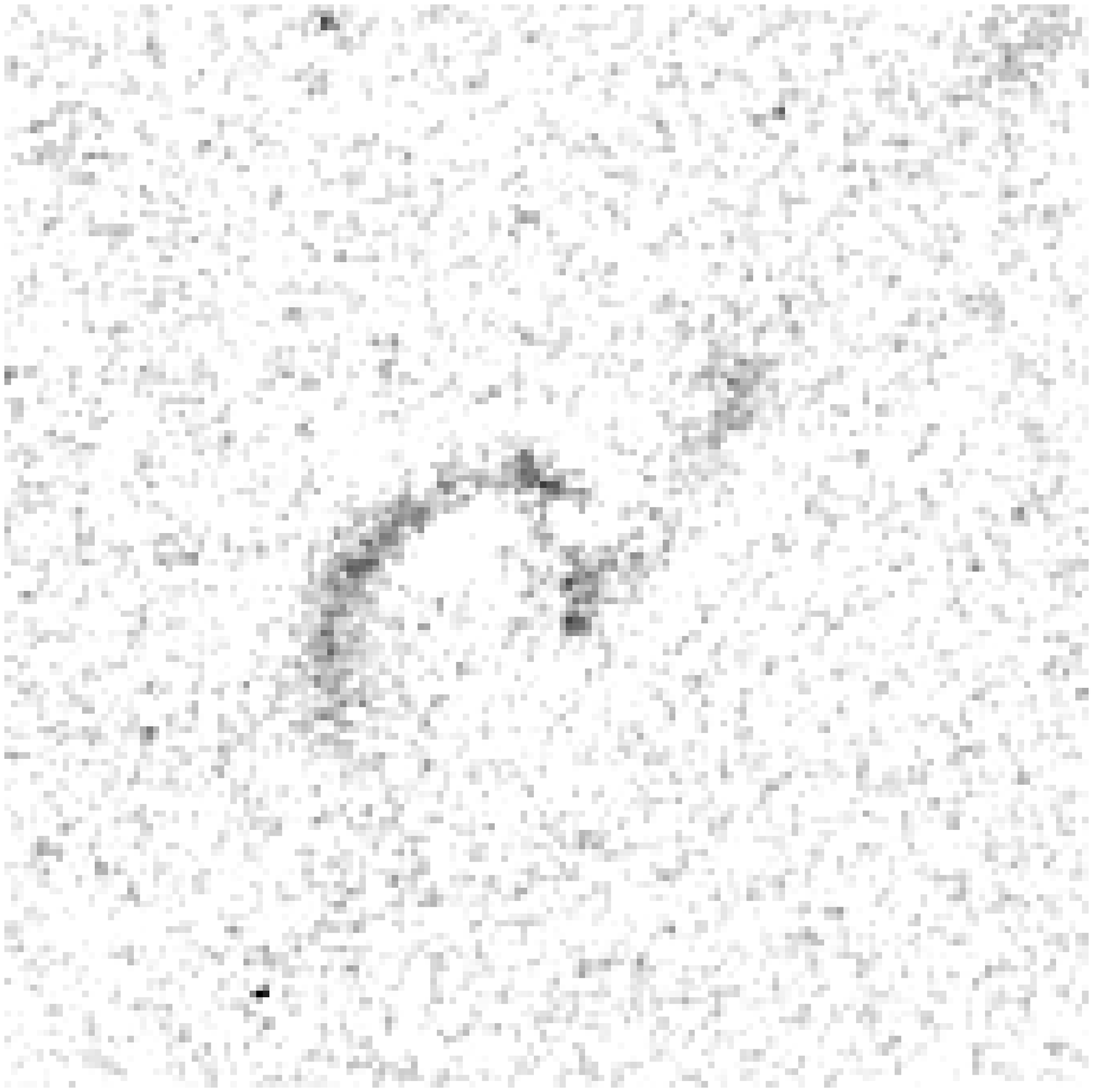}}&
\multicolumn{4}{c|}{}\\
\hline
\multicolumn{10}{c}{continued on next page $...$}\\
\end{tabular}
\end{table*}





\end{document}